\documentclass[useAMS,usenatbib]{mnras}
\bibpunct{(}{)}{;}{a}{}{,} 
\usepackage{graphicx, longtable, lscape, booktabs}
\usepackage{array, color, float, afterpage}
\usepackage{rotating,array,color,float,afterpage}

\usepackage[printwatermark]{xwatermark}

\newcolumntype{L}[1]{>{\raggedright\let\newline\\\arraybackslash\hspace{0pt}}m{#1}}
\newcolumntype{C}[1]{>{\centering\let\newline\\\arraybackslash\hspace{0pt}}m{#1}}
\newcolumntype{R}[1]{>{\raggedleft\let\newline\\\arraybackslash\hspace{0pt}}m{#1}}

\newcommand{\kms}{\,km\,s$^{-1}$}

\DeclareMathAlphabet{\mathsc}{OT1}{cmr}{m}{sc}
\def\testbx{bx}%
\DeclareRobustCommand{\ion}[2]{%
\relax\ifmmode
\ifx\testbx\f@series
{\mathbf{#1\,\mathsc{#2}}}\else
{\mathrm{#1\,\mathsc{#2}}}\fi
\else\textup{#1\,{\mdseries\textsc{#2}}}%
\fi}

\title[The \ion{H}{i} intergalactic medium]{The evolution of the low-density
      \ion{H}{i} {\bf intergalactic medium} from $z$\,$=$\,3.6 to 0: Data, transmitted flux and \ion{H}{i} column 
      density{\thanks{Based on observations made with the NASA/ESA Hubble Space Telescope, obtained at the 
      Space Telescope Science Institute, which is operated by the Association of Universities for Research 
      in Astronomy, Inc., under NASA contract NAS 5-26555.}}
      \thanks{Based on data obtained from the ESO Science Archive Facility under various request numbers.} 
      \thanks{Some of the data presented herein were obtained at the W. M. Keck 
      Observatory, which is operated as a scientific partnership among the California Institute of Technology, 
      the University of California and the National Aeronautics and Space Administration. The Observatory 
      was made possible by the generous financial support of the W. M. Keck Foundation.}
     }
\author[Kim et al.]{T.-S. Kim$^{1, 2}$, B. P. Wakker$^{1}$, F. Nasir$^{3, 4}$,
R. F. Carswell$^{5}$, B. D. Savage$^{1}$,
\newauthor 
J. S. Bolton$^{3}$, A. J. Fox$^{6}$, M. Viel$^{7, 8, 9, 2}$, M. G. Haehnelt$^{5}$, J. C. Charlton$^{10}$ and
\newauthor
B. E. Rosenwasser$^{1}$ \\
$^{1}$ Department of Astronomy, University of Wisconsin, 475 North Charter Street,
       Madison, WI 53706, USA \\
$^{2}$ Osservatorio Astronomico di Trieste, Via G. B. Tiepolo, 11, 34143, Trieste, Italy \\
$^{3}$ School of Physics and Astronomy, University of Nottingham,
University Park, Nottingham NG7 2RD \\
$^{4}$ Department of Physics and Astronomy, University of California Riverside, 900 University Avenue, Riverside, 
   CA 92507, USA \\
$^{5}$ Institute of Astronomy, Madingley Road, Cambridge CB3 0HA \\
$^{6}$ AURA for ESA, Space Telescope Science Institute, 3700 San Martin Drive, Baltimore, MD 21218, USA \\
$^{7}$ SISSA, International School for Advanced Studies, Via Bonomea 265, 34136, Trieste, Italy \\
$^{8}$ INFN, Sezione di Trieste, Via Valerio 2, 34127 Trieste, Italy \\
$^{9}$ IFPU, Institute for Fundamental Physics of the Universe, Via Beirut 2, 34014 Trieste, Italy \\
$^{10}$ 525 Davey Lab, Department of Astronomy and Astrophysics, Penn State University, University Park, PA 16802, USA}

\begin{document}

\date{Draft as of 2020 February 17}

\pagerange{\pageref{firstpage}--\pageref{lastpage}} \pubyear{2020}

\maketitle

\label{firstpage}

\begin{abstract}
  We present a new, uniform analysis of the \ion{H}{i} transmitted flux ($F$)
  and \ion{H}{i} column density ($N_{\mathrm{\ion{H}{i}}}$) distribution
  in the low-density IGM as a function of redshift $z$
  for 0\,$<$\,$z$\,$<$\,3.6 using 55 {\it HST}/COS FUV ($\Delta z$\,$=$\,7.2
  at $z$\,$<$\,0.5), five {\it HST}/STIS+COS NUV ($\Delta z$\,$=$\,1.3 at  $z$\,$\sim$\,1)
  and 24 VLT/UVES and Keck/HIRES ($\Delta z$\,$=$\,11.6 at 1.7\,$<$\,$z$\,$<$\,3.6)
  AGN spectra. We performed a 
  consistent, uniform Voigt profile analysis to combine spectra taken with different 
  instruments, to reduce systematics and to remove metal-line contamination. 
  We confirm previously known conclusions on firmer quantitative 
  grounds in particular by improving the measurements at $z$\,$\sim$\,1.
  Two flux statistics at 0\,$<$\,$F$\,$<$\,1, the mean \ion{H}{i} flux and the 
  flux probability distribution function (PDF), 
  show that considerable evolution occurs from $z$\,$=$\,3.6 to $z$\,$=$\,1.5, 
  after which it slows down to become effectively stable for $z$\,$<$\,0.5. However, 
  there are large sightline variations. For the \ion{H}{i} column density distribution function 
  (CDDF, $f$\,$\propto$\,$N_{\ion{H}{i}}^{-\beta}$) at 
  $\log (N_{\mathrm{\ion{H}{i}}}/1\,{\mathrm{cm}^{-2}})$\,$\in$\,[13.5, 16.0],
  $\beta$ increases as $z$ decreases from $\beta\!=\!1.60$ at $z$\,$\sim$\,3.4 to 
  $\beta$\,$=$\,1.82 at $z$\,$\sim$\,0.1. The CDDF shape at lower redshifts can be 
  reproduced by a small amount of clockwise rotation of a higher-$z$ CDDF with a slightly 
  larger CDDF normalisation. 
 The absorption line number per $z$ ($dn/dz$)
  shows a similar evolutionary break at $z$\,$\sim$\,1.5 as seen in the flux statistics.
  High-$N_{\mathrm{\ion{H}{i}}}$ absorbers evolve more rapidly than low-$N_{\mathrm{\ion{H}{i}}}$ 
  absorbers to decrease in number or cross-section with time.
  The individual $dn/dz$ shows a large scatter at a given $z$. The scatter increases toward
  lower $z$, possibly caused by a stronger clustering at lower $z$. 
\end{abstract}


\section{Introduction}

The small amount of neutral hydrogen (\ion{H}{i}) in the diffuse, 
warm ($\sim\!10^{4}$\,K), 
highly ionised intergalactic medium (IGM) produces a rich series of narrow absorption
lines blueward of the Ly$\alpha$ emission line in the spectra of AGN, also known as 
the Ly$\alpha$ forest\footnote{Although the metal-enriched forest  
likely originates in the circumgalactic medium (CGM), loosely defined as
any gas inside one or two virial radii of galaxies, the metal-free \ion{H}{i}
forest cannot be unambiguously identified as either the IGM or the CGM. 
Following the traditional convention,
we use the ``IGM" to describe any \ion{H}{i} lines with \ion{H}{i} column density
less than $10^{17}$\,cm$^{-2}$ regardless of associated metals.}.
Combined with theory and state-of-art cosmological, hydrodynamic
simulations, the evolution of the Ly$\alpha$ forest
over cosmic time provides some of the most powerful
cosmological and astrophysical constraints as 1) hydrogen is the
most abundant element and a mostly unbiased basic building block of stars and
galaxies, 2) the forest is the largest reservoir of
baryons at all epochs, 3) it traces
the underlying dark matter in a simple manner, thus outlining the
skeleton of the large-scale structure, 4) its thermal state provides clues on 
the reionisation history, and 5)
it contains information on galaxy formation and evolution through the
gas infall from the surrounding IGM and galactic feedback
\citep{sargent80, cen94, weymann98, schaye01,
lehner07, dave10, shen12, ford13, danforth16, martizzi19}.

The physics of the Ly$\alpha$ forest is largely determined by 
a combination of the Hubble expansion, the changes in the ionising UV 
background radiation field (UVB)
and the formation and evolution of the 
large-scale structure and galaxies. 
The Hubble expansion cools the gas adiabatically and decreases the gas density and the
recombination rate. This process is fairly well-constrained by the cosmological
parameters from WMAP and Planck observations \citep{jarosik11, planck16}. 

On the other hand,
the UVB assumed to originate primarily from AGN and in some degree
also from star-forming galaxies
photoionises and heats the IGM. If the intensity of the UVB
decreases, the \ion{H}{i} fraction increases.
Unfortunately, the UVB and its evolution
are less well constrained both theoretically and observationally. 
The relative contributions
from AGN and galaxies are poorly known as a function of redshift, in part
since the escape fraction of \ion{H}{i} ionising photons and the amount of dust attenuation of
galaxies is uncertain 
and since the AGN spectral energy distribution including both obscured and unobscured AGN
is poorly constrained. The
process of the photoionisation and recombination of the integrated UV emission
through the clumpy, opaque IGM is also complex
\citep{bolton05, 
faucher08b, haardt12, kollmeier14, khaire19, puchwein19, faucher20}.
At the same time, outflows from star formation and AGN activity change the
dynamical, chemical
and thermal states of galaxy halos and the surrounding IGM, slowing down the gas infall
\citep{dave10, steidel10, suresh15}. 
In addition, structure evolution is expected to create collisionally-ionised
hot gas known
as the Warm-Hot Intergalactic Medium (WHIM) with
temperature $\sim\!10^{5-7}$\,K through gravitational shock heating.
The WHIM becomes a more dominant phase at $z\!<\!1$ and could hide
a large fraction of missing baryons 
\citep{fukugita98, cen99, savage14, haider16}.

All of these physical processes leave their footprints on the evolutionary properties of the
diffuse IGM in the expanding universe through the shape and number of absorption profiles. 
The \ion{H}{i} column density 
$N_{\mathrm{\ion{H}{i}}}$ is determined by a combination of the
neutral fraction of photoionised hydrogen, the gas density and the UVB, while 
the absorption line width constrains the 
temperature and non-thermal turbulent motion of the IGM.

At 1.5\,$<$\,$z$\,$<$\,3.6, the evolution of the Ly$\alpha$ forest is well-established observationally 
from the Voigt profile fitting analysis of high-resolution and high signal-to-noise (S/N) 
ground-based 
optical QSO spectra taken with instruments such as the HIRES (HIgh-Resolution Echelle Spectrometer,
\citet{vogt94, vogt02})
on Keck I and the UVES (UV-Visible Echelle Spectrograph, \citet{dekker00})
on the VLT (Very Large Telescope), as the \ion{H}{i} absorption lines
at $N_{\mathrm{\ion{H}{i}}}\!\le\!10^{17}$\,cm$^{-2}$ are usually fully
resolved.    

\begin{figure}
\includegraphics[width=8.5cm]{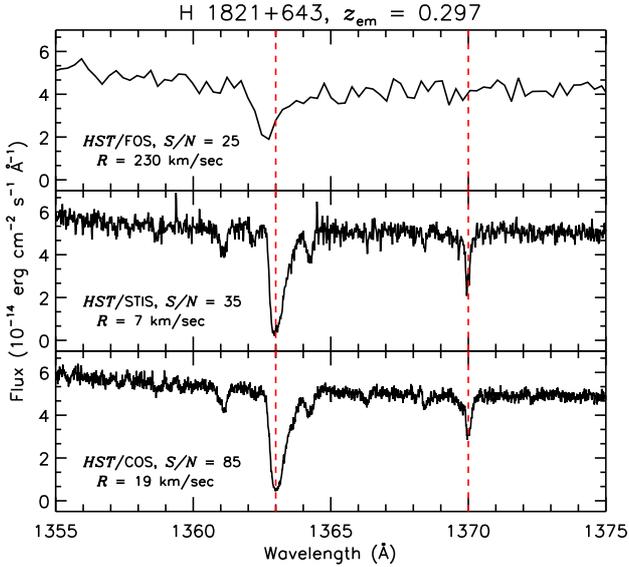}
\vspace{-1.0cm}
\caption{Comparisons of H\,1821+643 spectra taken with {\it HST}/FOS (upper panel,
from the {\it HST} archive for high-level products),
{\it HST}/STIS (middle panel, \citet{wakker09}) and {\it HST}/COS (lower panel, taken at
Lifetime Position 1). 
The low-resolution FOS spectrum does not show an
asymmetric profile of the \ion{H}{i} Ly$\alpha$ at
1363\,\AA\/ as convincingly as STIS E140M and COS G130M 
spectra. Two weak absorption lines with $\log N_{\mathrm{\ion{H}{i}}}\!\sim\!12.65$ 
are clearly present at 1366.2\,\AA\/ and 1368.4\,\AA\/ 
in the higher-S/N COS spectrum.
Being usually much narrower than \ion{H}{i}, most metal lines 
are not resolved at the COS resolution, as seen in the profiles of
Galactic ISM \ion{Ni}{ii} $\lambda$1370.13 from STIS and COS. 
}
\label{fig1}
\end{figure}

At $z$\,$<$\,1.5, the \ion{H}{i} Ly$\alpha$ can be observed
only in the UV region from space due to the atmospheric cutoff at
$\sim$3050\,\AA\/. 
Before the installation of COS (Cosmic Origins Spectrograph) onboard
{\it HST} in 2009, 
the low sensitivity of available UV spectrographs such as {\it HST}/STIS (Space Telescope Imaging
Spectrograph) had seriously limited the sample size and data quality, hindering
a consistent analysis of the IGM combined at $z$\,$>$\,1.5 from optical data and at 
$z$\,$<$\,1.5 from UV data
\citep{weymann98, janknecht06, lehner07}. 
With its factor of $\sim$10 higher throughput than STIS, COS has opened a new era 
for the low-$z$ IGM study
from a unprecedented large number of good-quality AGN spectra \citep{danforth16}.
Although the COS G130M/G160M grating
has a factor of 3 lower resolution ($\sim$\,19\,\kms\/) than the UVES/HIRES resolution, 
most low-$z$ \ion{H}{i} lines are 
resolved at the COS resolution (Fig.~\ref{fig1}) and line blending is not as problematic 
as at $z$\,$>$\,1.

Here in the first of a series from our ongoing observational study on the redshift evolution 
of the IGM,
we present the properties of the transmitted flux $F$ at 0\,$<$\,$F$\,$<$\,1
and \ion{H}{i} column density
$N_{\mathrm{\ion{H}{i}}}$ at $N_{\mathrm{\ion{H}{i}}}$\,$\in$\,[13.5, 17]
of the low-density intergalactic \ion{H}{i} 
from $z$\,$=$\,3.6 to $z$\,$=$\,0, i.e. since the universe was 1.8 Gyr old. 
We constructed a high-quality IGM sample from three public archives:
55 {\it HST}/COS FUV G130M/G160M AGN
spectra covering the Ly$\alpha$ forest
at $z$\,$<$\,0.47, two QSO spectra from the {\it HST}/STIS E230M archive supplemented
with our new observations of three QSOs with the {\it HST}/COS NUV G225M grating at 
$z$\,$\sim$\,1 
and 24 VLT-UVES/Keck I-HIRES QSO spectra 
at 1.7\,$<$\,$z$\,$<$\,3.6\footnote{Being the most powerful subclass of AGN,
QSOs are the only AGN observable at high redshifts. On the other hand,
the COS data set includes all the AGN subclasses including Seyfert galaxies.}.

We have performed our own consistent, uniform in-depth Voigt profile fitting 
analysis to the three data sets, instead of compiling fitted line parameters from
literature, cf. \citet{tilton12}. Although time-intensive, this approach is the only viable option to 
reduce any systematics, to account for
the different spectral characteristics of each spectrograph and to remove 
metal contamination. One of our primary aims is to provide the fundamental
measurements of the low-density IGM from the self-consistent analysis for theoreticians
to test cosmological simulations and theories on structure/galaxy formation and evolution.

We produced two sets of the fitted parameters:
one using only the Ly$\alpha$ (the Ly$\alpha$-only fit) as most simulations 
use the Ly$\alpha$ forest region and another using all the available
Lyman series (the Lyman series fit) 
to derive reliable
line parameters of saturated Ly$\alpha$ lines.
Although the redshift coverage is not continuous and the sample
size at $z$\,$\sim$\,1 is rather small,
the analysed redshift range is the best compromise within the capabilities of
currently available ground-based and space-based spectrographs.

This paper is organised as follows.
Our data sets are presented in Section~\ref{sect2}. 
The Voigt profile fitting technique
and its caveats are discussed in Section~\ref{sect3}. The \ion{H}{i} continuous flux
statistics are found in Section~\ref{sect4}. The distribution of \ion{H}{i} column densities is 
discussed in Section~\ref{sect5}. 
We summarise our results in Section~\ref{sect6}. All the long tables are published 
electronically on the MNRAS webpage.  
Throughout this study, the cosmological parameters are assumed to be the matter
density $\Omega_{\mathrm{m}}$\,$=$\,0.3, the cosmological constant
$\Omega_{\mathrm{\Lambda}}$\,$=$\,0.7, and the
current Hubble constant $\mathrm{H_{0}}$\,$=$\,100\,$h$\,\kms\/Mpc$^{-1}$
with $h$\,$=$\,0.7.
The logarithm $N_{\ion{H}{i}}$ is defined as
$\log N_{\ion{H}{i}}$\,$=$\,$\log (N_{\ion{H}{i}} / 1\,{\mathrm{cm}}^{-2})$.
All the quoted S/N ratios are per resolution element.
The atomic parameters are taken from the
atomic parameter file in the Voigt profile fitting package
VPFIT \citep{carswell14}, 
with some unlisted values from the NIST (National Institute of Standards and Technology)
Atomic Spectra Database.
We also use the terms ``absorbers", ``components" and ``absorption lines" 
interchangeably.

\section{Data}
\label{sect2}

\subsection{General description of the analysed data}

The most physically meaningful analysis of absorption spectra is to 
decompose absorption lines into discrete components to derive 
column densities and line widths, assuming 
the profile shape to be the Voigt function. The commonly used
curve-of-growth analysis from the equivalent width measurement
is straightforward with the mathematically well-characterised
associated error \citep{ebbets95}. However, its derived column density is
degenerate with the absorption line width for a single-line transition, such as 
typical IGM \ion{H}{i} Ly$\alpha$ with $\log N_{\ion{H}{i}}$\,$\le$\,13.5 for which Ly$\beta$ 
cannot be detected in COS spectra with $S/N$\,$\le$\,25. Since about 60\% of
IGM \ion{H}{i} lines with $\log N_{\ion{H}{i}}$\,$\in$\,[13, 15] at $z$\,$\sim$\,0.2
have $\log N_{\ion{H}{i}}$\,$\le$\,13.5, inability of constraining the line width, thus
the column density in some degree, is a serious
drawback of the curve-of-growth analysis. Moreover, deblending of absorption
complexes is not straightforward
in the curve-of-growth analysis. High-$z$ IGM spectra suffer from severe blending 
and measuring the equivalent width in high-resolution UVES/HIRES spectra is 
almost impossible and meaningless.

The Voigt profile fitting analysis requires high-quality spectra in which 
absorption lines are resolved and deblending is possible.  
In order to achieve a data quality adequate enough for the profile fitting analysis,
we have built the three IGM data sets by selecting 
good-quality AGN spectra publicly available as of the 
end of 2017 from {\it HST}, {\it FUSE}, VLT and Keck archives.
Due to the rapid increase of the number of absorption lines with $z$, 
it is essential to have high-resolution, high-S/N spectra that allow for deblending 
at $z$\,$>$\,1.5. At lower redshifts, high resolution
is not as crucial due to much less blending, but a high S/N is still 
required to place a reliable
continuum and to obtain robust
fitted line parameters. Our main AGN selection criteria are:

\begin{enumerate}

\item Sightlines without damped
Ly$\alpha$ systems (DLA, $\log N_{\mathrm{\ion{H}{i}}}$\,$\ge$\,20.3) in the
Ly$\alpha$ forest region 
and only a few Lyman limit systems ($\log N_{\mathrm{\ion{H}{i}}}$\,$\ge$\,17.2)
in the entire spectrum in order to maximise useful 
wavelength regions. 

\item Spectra covering higher-order \ion{H}{i} Lyman lines, at least 
Ly$\beta$, to obtain a reliable line parameter for saturated Ly$\alpha$ lines.
Available {\it FUSE} spectra were included to cover the corresponding Lyman 
series of COS Ly$\alpha$. 

\item For COS FUV, STIS and UVES/HIRES spectra, the S/N cut is 
set to be $\ge$\,18, $\ge$\,18 and
$\ge$40 per resolution element in a large fraction of forest regions. 
This rather arbitrary S/N cut is a compromise between having well measurable
lines and as large a sample as possible.

\item To increase the sample size at $z$\,$\sim$\,1, we relax the S/N cut and include 
our three new COS NUV QSO spectra obtained through HST GO program 14265.
Two have $S/N$\,$\sim$\,15--18, while
one has  $S/N$\,$\sim$\,10--15. 
Since high-order Lyman regions of the three sightlines are only partly observed,
we use these spectra only for the Ly$\alpha$-only analysis. The lower-S/N
increases the lowest reliable value for $N_{\ion{H}{i}}$ 
and leads to a unreliable measurement of
the transmitted flux (see Section~\ref{sect4.2}). However, including
the three COS NUV spectra does not change our conclusions. 

\item For COS FUV/NUV spectra, a region with S/N lower than each S/N cut 
is discarded if it is longer than $\sim$5\,\AA\/, so as not to compromise
the reliable Voigt profile fitting and flux statistics. 

\item The forest region with $S/N$\,$>$\,18 of
the COS FUV spectra is required to be $\ge$\,100\,\AA\/ wide. This limits the emission redshift to be
$z_{\mathrm{em}}$\,$>$\,0.1, for which the possible forest coverage is 
$\ge$\,120\,\AA. Considering that the forest is 
$\sim$550\,\AA\/ long at $z$\,$\sim$\,2.5, such a small wavelength coverage 
makes cosmic variance a major issue. 
To avoid confusion with high-order Lyman lines in the
FUV spectra, the maximum forest $z$ is set to be 0.47.

\item No broad absorption line (BAL) AGN. 
Mini-BALs are included with the affected wavelength region excluded.  

\end{enumerate}

\begin{figure}

\includegraphics[width=8.5cm]{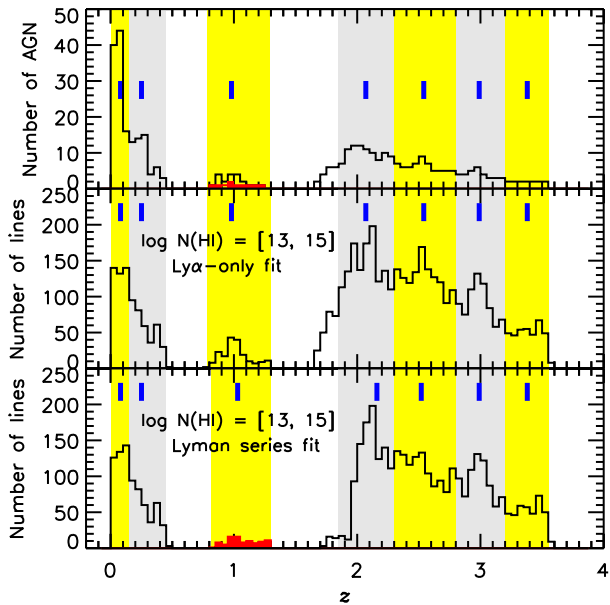}\\

\vspace{-0.5cm}

\caption{{\it Upper panel:} Number of AGN covering the Ly$\alpha$ forest using
a $\Delta z$\,$=$\,0.05 bin. Any
excluded regions due to a low S/N, Milky Way ISM contamination or 
detector gap are not counted in the number of AGN and IGM \ion{H}{i}
lines.
The red histogram is from the two STIS E230M spectra used for both Ly$\alpha$-only
and Lyman series fits. The thick vertical blue lines mark
the median $z$ of each redshift bin used for the Ly$\alpha$-only fit. Each redshift bin
is shaded as gray and yellow alternatively to clearly distinguish from each other.
These bins are chosen to
exclude the $z$ range at which the Lyman series fit does not contain enough lines.
{\it Middle and lower panels:} Number of \ion{H}{i} lines from the Ly$\alpha$-only
and Lyman series fits, respectively.}
\label{fig2}
\end{figure}

The COS FUV (1100--1800\,\AA), COS NUV (2225--2525\,\AA), and 
STIS NUV E230M (1850--3050\,\AA) 
spectra contain many Galactic ISM lines, 
such as \ion{Si}{ii} $\lambda\lambda$\, 1260.42, 1304.37, 1526.70,
\ion{C}{ii} $\lambda$\,1334.53, \ion{Mg}{ii} $\lambda\lambda$, 2796.35, 2803.53,
and \ion{Fe}{ii} $\lambda\lambda$\,1608.45, 2382.76, 2600.17. 
The profile fit can easily reveal typical IGM lines blended with ISM lines, if multiple transitions
of the same ISM ion are available and if some of the clean transitions are not saturated.
However, broad and/or weak blended IGM lines can not be always validated when the
spectrum has a low resolution, low S/N or fixed pattern noise.
The most noticeable ISM line in the NUV region of our interest is multiple \ion{Fe}{ii} including
non-saturated transitions so that blended IGM lines above the detection limit 
are easily detected. 
However, the COS FUV region contain many single/multiple ISM lines as well as
geocoronal emission lines. Therefore, 
regions contaminated with strong and medium-strength ISM lines are excluded in our IGM study, 
regardless of available multiple transitions of the same ion.

Our final sample consists of 24 UVES/HIRES QSOs covering the forest at 1.67\,$<$\,$z$\,$<$\,3.56
with the total analyzed $z$ range $\Delta z$\,$=$\,11.6, 
five STIS E230M and COS NUV QSOs at 0.76\,$<$\,$z$\,$<$\,1.30 with $\Delta z$\,$=$\,1.3
and 55 COS FUV AGN at 0.00\,$<$\,$z$\,$<$\,0.47 with $\Delta z$\,$=$\,7.2. 
The upper panel of Fig.~\ref{fig2} shows the number of AGN per unit $z$. 
The thick vertical line notes the median redshift of the seven redshift bins used in this 
study from the Ly$\alpha$-only fit:
$z$\,$\in$\,[0.00, 0.15] ($\tilde{z}$\,$=$\,0.08), [0.15, 0.45] ($\tilde{z}$\,$=$\,0.25), 
[0.78, 1.29] ($\tilde{z}$\,$=$\,0.98), [1.85, 2.30] ($\tilde{z}$\,$=$\,2.07), [2.30, 2.80] ($\tilde{z}$\,$=$\,2.54),  
[2.80, 3.20] ($\tilde{z}$\,$=$\,2.99) and  [3.20, 3.55] ($\tilde{z}$\,$=$\,3.38),
respectively. Median 
redshift of the seven redshift bins from the Lyman series fit is
slightly different as this requires a coverage of the higher-order Lyman lines:
$z$\,$\in$\,[0.00, 0.15] ($\tilde{z}$\,$=$\,0.08), [0.15, 0.45] ($\tilde{z}$\,$=$\,0.25), 
[0.82, 1.29] ($\tilde{z}$\,$=$\,1.03), [1.85, 2.30] ($\tilde{z}$\,$=$\,2.12), [2.30, 2.80] 
($\tilde{z}$\,$=$\,2.52),  
[2.80, 3.20] ($\tilde{z}$\,$=$\,2.99) and  [3.20, 3.55] ($\tilde{z}$\,$=$\,3.38),
respectively. At $z$\,$>$\,1.5,  a sightline with less than $\sim$100\,\AA\/-long 
in a redshift bin is excluded to reduce a sightline variation, since each $z$ bin samples
a wavelength range with $\ge$400\,\AA\/.
The middle and lower panels show the number of \ion{H}{i} lines at
$\log N_{\mathrm{\ion{H}{i}}}\!\in\![13, 15]$ from
the Ly$\alpha$-only fit and Lyman series fits, respectively. 
The steep decrease of the number of \ion{H}{i} lines from the Lyman series fit at $z$\,$\sim$\,1.95 
is caused by the atmospheric cutoff at 3050\,\AA\/ in the optical spectra without a corresponding
UV spectrum.

\begin{table*}
\caption{Analysed UVES/HIRES QSOs}
\label{tab1}
{\small{
\begin{tabular}{llccccccc}
\hline
\noalign{\smallskip}
QSOs & $z_{\mathrm{em}}^{a}$ & $z_{\mathrm{Ly\alpha}}$ & 
    $\lambda\lambda_{\mathrm{Ly\alpha}}$ &
    $\lambda\lambda_{\mathrm{Ly\alpha\beta}}^{\mathrm{b}}$ &
    S/N$^{\mathrm{c}}$ &  
    $\Delta X_{\mathrm{Ly\alpha}}$ & $\Delta X_{\mathrm{Ly\alpha\beta}}^{\mathrm{d}}$ & Instrument \\
     &   &    & (\AA\/)  & (\AA\/) & p.r. &   &     &  \\
\noalign{\smallskip}
\hline
\noalign{\smallskip}

HE\,1341--1020$^{\mathrm{e}}$  & 2.1356$^{\mathrm{f}}$ & 1.667--2.083 &   3242.0--3748.0 &   3609.0--3748.0 & 
      55--90 & 1.2289 & 0.3487 & UVES \\
Q\,1101--264$^{\mathrm{g}}$ & 2.1413 & 1.659--1.795 & 3233.0--3398.0 &   no coverage &
      45--90 &  0.3875 & no value & UVES \\
      &        & 1.882--2.090 &  3503.0--3756.0  &  3607.0--3756.0 & 65--135 & 0.6294 & 0.3740 &  \\
Q\,0122--380    & 2.1895$^{\mathrm{h}}$ & 1.700--2.134 &   3282.0--3810.0 &   3619.0--3810.0 & 
     40--120 & 1.2941 &   0.4821 &  UVES \\
PKS\,1448--232  & 2.2197 &  1.716--2.164 &   3302.0--3846.0 &   3615.0--3846.0 & 
     45--90 & 1.3399 &   0.5845 &  UVES  \\
PKS\,0237--23$^{\mathrm{i}}$  & 2.2219$^{\mathrm{f}}$ & 1.735--2.169 &   3325.0--3853.0 &   3615.0--3853.0 &
     77--137 & 1.3039 &   0.6026 &  UVES \\
J\,2233--6033$^{\mathrm{e}}$ & 2.2505 & 1.741--2.197 &   3332.0--3886.0 &   3332.0--3886.0$^{\mathrm{j}}$ & 
     35--56 &   1.3729 &   &  UVES, STIS$^{\mathrm{j}}$  \\
HE\,0001--2340$^{\mathrm{k}}$  & 2.2641 & 1.752--2.143 &   3346.0--3821.0 &   3622.0--3821.0 & 55--130 &  1.1720 &   0.5028 & UVES \\
Q\,0109--3518    & 2.4047 & 1.873--2.348 &   3492.6--4070.0 &   3615.0--4070.0 & 82--110 &  1.4725 &   1.1720 & UVES  \\
HE\,1122--1648  & 2.4050 & 1.891--2.348 & 3514.0--4070.0 &   3615.0--4070.0 & 80--205 & 1.4205 &   1.1720 & UVES   \\
HE\,2217--2818  & 2.4134 & 1.886--2.355 &   3509.0--4078.2 &   3613.0--4078.2 & 85--140 &  1.4545 &   1.1988 &  UVES  \\
Q\,0329--385      & 2.4350 & 1.896--2.378 &   3521.0--4106.0 &   3617.0--4106.0 & 50--80 &  1.4996 &   1.2632 & UVES \\
HE\,1158--1843$^{\mathrm{e}}$ &  2.4478 & 1.940--2.391 &   3574.5--4122.0 &   3621.0--4122.0 &      &   1.4113 &   1.2962 & UVES \\
HE\,1347--2457  & 2.6261$^{\mathrm{l}}$ & 2.058--2.564 &   3717.5--4333.0 &   &  71--116 & 1.6297 &   & UVES  \\
Q\,0453--423$^{\mathrm{e, m}}$ & 2.6569 & 2.086--2.260 & 3752.0--3962.5 &   & 70--137 &   0.5436 &  &  UVES \\
                       &                                    & 2.347--2.593 & 4069.0--4368.4 &   & 85--151 &    0.8151 &  &  \\
PKS\,0329--255  & 2.7041$^{\mathrm{n}}$ & 2.134--2.642 &   3809.4--4427.0 &  & 40--80 &  1.6574 &  &  UVES  \\
Q\,0002--422    & 2.7676  &  2.183--2.705 &   3870.0--4504.0 &   &  66-145 &    1.7179  &   & UVES  \\
HE\,0151--4326$^{\mathrm{e}}$  & 2.7810 & 2.206--2.631 &   3897.0--4414.0 &   &  95--170 &  1.3949 &  &  UVES \\
HE\,2347--4342$^{\mathrm{e}}$  & 2.8740$^{\mathrm{f}}$ & 2.333--2.812 & 4052.4--4634.0 &    &  188-278 & 1.6098 &  &  UVES \\
HE\,0940--1050  & 3.0836 & 2.452--3.014 & 4197.0--4880.0 &   &  103--145 &  1.9382 &   &  UVES \\
Q\,0420--388${\bf ^{\mathrm{o}}}$      & 3.1152${\bf ^{\mathrm{p}}}$ & 2.480--3.044 &   4231.0--4916.0 &  4455.0--4916.0 & 103--210 & 1.9523  & 1.3321 &  UVES  \\
Q\,0636+6801     & 3.1752 & 2.525--3.097 &   4285.0--4981.0 &  4532.0--4981.0 &  65--105 & 1.9981 & 1.3084 &  HIRES  \\
PKS\,2126--158   & 3.2796 & 2.684--3.208 &   4479.0--5115.0 &   & 100--250 & 1.8618 &  & UVES  \\
Q\,1422+2309      & 3.6288 & 2.919--3.552 & 4764.0--5533.3 &  &  122--165 &  2.3412 &  &  HIRES  \\
Q\,0055--269       & 3.6563 & 2.936--3.562 & 4785.0--5546.0 &  &  80--140 &  2.3201 &  &  UVES  \\

\noalign{\smallskip}
\hline
\end{tabular}
}}

\begin{flushleft}
{\small
{\setlength{\itemsep}{1pt}
{\bf Notes -- a:} The redshift is measured from the observed Ly$\alpha$ emission line of the QSO.
{\bf b:} The Ly$\alpha$ forest region having a corresponding Ly$\beta$. When left blank,
    it is the same as $\lambda\lambda_{\mathrm{Ly\alpha}}$.
{\bf c:} S/N per resolution element.
{\bf d:} The absorption line path length corresponding the Ly$\alpha\beta$ forest region.
    When left blank, it is the same as $\Delta X_{\mathrm{Ly\alpha}}$.
{\bf e:} Mini-BAL (broad absorption line) QSO. 
{\bf f:} Due to the intrinsic absorbers around the Ly$\alpha$ emission line of
the QSO, the redshift is less accurate.
{\bf g:} A sub-DLA at $z\!=\!1.839$ in the Ly$\alpha$ region.
{\bf h:} The emission feature
is rather flat, in addition to several intrinsic absorption lines. The redshift is set to be the position of the highest flux. 
{\bf i:} A sub-DLA at $z\!=\!1.673$ in the Ly$\beta$ region.
{\bf j:} The publicly available, science-ready STIS E230M spectrum \citep{savaglio99}
   covers a high-order Lyman region at 2550--3057\,\AA.
{\bf k:} A sub-DLA at $z\!=\!2.187$ in the Ly$\alpha$ region.
{\bf l:} The Ly$\alpha$ emission is slightly double-peaked. The redshift is set to the wavelength
   of the highest flux. 
{\bf m:} A sub-DLA at $z\!=\!2.305$ in the Ly$\alpha$ region.
{\bf n:} The emission feature
is very flat with several intrinsic absorption lines. The redshift is set to be the center of the flat emission feature.
{\bf o:} A sub-DLA at $z\!=\!3.087$ in the Ly$\alpha$ region causes the flux to be zero at $\le$3754\,\AA.
{\bf p:} As the right wing of the sub-DLA at $z\!=\!3.087$
   covers the Ly$\alpha$ emission feature in addition to several intrinsic absorbers, the redshift is not accurate.
}}
\end{flushleft}
\end{table*}

All the analysed spectra are in the heliocentric velocity frame. 
In order to avoid the proximity effect,
the region of 5,000 \kms\/ blueward of the Ly$\alpha$ emission 
was excluded. When a sub-DLA with 
$\log N_{\mathrm{\ion{H}{i}}}$\,$\in$\,[19.0, 20.3] is present in the Ly$\alpha$
forest region, a region of $\pm$50\,\AA\ centred at the sub-DLA was
discarded, as the low-density \ion{H}{i} around sub-DLAs is not likely to represent
the typical IGM due to a strong influence by the galaxy producing the sub-DLA.


\subsection{UVES and HIRES data}

Table~\ref{tab1} lists the 24 QSOs observed with the UVES at the VLT or
with the HIRES at Keck I, along with their emission
redshift, analysed absorption redshift ranges and S/N per
resolution element in the Ly$\alpha$ forest region. 
The UVES spectra are the same ones analysed by \citet{kim07, kim13, kim16a},
while the HIRES spectra are the same ones described by \citet{boksenberg15}.
The UVES and HIRES spectra were sampled at 0.05\,\AA\/ and 0.04\,\AA\/,
respectively. Their resolution is about 6.7\,\kms\/. 
Although the S/N differs from QSO to QSO and even varies along the same QSO, 
the practical $N_{\mathrm{\ion{H}{i}}}$ detection limit is 
$\log N_{\mathrm{\ion{H}{i}}} \!\sim$\,12.5.

Table~\ref{tab1} also lists the absorption distance path length,
$\Delta X$, which accounts for 
comoving coordinates at a given $z$ for the adopted cosmology:
\begin{equation}
\label{eq1}
\Delta X = \int dX = \int \frac{H_{0}}{H(z)} (1+z)^2 dz,
\end{equation}

\noindent
where $H(z) = 100\, h\, [\Omega_{\mathrm{m}} (1+z)^{3} +
(1 - \Omega_{\mathrm{m}} - \Omega_{\mathrm{\Lambda}}) (1 + z)^{2}
+ \Omega_{\mathrm{\Lambda}}]^{\frac{1}{2}}$ \citep{bahcall69}.

\subsection{{\it HST}/STIS data}


\begin{table*}
\caption{Analysed STIS/COS NUV QSOs}
\label{tab2}
{\small{
\begin{tabular}{llc cc ccccl}
\hline
\noalign{\smallskip}
QSOs & $z_{\mathrm{em}}^{\mathrm{a}}$ & $z_{\mathrm{Ly\alpha}}^{\mathrm{b}}$ &
    $\lambda\lambda_{\mathrm{Ly\alpha}}^{\mathrm{b}}$ & 
    $\lambda\lambda_{\mathrm{Ly\alpha\beta}}^{\mathrm{c}}$ & 
    Resolving & S/N &
    $\Delta X_{\mathrm{Ly\alpha}}^{\mathrm{d}}$ & Inst & Program ID \\
     &    &   &   (\AA\/)  &  (\AA\/) &  power  & p.r.  &   &    &      \\
\noalign{\smallskip}
\hline
\noalign{\smallskip}

PG\,1718+481   &  1.0832  & 0.783--1.047 & 2167.0--2489.0 & 2207.0--2489.0 & 30,000 &
    18--26 & 0.5793 (0.5113) & STIS & 7292 \\
HE\,1211$-$1322 & 1.121 & 0.835--1.076 & 2231.0--2524.0 & no coverage & 24,000 & 10-15 & 0.5113    &  COS & 14265 \\
HE\,0331$-$4112 & 1.124 &  0.832--1.076 & 2226.5--2524.0 & no coverage & 24,000 & 13--18 & 0.4935   &  COS & 14265 \\
HS\,2154+2228 & 1.298 & 0.831--1.076 & 2225.5--2524.0 & no coverage & 24,000 & $\sim$18 & 0.5172  &  COS & 14265 \\
PG\,1634+706  &  1.3340 & 0.981--1.295 & 2402.5--2789.0 &  2402.5--2789.0  & 30,000 &
    34--46 & 0.7612 (0.7612) & STIS & 7292/8312\\

\noalign{\smallskip}
\hline
\end{tabular}
}}

\begin{flushleft}
{\small \setlength{\itemsep}{1pt}
{\bf Notes -- a:} The redshift with a four decimal place is measured from the Ly$\alpha$ emission line of the QSO,
while the one with a three decimal place is from Simbad. 
{\bf b:} The Ly$\alpha$ forest region. 
{\bf c:} The Ly$\alpha$ forest region covering the corresponding Ly$\beta$. The COS NUV spectra are used only
for the Ly$\alpha$-only fit. 
{\bf d:} The number in parentheses is $\Delta X$ for the Ly$\alpha\beta$ region. The excluded region due to
a very-low S/N of the COS NUV spectra are taken into account.
}
\end{flushleft}
\end{table*}


Due to the low efficiency of STIS E230M, the {\it HST} archive offers only one good-quality
AGN spectrum covering the forest at $z$\,$\sim$\,1, QSO PG\,1634+706.
The spectrum has $S/N$\,$\sim$\,40, comparable to UVES/HIRES data. In order to increase 
our sample at $z$\,$\sim$\,1,  
PG\,1718+481 with the second highest S/N ($\sim$20) is also included (Table~\ref{tab2}).
These spectra are same as those analysed by \citet{wakker09}.
The resolution is $\sim$\,10\,\kms\/, if
the slightly non-Gaussian line spread function (LSF) is approximated as a Gaussian (see more details 
in Section~\ref{sect3.2}). The typical detection limit is $\log N_{\mathrm{\ion{H}{i}}}$\,$\sim$\,13.0.
The pixel size of the final combined STIS spectra continuously increases toward longer wavelengths,
$\sim$0.034\,\AA\/ per pixel at $\sim$2100\,\AA\/ and
$\sim$0.039\,\AA\/ per pixel at $\sim$2400\,\AA\/.

\subsection{{\it HST}/COS NUV data}
\label{sect2.3}

The three selected QSOs observed with the COS NUV G225M grating are part of our observing program 
(HST GO 14265) to study the IGM at $z$\,$\sim$\,1 (Table~\ref{tab2}). The observations were
obtained in TIME-TAG mode in 2015--2016. The central wavelength setting was setup to produce
a continuous wavelength coverage at $\sim$2226--2524\,\AA\/. To increase the S/N of individual 
extractions, we ran the COS data reduction pipeline CalCOS version 3.3.4 with a 12-pixel-wide 
extraction box instead of the CalCOS default 57-pixel extraction box. 

Coadding mis-aligned absorption lines due to wavelength
calibration errors produces absorption lines artificially broader and smoother.
While UVES, HIRES and STIS have a wavelength uncertainty less than 1\,\kms\/,
the CalCOS wavelength calibration uncertainty is quoted as $\sim$15\,\kms\/ 
\citep{dashtamirova19}. 
In general, the Cal-COS wavelength uncertainty tends to vary 
with wavelength and becomes larger at the edges of detector segments.
A custom-built semi-automatic IDL program was developed to improve the CalCOS
wavelength calibration and to coadd the individual CalCOS extractions 
\citep[see their Appendix for details on the COS wavelength re-calibration procedure]{wakker15}. 
We first recalibrate the CalCOS wavelength on a relative scale 
better than $\sim$5\,\kms\/ between the same absorption features 
by cross-correlating 
the strong, clean Galactic ISM or IGM lines in all the available, individual extractions of the same QSO
in the {\it HST} COS/STIS and {\it FUSE} archives. 
The absolute wavelength calibration was further performed using Galactic 21 cm emission toward
the QSO by aligning this with the interstellar lines \citep{wakker15}.
Since the majority of individual extractions have low
S/N, it is not always straightforward to align weak/moderate-strength lines in 
the presence of fixed pattern noise, with the wavelength calibration uncertainty
at 5--10\,\kms. For strong lines, our wavelength
recalibration has uncertainty better than 5\,\kms\/ in general.
However, when absorption lines fall on the edge of the COS detector, their
wavelength uncertainty can be at 10--15\,\kms\/ occasionally.

The final coadded spectrum is sampled at $\sim$0.034\,\AA\/ per pixel, slightly smaller at longer
wavelengths. The resolution is 
$\sim$12\,\kms\/ with a time-independent non-Gaussian LSF. While the non-Gaussian LSF has
an extended wing, the FWHM ($\sim$10.5\,\kms) at the core is comparable to the one of STIS spectra. 
Unfortunately, the two QSOs, HE\,1211$-$1322 and HE\,0331$-$4112, 
had become fainter at the time of observations compared to earlier low-resolution spectra,
causing a lower S/N than the expected 
$S/N\!\sim\,18$. The region having a much lower S/N than quoted in Table~\ref{tab2} is discarded
to keep the spectral quality as high as the data allow. 
The typical COS NUV $N_{\mathrm{\ion{H}{i}}}$ limit is $\log N_{\mathrm{\ion{H}{i}}}\!\sim\!13.0$.

The Ly$\beta$ region and the higher-order Lyman regions are not observed 
or are only observed in part by other UV spectrographs. Since one of the selection criteria 
is the coverage of the Ly$\beta$ forest, 
the three NUV G225M spectra are only used for the Ly$\alpha$-only analysis.

\begin{table*}
\caption{Analysed COS FUV AGN}
\label{tab3}

{\scriptsize{
\begin{tabular}{l ccc l ccc cl}

\hline \\[-0.2cm]

AGN &  $z_{\mathrm{em}}^{\mathrm{a}}$ &  $z_{\mathrm{Ly\alpha}},^{\mathrm{b}}$ & 
   $\lambda\lambda_{\mathrm{Ly\alpha}} (\lambda\lambda_{\mathrm{Ly\alpha\beta}})^{\mathrm{b}}$ & 
   Others$^{\mathrm{c}}$ & Excluded$^{\mathrm{d}}$ & S/N$^{\mathrm{e}}$ & $\Delta X_{\mathrm{Ly\alpha}}^{\mathrm{f}}$ & 
   LP$^{\mathrm{g}}$ & Prog.  \\
   &   & $z_{\mathrm{Ly\alpha\beta}}$ &  (\AA\/) &  &  region (\AA\/) & p.r. & $\Delta X_{\mathrm{Ly\alpha\beta}}$ &  & ID \\[2pt]

\hline

PKS\,2005--489  & 0.0711  & 0.003--0.053 & 1219.0--1280.5 & F(17), D16 &   & 26--31 & 0.0505 & LP1 & 11520 \\
PG\,0804+761    & 0.1002  & 0.002--0.082 & 1218.0--1315.0 & F(28), D16 &   & 45--60 & 0.0804 & LP1 & 11686 \\
RBS\,1897          & 0.1019  & 0.003--0.083 & 1219.0--1317.0 & F(11) &   & 31--53 & 0.0812 & LP1 & 11686 \\
1H\,0419--577    & 0.1045  & 0.003--0.086 & 1219.0--1320.0 & F(9), D16 &  & 33--86 & 0.0822 & LP1 & 11686, 11692 \\
PKS\,2155--304$^{\mathrm{h}}$ & $\sim$0.1103$^{\mathrm{i}}$ & 0.003--0.092 & 1219.0--1327.0 & F(38), D16 &  & 30--42 & 0.0909 & LP2 & 12038 \\
Ton\,S210           & 0.1154 & 0.002--0.096 & 1218.0--1332.5 &  F(27), D16 &   & 35--50 & 0.0943 & LP1 & 12204  \\
HE\,1228+0131 & 0.1168$^{\mathrm{j}}$  & 0.002--0.097 & 1218.0--1333.5  &  F(7), D16 &   & 40--72 & 0.0979 & LP1 & 11686  \\
Mrk\,106            & 0.1233  & 0.003--0.105 & 1219.0--1343.0  &  F(10), D16 &  &  22--33 & 0.1006 & LP1 & 12029 \\
IRAS\,Z06229--6434 & 0.1290 & 0.003--0.110 & 1219.0--1349.5  &  F(7), D16 & 1272.3--1292.0  & 30--37 & 0.0866 & LP1 & 11692  \\
Mrk\,876            & 0.1291 & 0.002--0.110 & 1218.0--1350.0 & F(35), D16 &     & 58--62 & 0.1093 & LP1 & 11686, 11524 \\
PG\,0838+770  &  0.1312 & 0.003--0.112 & 1219.5--1352.0 & F(10), D16 &    & 21--40 & 0.1094 & LP1 & 11520 \\ 
PG\,1626+554  & 0.1316 &  0.002--0.113 & 1218.0--1353.0 & F(15), D16 &    & 20--35 & 0.1109 & LP1 & 12029  \\
RX\,J0048.3+3941 & 0.1344 & 0.003--0.115 & 1219.0--1356.0 & F(20), D16 &  & 20--36 & 0.1136 & LP1 & 11686 \\
PKS\,0558--504 &  0.1374 & 0.002--0.118 & 1219.0--1359.0 &  F(25) & 1273.4--1300.4 & 18--23 & 0.0926 & LP1 & 11692 \\
PG\,0026+129$^{\mathrm{h}}$ & 0.1452 & 0.003--0.126 & 1219.0--1369.0 & F(7), D16 & 1270.6--1300.5 & 18--23 & 0.0980 & LP1 & 12569 \\
PG\,1352+183  & 0.1508 & 0.002--0.131 & 1218.0--1375.5 & F(4) & 1273.1--1291.0 & 20--37 & 0.1191 & LP2 & 13448 \\
PG\,1115+407 & 0.1542$^{\mathrm{j}}$ & 0.002--0.135 & 1218.0--1380.0 & F(3), D16 &   & 20--34 & 0.1371 & LP1 & 11519 \\
PG\,0052+251 &  0.1544 & 0.003--0.134 & 1219.0--1379.0 & F(3) &   & 19--33 & 0.1371 & LP3 & 14268 \\
PG\,1307+085$^{\mathrm{h}}$ & 0.1544 & 0.003--0.135 & 1219.0--1380.0 & F(6), D16 & 1295.3--1325.4  & 20--26 & 0.1143 & LP1 & 12569 \\
3C\,273$^{\mathrm{h}}$ & 0.1565 & 0.002--0.135 & 1218.0--1382.0 & F(38), D16 &   & 48--82 & 0.1429 & LP1 & 12038 \\
IRAS\,F09539--0439$^{\mathrm{h}}$ & 0.1568 & 0.003--0.138 & 1219.0--1383.0 & D16 & 1273.4--1288.1  
       & 18--27 &  0.1265 & LP1 &  12275 \\
       &    &  (0.065--0.138)  & (1295.0--1383.0) &   &   &    &   (0.0763)  &   &  \\
Mrk\,1014$^{\mathrm{h}}$ & 0.1631 & 0.003--0.143 & 1219.0--1390.0 &  D16 & 1300.9--1325.1 & 18--22 & 0.1298 & LP1 & 12569 \\
HE\,0056--3622 & 0.1631$^{\mathrm{j}}$ & 0.002--0.143 & 1218.0--1390.0 & D16 & 1274.0--1294.0 & 24--37 & 0.1306 
      & LP1 & 12604 \\
       &   &  (0.045--0.143) & (1270.0--1390.0) &   &  &  &  (0.0882)  &  &   \\
IRAS\,F00040+4325 &  0.1636 & 0.003--0.144 & 1219.0--1391.0 & F(5) &  & 18--34 & 0.1481 & LP3 & 14268 \\
PG\,1048+342          & 0.1667 & 0.002--0.148 & 1218.0--1395.0  &  F(4), D16 &  & 18--33 & 0.1537 & LP1 & 12024 \\
PG\,2349--014$^{\mathrm{h}}$ & 0.1740 & 0.003--0.154 & 1219.0--1403.4  & F(6), D16 & 1295.2--1325.5 & 18--24 & 0.1378 & LP1 & 12569 \\ 
PG\,1116+215           & 0.1749 & 0..002--0.156 & 1218.0--1405.0 & F(25), D16 & 1301.0--1307.5 & 33--50 & 0.1620 & LP1 & 12038 \\
RBS\,1768                & 0.1831 & 0.003--0.164 & 1219.0--1415.0 &  & 1294.5--1311.0 & 23--31 & 0.1623 & LP2 & 12936 \\
PHL\,1811    & 0.1914$^{\mathrm{k}}$ & 0.006--0.171 & 1223.0--1424.0 &  F(24), D16 &  & 33--56 & 0.1786 & LP1 & 12038 \\
PHL\,2525    & 0.2004 & 0.014--0.180 & 1233.0--1435.0 & F(7), D16 & 1270.6--1292.0  & 18--25 & 0.1634 & LP2 & 12604 \\
RBS\,1892 & 0.2005 & 0.013--0.180 & 1231.0--1435.0 & D16 & 1276.0--1306.9 & 20--28 & 0.1612 & LP2 & 12604 \\
    &   &   (0.084--0.180)  &  (1318.0--1435.0) &   &   &   &  (0.1133)  &    &   \\
PG\,1121+423        & 0.2240 & 0.032--0.203 & 1255.0--1463.0 & D16 &   & 18--27 & 0.1699  & LP1 & 12024 \\
1H\,0717+714 & $\sim$0.2314$^{\mathrm{i}}$ & 0.039--0.211 & 1263.5--1472.0 & F(18), D16 &   & 28--52 & 0.1960 & LP1 & 12025 \\
PG\,0953+415 & 0.2331$^{\mathrm{j}}$ & 0.042--0.221 & 1267.0--1484.0  & F(25), D16 &    & 32--52 & 0.1954 & LP1 & 12038 \\
RBS\,567   & 0.2412 & 0.078--0.221 & 1310.5--1484.0 & D16 &   & 18--25 & 0.1727 & LP1 & 11520 \\
3C\,323.1   & 0.2649 & 0.073--0.244 & 1304.8--1512.0 &    &    & 18--37 & 0.2091 & LP1 & 12025 \\
PG\,1302--102 & 0.2775 & 0.078--0.255 & 1310.0--1526.0  &  F(20), D16 &   & 25--34 & 0.2203 & LP1 & 12038 \\
4C\,25.01  &   0.2828$^{\mathrm{j}}$ & 0.084--0.261 & 1318.0--1533.5  &  & 1387.0--1435.5 & 18--24 & 0.1691 & LP3 & 14268 \\
Ton\,580    & 0.2901 & 0.090--0.268 & 1325.5--1542.0 & D16 &   & 20--27 & 0.2232 & LP1 & 11519 \\
H\,1821+643 & 0.2967$^{\mathrm{j}}$ & 0.099--0.201$^{\mathrm{l}}$ & 1336.0--1460.0$^{\mathrm{l}}$ & F(25), D16 &   & 35--80 & 0.1254 & LP1 & 12038 \\
PG\,1001+291 & 0.3283 & 0.121--0.298 & 1363.0--1578.0 & F(5), D16 &   & 20--27 & 0.2315 & LP1 & 12038 \\
PG\,1216+069 &  0.3322 & 0.124--0.310 & 1366.5--1592.0 & F(4), D16 &  & 20--33 & 0.2450 & LP1 & 12025  \\
3C\,66A  & $\sim$0.3347$^{\mathrm{i}}$ & 0.128--0.281 & 1371.5--1557.0 & F(2), D16 &   & 20--27 & 0.1983 & LP2 & 12863, 12612 \\
RBS\,877  & $\sim$0.3373$^{\mathrm{i}}$ & 0.129--0.267 & 1373.0--1540.0 &   &  & 18--23 & 0.1769 & LP1 & 12025 \\
RBS\,1795 & 0.3427 & 0.133--0.320 & 1377.5--1605.0 & F(4), D16 &  & 18--33 & 0.2499 & LP1 & 11541 \\ 
MS\,0117.2--2837 & 0.3487$^{\mathrm{j}}$ & 0.139--0.326 & 1385.0--1612.0 &  D16 &  & 18--37 & 0.2502 & LP1 & 12204 \\
PG\,1553+113 & $\sim$0.4131$^{\mathrm{i}}$ & 0.193--0.389 & 1450.0--1689.0 & F(15), D16 &   & 22--40 & 0.2776 & LP1 & 11520, 12025  \\
CTS\,487 & 0.4159 & 0.194--0.300 & 1452.0--1580.0 &  &   & 18--20 & 0.1422 & LP2 & 13448  \\
PG\,1222+216     & 0.4333 & 0.210--0.409 & 1471.0--1713.0  & D16 &   & 21--40 & 0.2877 & LP2 & 12025 \\
HE\,0153--4520   & 0.4496 & 0.223--0.426 & 1487.0--1733.0 & F(5), D16 & 1580.0--1614.0 & 18--36 & 0.2542 & LP1 & 11541 \\
PG\,0003+158     & 0.4504 & 0.224--0.426 & 1488.0--1734.0 &  D16 & 1593.0--1617.0 & 20--27 & 0.2670 & LP1 & 12038 \\
PG\,1259+593     & 0.4762 & 0.245--0.452 & 1514.0--1765.0 & F(25), D16 &   & 22--36 & 0.3078 & LP1 & 11541 \\
HE\,0226--4110  & 0.4934 & 0.261--0.456 & 1533.0--1770.0 &  F(28), D16 &   & 23--31 & 0.2973 & LP1 & 11541 \\
PKS\,0405--123 & 0.5726 & 0.327--0.466 & 1613.0--1782.5 &  F(23), D16 &   & 27--45 & 0.2193 & LP1 & 11541, 11508 \\
PG\,1424+240  & $\sim$0.6035$^{\mathrm{i, m}}$ & 0.354--0.439 & 1645.5--1749.0 &  &  & 25--30 & 0.1330  & LP1 & 12612 \\
\hline

\end{tabular}
}}

\begin{flushleft}
{\footnotesize{
{\bf Notes -- a:} The redshift is measured from the observed Ly$\alpha$ emission line of the QSO in
   the COS FUV spectra. Otherwise, the redshift is taken from NED or Simbad.
{\bf b:} If the $z$/wavelength range of the Ly$\alpha\beta$ region
in COS and/or FUSE spectra is different from the Ly$\alpha$ region, it is listed in parenthesis in the next row. 
{\bf c:} F--An available {\it FUSE} spectrum is used to cover the high-order Lyman lines. The number in
parenthesis is a S/N per resolution element at $\sim$1050\,\AA.
D16--The AGN is also included in the low-$z$ COS IGM
study in D16, although our adopted AGN naming is often different.
{\bf d:} The wavelength regions with $S/N\!<\!18$ and/or the unobserved regions due to
a detector gap. Excluded regions due to the Galactic ISM contamination are not listed. These are
\ion{Si}{ii} $\lambda$\,1260.42, 1304.37, 1526.70, \ion{O}{i}  $\lambda$\,1302.16, \ion{C}{ii} $\lambda$\,1334.53,
\ion{Fe}{ii} $\lambda$\,1608.45 and \ion{Al}{ii} $\lambda$\,1670.78. 
When \ion{Fe}{ii} $\lambda$\,1608.45 is not saturated 
and \ion{Fe}{ii} $\lambda\lambda$\,1144.93, 1143.22, 1142.36 are covered in G130M, the region at 
$\sim$1608\,\AA\ is included.
{\bf e:} S/N per resolution element.
{\bf f:} The number in parentheses in the next row is $\Delta X$ of the Ly$\alpha\beta$ region.
{\bf g:} The COS FUV Lifetime Position: LP1 -- before July 22, 2012, LP2 -- from July 23, 2012 to
February 8, 2015, LP3 -- from February 9, 2015 to October 1, 2017, LP4 -- since October 2, 2017.
{\bf h:} Only the G130M spectrum was obtained.
{\bf i:} The AGN is a BL Lac type, showing no conspicuous emission peak. The emission redshift is set to be
that of the Ly$\alpha$ absorption at the highest redshift.
{\bf j:} Due to strong intrinsic absorbers on top of the emission peak, the redshift is slightly uncertain.
{\bf k:} The emission peak is relatively flat. The redshift is set to be at the highest flux around the peak.
{\bf l:} The S/N ratio changes abruptly in the forest region: $S/N$\,$\ge$\,35 at $\le$\,1460~\AA\/
and $\sim$14--17 at $\ge$\,1460~\AA\/. To satisfy our S/N selection criteria, only the forest region at 
$\le$\,1460~\AA\/ is included. 
{\bf m:} Both NED and Simbad list its redshift as 0.16. However, the STIS E230M spectrum shows that it
is a BL Lac type and the redshift is higher than 0.604 from the Ly$\alpha$ absorption features.
}}

\end{flushleft}
\end{table*}


\subsection{{\it HST}/COS FUV data}
\label{sect2.4}

We select 55 COS G130M/G160M (1100--1800\,\AA) AGN spectra (Table~\ref{tab3}).
We note that 44 out of our 55 COS AGN are also included in the 
COS IGM sample of \citet[D16 hereafter]{danforth16}. 
However, our analysis methods are different and there is a difference in line identifications
and fitted line parameters for $\sim$30\% of the lines (see more details in Section~\ref{sect3.4}).

All the raw, individual COS exposures were reduced with CalCOS versions 3.0 or 3.1 with
the flat-field correction on. Similar to the treatment of COS NUV data as outlined in 
Section~\ref{sect2.3}, we re-calibrated the CalCOS wavelength to an uncertainty better 
than $\sim$5--10\,\kms\/
\citep[see their Appendix for details]{wakker15}
and coadded the individual extractions sampled at 0.00997\,\AA\/ 
(0.01223\,\AA\/) per pixel for the G130M (G160M) grating.
Since COS spectra are highly oversampled, we binned the final coadded spectrum by 3 pixels,
sampled at 0.02991\,\AA\/ (0.03669\,\AA\/) per pixel 
for the G130M (G160M) grating. The resolving power of each individual extraction
is quoted as $R$\,$\sim$\,18,000 to 20,000, which corresponds to 15 to 17 \kms\
for a Gaussian LSF.
However, the COS FUV LSF shows the time-dependent non-Gaussianity
and the resolving power degraded with time. The spectral resolution can
be approximated to $\sim$19 \kms\ for the COS non-Gaussian LSF at Lifetime Position 1
(see more details in Section~\ref{sect3.2}). 
The wavelength regions contaminated by strong Galactic ISM lines are discarded.
The typical COS FUV $N_{\mathrm{\ion{H}{i}}}$ limit 
is $\log N_{\mathrm{\ion{H}{i}}}$\,$\sim$\,13.0.

\begin{figure}

\includegraphics[width=8.5cm]{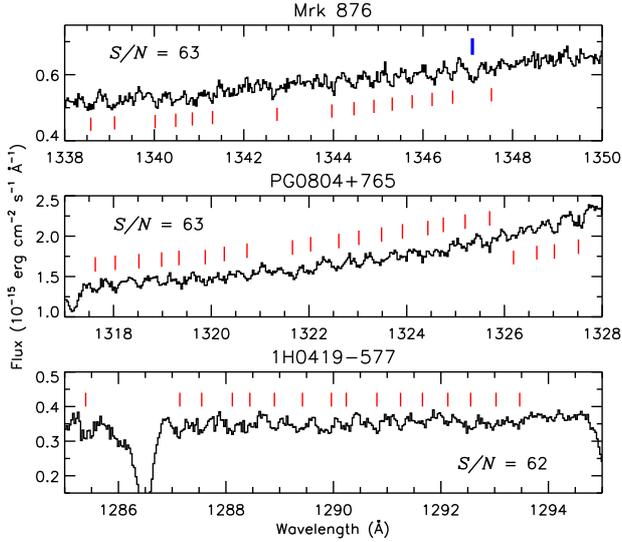}\\
\vspace{-0.3cm}
\caption{Examples of sawtooth-line 
MCP Hex fixed pattern noise (FPN) in coadded COS spectra. 
The red vertical ticks spaced at $\sim$0.4\,\AA\ mark the position of Hex FPN.
The blue thick tick in the upper panel marks the position of the Milky Way
\ion{Cl}{i} $\lambda$1347.23.
When assumed as a \ion{H}{i} Ly$\alpha$, FPN has 
$\log N_{\mathrm{FPN}}$\,$\sim$\,12.5. If a weak IGM \ion{H}{i} absorption
falls on FPN, the apparent $N_{\mathrm{\ion{H}{i}}}$ increases.
}
\label{fig3}
\end{figure}

The CalCOS flat-field correction corrects 
strong wire grid shadow features greater than $\sim$20\% in intensity, but 
not weak ($\le$10\% in intensity) fixed pattern noise (FPN) produced by 
the hexagonal pattern of the fiber bundles in the COS FUV micro-channel plate known as
MCP Hex \citep{dashtamirova19}. MPC Hex is supposed to be fixed in the detector pixel
space, but not in the wavelength space. In practice, the position of MCP Hex 
and its intensity change along the pixel space. This sometimes produces false, equally-spaced 
weak absorption-like features in the high-S/N region of the coadded spectrum 
(Fig.~\ref{fig3}). It is the most conspicuous 
when a high-flux Ly$\alpha$ emission region falls on the longer-wavelength edge of
Segment B of the detector.
Due to FPN, the noise is not Gaussian
and the conventional way to quote noise as the reciprocal of 1 r.m.s. of the unabsorbed
region underestimates true noise \citep{keeney12}. 
Since only an individual extraction with $S/N\,\ge$\,12 
shows distinct FPN and the majority of our individual extractions has a lower S/N,
we did not correct for MCP Hex \citep{fitzpatrick94, savage14, wakker15}.

\subsection{{\it FUSE} data}

Available {\it FUSE} spectra (917--1187\,\AA\/) were used to obtain a reliable column 
density of saturated COS FUV \ion{H}{i} Ly$\alpha$ lines, 
since {\it FUSE} spectra cover high-order Lyman lines at $z$\,$\le$\,0.12. 
The 8th column of Table~\ref{tab3} lists whether the COS AGN has a corresponding
{\it FUSE} spectrum. The {\it FUSE} spectra used in this study are the same ones
analysed by \citet{wakker06}. 
They are sampled at $\sim$0.0066\,\AA\/ 
per pixel, weakly dependent on the wavelength. As they are oversampled,
we binned the {\it FUSE} spectra by 3, 5 or 7 pixels to increase the S/N.
The S/N in general increases toward longer wavelengths, 
i.e. more reliable Ly$\beta$ profiles than Ly$\gamma$ profiles. 
The S/N per resolution element at $\sim$1050\,\AA\ is listed in parenthesis in the 
5th column of Table~\ref{tab3}. 
Since wavelength regions with $S/N$\,$<$\,5 are not very useful 
to deblend saturated lines reliably, we excluded these low-S/N regions in our Lyman series fit.
The 4th column in Table~\ref{tab3} accounts for this exclusion. AGN with low-$S/N$ {\it FUSE} spectra
but with a low-$z$ limit $z$\,$\sim$\,0.002 ($\sim$1218\,\AA\/) do not have a saturated Ly$\alpha$ 
(no need for {\it FUSE} spectra) or have a higher S/N in {\it FUSE} Ly$\beta$ regions of interest than 
the S/N at $\sim$1050\,\AA\/ as quoted in Table~\ref{tab3}.
The resolution varies from AGN to AGN, usually ranging from $\sim$20\,\kms\/
above 1000\,\AA\/ to $\sim$25--30\,\kms\/ below 1000\,\AA. 
For 3C273, its {\it FUSE} observations were taken
in the early operation days when the telescope suffered from a focusing problem.
This degraded the resolution to $\sim$30\,\kms\/ at 1100\,\AA\/ and 
to $\sim$60\,\kms\/ at 930\,\AA\/. The wavelength uncertainty is about 5--10\,\kms. However,
if the Galactic molecular hydrogen with numerous transitions is detected, the wavelength 
uncertainty can be $\le$5\,\kms.

\section{Voigt profile fitting analysis}
\label{sect3}

\begin{figure}

\includegraphics[width=8.7cm]{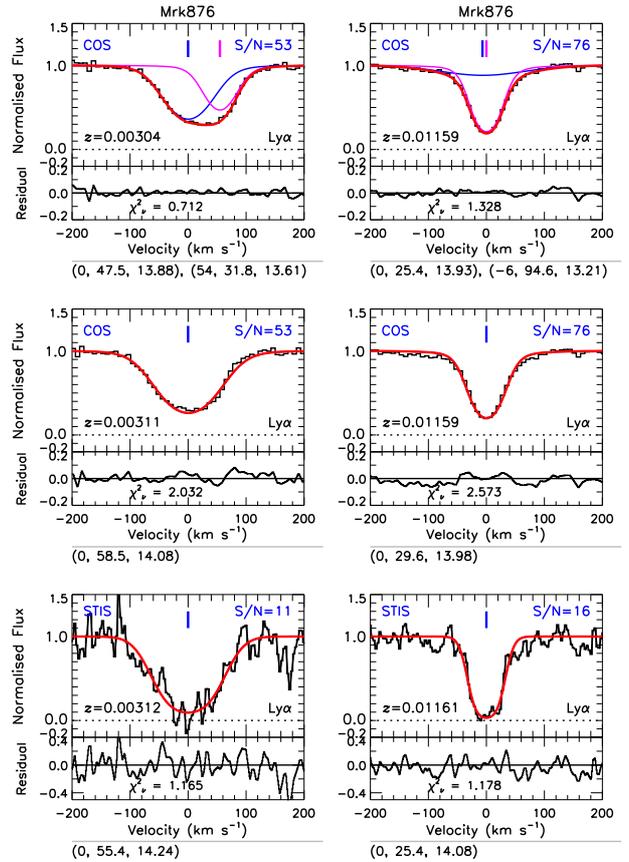}\\
\vspace{-0.6cm}
\caption{Effect of the S/N on the {\tt VPFIT} Voigt profile fitting analysis. {\it Left panel:} 
The velocity plot (the relative velocity vs normalised flux) of the $z$\,$=$\,0.0030 absorber
toward Mrk\,876. The velocity centre is set to the redshift of the strongest \ion{H}{i}
component. 
The observed spectrum is shown as a black histogram, while
the red profile is the generated spectrum using all the fitted components. Blue and
magenta profiles with the ticks are the individual fitted components. 
The top and middle panels show a fiducial 2-component fit and a
one-component fit for the COS spectrum, respectively. The bottom panel 
displays a one-component
fit for the STIS E140M spectrum. The noisy STIS spectrum allows a single-component fit 
with a good $\chi^{2}_{\nu}$, while the higher-S/N COS spectrum requires a two-component fit.
The lower part of each panel shows the residual of the fitted
components with the normalised $\chi^{2}_{\nu}$.
The three numbers in parentheses at the bottom of each panel
are the relative velocity
in \kms, the $b$ parameter in \kms\ and the logarithmic column density, respectively.
{\it Right panel:}
Another absorber at $z$\,$=$\,0.0116 toward Mrk\,876.
}
\label{fig4}
\end{figure}

\subsection{The Voigt profile fitting analysis}
\label{sect3.1}

From the profile fitting of identified lines, three line parameters are obtained,
the redshift $z$,
the column density $N$ in cm$^{-2}$
and the line width or the Doppler parameter $b$ in $\textrm{km s}^{-1}$.
For thermal broadening, the $b$ parameter ($=\sqrt{2}\sigma$, where $\sigma$ is 
the standard deviation) is related to the gas temperature $T$ in K by
$b$\,$=$\,$\sqrt{2k_{\mathrm{B}}T/m_{\mathrm{ion}}}$, where $k_{\mathrm{B}}$ is the
Boltzmann constant and $m_{\mathrm{ion}}$ is the atomic mass of ions.


We have performed the profile analysis to all the AGN spectra in this study
using {\tt VPFIT} version 
10.2{\footnote{Carswell et al.:
http://www.ast.cam.ac.uk/$\sim$rfc/vpfit.html}} 
with the {\tt VPFIT} 
continuum adjustment option on \citep{carswell14}. We remind readers that 
the publicly available {\tt VPFIT} code has been extensively tested  
by the IGM community 
over three decades, including comparisons to curve-of-growth fit results.
Our already published UVES and HIRES spectra \citep{kim07, kim13, kim16a} 
were also refit with {\tt VPFIT} v10.2 to be
consistent with the new COS and STIS fits. 
While the new fits overall do not change significantly from the
previous ones, the errors 
produced by {\tt VPFIT} v10.2 tend to be larger when the components are
at absorption wings. Also note that the COS FUV spectra and line lists 
used in this study are 
updated from our previous ones analysed in \citet{viel17}.

Unfortunately, the Voigt profile fitting result is not unique
\citep{kirkman97, tripp08, kim13}. The normalised $\chi^{2}_{\nu}$ criterion does not 
always guarantee a good actual fit, as illustrated in Fig.~\ref{fig4}. 
The number of fitted components is
more sensitive to S/N than the spectral resolution since both STIS and COS
spectra resolve the IGM \ion{H}{i} lines.
As S/N increases, a fitting program often tends
to include more narrow, weak components to reproduce small
fluctuations. Although additional components 
added to improve $\chi^{2}_{\nu}$ are in general weak, 
$\log N_{\mathrm{\ion{H}{i}}}\,\le\!13.5$,
an actual change in the fitted parameters depends on S/N and differs for
each absorption complex. Despite the non-uniqueness, 
our fitting analysis uses the same program to fit similar-quality spectra within
each data set. Any judgmental calls and systematics would be
repeated in similar ways. Therefore, our final combined fitted parameters 
from different spectrographs
can be considered consistent and uniform within our own data sets.

\subsection{The COS FUV line spread function}
\label{sect3.2}

The profile fitting technique requires an instrumental line spread function (LSF)
to convolve with the model fit profile. The LSFs of UVES, HIRES, STIS and COS
NUV spectra are 
straightforward and well-characterised \citep{vogt94, dekker00, riley18, dashtamirova19}. 

The COS FUV LSF is more complicated and changes with wavelength and 
time.
The COS optics do not correct for the mid-frequency wavefront
errors due to polishing irregularities in the {\it HST} primary and secondary mirrors.
This causes the non-Gaussian COS FUV LSF with an extended wing and a broader 
and shallower core. This is stronger at shorter wavelengths and 
in particular evident for strong, saturated absorption lines
\citep{kriss11, keeney12}. 
The non-Gaussianity produces a broader and shallower line, with the bottom 
of saturated lines not reaching to a zero flux. 
Therefore, the flux statistics directly obtained from observed COS spectra
cannot be compared with the one from STIS, UVES and HIRES spectra. 
The non-Gaussian LSF also increases the $N_{\mathrm{\ion{H}{i}}}$
detection limit compared to the same Gaussian resolving power.

In addition, the COS FUV detector loses its sensitivity from accumulated exposures known
as gain sag. To avoid gain sagged regions, 
the position of the science spectrum on the FUV detector has been moved to a different
Lifetime Position (LP) periodically in the cross-dispersion direction, as noted in the
9th column of Table~\ref{tab3}. At the later lifetime positions,
the COS FUV LSF has a broader core and more extended non-Gaussian wings
\citep{dashtamirova19}. Both non-Gaussianity and LP change reduce the 
resolving power as a function of wavelength and time: at 1300\,\AA,
the resolving power at LP3 decreases $\sim$12\% from LP1. 
Note that $\sim$80\% of our COS sample is taken at LP1.

\subsection{Voigt profile fitting procedure}
\label{sect3.3}

Our fitting approach is: 

\begin{enumerate}

\item The COS FUV/NUV LSF is taken from
the {\it HST}/COS Spectral Resolution 
homepage\footnote{http://www.stsci.edu/hst/cos/performance/spectral\_resolution\/},
taking account of the Lifetime Position of the FUV LSF. The STIS E230M LSF is taken from
the {\it HST}/STIS Spectral Resolution 
homepage\footnote{http://www.stsci.edu/hst/stis/performance/spectral\_resolution}.

\item The error array is scaled to satisfy that the r.m.s. of the unabsorbed region
is similar to the average of the errors in the same region, as the rebinning and
interpolation during the data reduction often overestimates the error. 

\item The appropriate good-fit $\chi^{2}_{\nu}$  
is set to be $\sim$1.3, as the average error array does not always 
correspond to the r.m.s. of the science array and noise is not often Gaussian.

\end{enumerate}

We followed the standard approach for absorption line analysis
\citep{carswell02, kim07}. First, the entire spectrum was divided into several regions.
The number of divided regions is dependent on an apparent underlying 
continuum shape. When the continuum varies smoothly, divided regions are $\sim$100\,\AA-long.
However, when the continuum varies rapidly such as a region around the Ly$\alpha$ emission
or the Ly$\beta$+\ion{O}{vi} emission, the length of divided regions
is adopted to accommodate the rapid change of the continuum, 5--30\,\AA.
For COS, STIS and {\it FUSE} spectra, an initial continuum fit was obtained by 
iterating a cubic spline polynomial fit for each region, rejecting deviant regions
at $|(\mathrm{flux}-\mathrm{fit})/\mathrm{fit}|$\,$>$\,0.025 \citep{songaila98}. 
The used fit order is between three and seven, depending
on a underlying continuum shape. 
The continua of each region were joined to form an initial continuum of the entire spectrum.
Any disjointed continua at the joined regions are adjusted manually as well as the
global continuum after visual inspection, which often gives a better continuum placement.
For UVES/HIRES spectra, we used the same normalised spectra analysed
by \citet{kim13}, which follows the same procedure to obtain a localised initial continuum 
except using the {\tt CONTINUUM/ECHELLE} command in IRAF.

Second, all possible metal lines were searched
for. We started from the most common metals found in the IGM 
(such as \ion{C}{iv}, \ion{Si}{iv} and \ion{O}{vi} doublets, \ion{C}{ii} and \ion{Si}{ii} multiplets,
and \ion{Si}{iii} and \ion{C}{iii} singlets)
at their expected position for each \ion{H}{i}, regardless of $N_{\mathrm{\ion{H}{i}}}$.
If any of these common metal lines
are detected, we searched for other less common metals, such as \ion{Fe}{ii}, \ion{Mg}{ii}
and \ion{Al}{ii}. We also used empirically known facts, such as that \ion{Mg}{ii} is not associated
with low-$N_{\mathrm{\ion{H}{i}}}$ lines.  
When metals were found, they were fit first, using the same
$z$ and $b$ values for the same ionic transitions. 
When metal lines were blended with \ion{H}{i}, these \ion{H}{i}
absorption regions were also included in the fit.
The rest of the absorption
features were assumed to be \ion{H}{i} and were fitted, including all the available
higher-order Lyman series, such as Ly$\beta$ and Ly$\gamma$. 
When $\chi^{2}_{\nu}$\,$\ge$\,1.5, additional components are added manually
and included only if they improve $\chi^{2}_{\nu}$ significantly. 
When lines are too narrow to be \ion{H}{i}, i.e. $b$\,$\le$\,10\,\kms, but without a robust line identification,
the identification is noted as ``??", but fitted assuming \ion{H}{i}. 
These lines are usually weak at $\log N_{\ion{H}{i}}$\,$\le$\,12.8. 
The contamination by these unidentified metals is negligible
at $z$\,$<$\,1, but can be around 2--3\% at $z$\,$\sim$3.
 
For each fit, we checked whether 
the initial continuum was appropriate for the available Lyman series
and different transitions by the same ion. When necessary,
a small amount of continuum adjustment was applied
to achieve $\chi^{2}_{\nu}$\,$\le$\,1.3.
The entire spectrum was re-fitted with this re-adjusted
continuum. In most cases, re-adjusting a local continuum makes it necessary
to increase a previous continuum slightly, especially below the Ly$\beta$ emission
where weak high-order Lyman absorptions at higher $z$ can depress the
continuum. This iteration has been performed several times until
the final fit of lines with
$\ge$\,3--4$\sigma$ significance was obtained at $\chi^{2}_{\nu}$\,$\le$\,1.3.
Due to un-removed fixed pattern noise and continuum uncertainties, we did not fit all the absorption
features at $\sim$3.5$\sigma$ such as closely spaced several weak absorption lines
as seen in Fig.~\ref{fig3}. Any noticeable velocity shifts caused by the COS wavelength
calibration uncertainty between the multiple transitions of the same ion are 
accounted for with the {\tt VPFIT} ``$<$$<$'' option.
The line identification and/or fitting are independently checked 
by B. P. Wakker for COS/STIS spectra and R. F. Carswell for
STIS/UVES/HIRES spectra, and are finalised by
T.-S. Kim. 

Since most IGM simulations analyse the Ly$\alpha$ forest without
incorporating high-order Lyman series, we also performed a fit using only Ly$\alpha$.
Note that even including all the
available high-order Lyman lines does not vouch for the completely resolved
profile structure of heavily saturated lines at $\log N_{\mathrm{\ion{H}{i}}}$\,$\ge$\,17--18, 
if severe line blending and intervening Lyman limits
leaves no clean high-order Lyman lines.  

The line parameters from {\tt VPFIT} include the uncertainty due to 
statistical flux fluctuations and fitting errors. However, they do not include 
the error due to the continuum placement uncertainty. 
For the Galactic ISM, the continuum uncertainty is often estimated simply by 
shifting a fraction of the r.m.s. of the continuum
\citep{savage91, sembach91} or by estimating all the uncertainties associated with 
a polynomial function fit to a continuum around an absorption line \citep{sembach92}.
In high-$z$ IGM spectra for which {\tt VPFIT} was initially developed, 
line blending is too severe to estimate a realistic local continuum around each absorption feature
and the flux calibration of high-resolution echelle spectra is not very reliable
due to a lack of well-calibrated high-S/N, high-resolution spectra of flux standard stars.
The continuum-adjustment $<$$>$ option in {\tt VPFIT} does not use a similar procedure.

\begin{figure}

\vspace{-0.1cm}
\includegraphics[width=8.5cm]{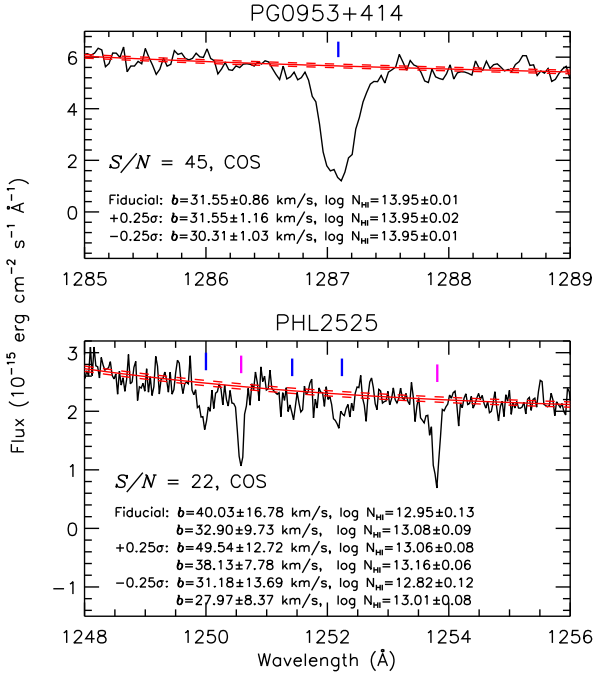}\\
\vspace{-0.7cm}

\caption{{\it Upper panel:} Red solid curve is our final continuum and the red dashed
curves are the continuum shifted by $\pm$0.25$\sigma$. The blue vertical tick marks
the typical IGM \ion{H}{i} lines. The line parameters for each continuum are noted in the
panel. {\it Lower panel:} Magenta ticks note the Galactic \ion{S}{ii}. The first 
(second) set of line parameters are for the \ion{H}{i}
line at 1251.4\,\AA\ and 1252.2\,\AA, respectively.}
\label{fig5}
\end{figure}

We estimated a continuum error by shifting $\pm$\,0.25$\sigma$ of our fiducial
continuum for 50 COS \ion{H}{i} absorption features as shown in Fig.~\ref{fig5}, 
since COS IGM \ion{H}{i}
features are not much affected by line blending. The $\pm$\,0.25$\sigma$ shift is decided
by visual inspection (see also \citet{sembach91, penton00, kim07}).
Obviously the $-$0.25$\sigma$ ($+$0.25$\sigma$) continuum returns a smaller (larger) $b$ and
$N_{\ion{H}{i}}$. Both sets of line parameters 
are {\it within the fiducial {\tt VPFIT} 1$\sigma$ fitting error}, with $b$ values being more sensitive to the
continuum. In general, the continuum error is $\le$5\% of the fitting
error when $\log N_{\ion{H}{i}}$\,$\ge$\,13.5 and $S/N$\,$\ge$\,30 (upper panel). 
The continuum error becomes larger for low S/N and $N_{\ion{H}{i}}$, 
especially for larger $b$ values. In the lower panel,
the continuum error of $b$ and $N_{\ion{H}{i}}$ 
is $\sim$25\% for \ion{H}{i} with $b$\,$\sim$\,40\,\kms\ and $\log N_{\ion{H}{i}}$\,$\sim$\,13.0 
at 1251.4\,\AA\ and is $\sim$15\%
with $b$\,$\sim$\,33\,\kms\ and $\log N_{\ion{H}{i}}$\,$\sim$\,13.0 at 1252.2\,\AA.
We remind that a large fraction of \ion{H}{i} at $\log N_{\ion{H}{i}}$\,$\le$\,13.1 can be spurious
if $S/N$\,$\le$\,20--25.

Although {\tt VPFIT} does not include a continuum error as in the ISM studies, its
fitting errors are calibrated with 
the curve-of-growth analysis and the associated error array. 
Weak and broad lines at lower S/N have larger associated error arrays and continuum uncertainties,
thus have larger fitting errors. Since our sample has mostly $S/N$\,$>$\,20 and our analysed
$N_{\ion{H}{i}}$ range in the absorption line statistics
is $\log N_{\ion{H}{i}}$\,$\ge$\,13.5, including the continuum fitting error 
will increase the fiducial fitting error by $\le$\,5--10\%.
Our main scientific goal is to quantify the observational estimates as uniformly as possible,
reducing a systematic bias. Since it is not clear how to define a reasonable continuum for 
highly-blended high-$z$ IGM spectra, we therefore used the {\tt VPFIT} fitting error without including 
the $\pm$0.25$\sigma$ continuum error in this work for consistency.

A profile fit of a single-line of \ion{H}{i} has been claimed to
overestimate the true line width by $\sim$1.5 compared to a curve-of-growth fit using
all available high-order Lyman lines in STIS, COS and {\it FUSE} spectra \citep{shull00, danforth10}.
We do not find such a tendency when we compare the Ly$\alpha$-only and Lyman series
fits for relatively clean, isolated and unsaturated \ion{H}{i} Ly$\alpha$ from
high-S/N, high-resolution optical UVES/HIRES spectra. 
Combined with large wavelength calibration uncertainties,
imperfect line spread function (LSF) and fixed pattern noise, an observed absorption profile
in lower-quality UV spectra does not necessarily show a Voigt-profile shape 
convolved with the true LSF.
We often find that the profile
shapes of Lyman lines, such as Ly$\alpha$ and Ly$\beta$ or Ly$\alpha$ and Ly$\gamma$,
are inconsistent in COS and {\it FUSE} spectra. The discrepancy of $b$ measurements between
the profile and curve-of-growth fits is likely to be caused by low-quality data 
or an inaccurate mathematical treatment
in some private profile fitting codes, not by the
fundamental inferiority of a profile fit to a curve-of-growth fit. 
We remind readers 
that the {\tt VPFIT} profile fit compromises all the absorption profile shapes included in the
fit as a function of S/N.
The {\tt VPFIT} fitting error can be used for reliability of fitted parameters.

\begin{figure}

\vspace{-0.1cm}
\includegraphics[width=8.5cm]{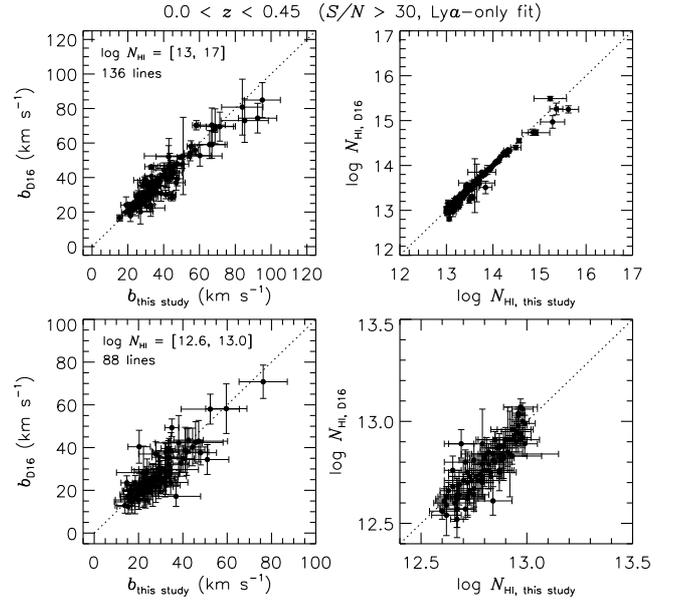}\\
\vspace{-0.7cm}

\caption{Comparisons of $b$ and $N$ of common \ion{H}{i} absorption lines
between D16 and our work from 
the 14 highest-S/N COS AGN 
(PG\,0804+761, 1H\,0419--577, PKS\,2155--304, TON\,S210, HE\,1228+0131,
Mrk\,876, IRAS\,Z06229$-$6434, 3C\,273, PG\,1116+215, PHL\,1811, 
1H\,0717+714, PG\,0953+415, H\,1821+643 and PKS\,0405--123). 
The dotted lines delineate the one-to-one correspondence. 
{\it Upper panels:} 136 common \ion{H}{i} components with  
$\log N_{\mathrm{\ion{H}{i}}}$\,$\in$\,[13, 17]. Only components from a similar
component structure in both studies are shown. 
{\it Lower panels:} 88 common \ion{H}{i} with $\log N_{\mathrm{\ion{H}{i}}}$\,$\in$\,[12.6, 13.0].
}
\label{fig6}
\end{figure}

\begin{figure}

\vspace{-0.1cm}
\includegraphics[width=8.5cm]{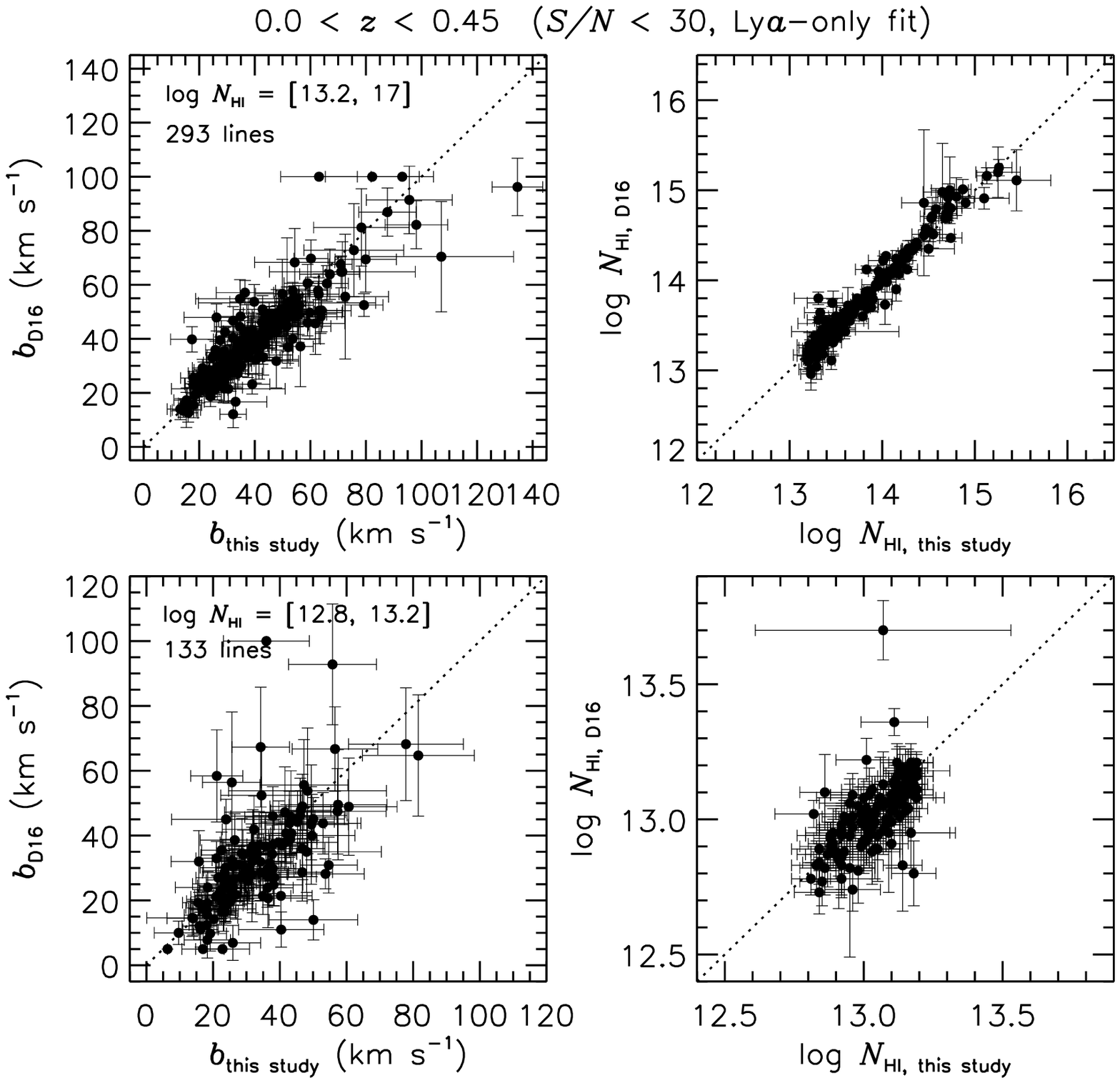}\\
\vspace{-0.7cm}

\caption{Comparisons of $b$ and $N$ of common \ion{H}{i}
from 30 COS AGN
with $S/N\!<\!30$ between D16 and this study.
All the symbols are the same as in Fig.~\ref{fig6}. Components with
$b$\,$=$\,100\,\kms\ without errors in D16 indicate highly uncertain.
{\it Upper panels:} Among our 337 secure \ion{H}{i} components,
87\% (293/337) shown have a similar component structure. 
About 6\% (19/337) has unaccounted blending by metals or mis-identified as \ion{H}{i}
in D16. The remaining \ion{H}{i} has a different component structure from D16.
{\it Lower panels:} Out of our 189 secure \ion{H}{i} components, 70\%
(133/189) are common with a similar component structure.
About 10\% (19/189) suffer from metal contamination or
are mis-identified in D16.
}
\label{fig7}
\end{figure}

\subsection{Comparisons with published line parameters}
\label{sect3.4}

Due to different data treatments and the non-uniqueness of the profile fit,
discrepancies between different studies are inevitable. 
The discrepancy introduces a
systematic uncertainty and can result in a contradictory result, especially for low-S/N data. 
Since only a few sightlines from UVES/HIRES spectra have published
line lists besides our own, we compare
the fit measurements exclusively using the D16
COS FUV line parameters. D16 sometimes misidentifies the
weak Galactic ISM lines such as \ion{Mg}{ii} $\lambda\lambda$\,1239.92, 1240.39 and
orphaned high-velocity components
as intergalactic \ion{H}{i} Ly$\alpha$ and does not fully account for 
contaminations by the ISM lines. 
Misidentification as metals and unaccounted metal contamination
affects $\sim$10\% of the D16 lines at 
their $\log N_{\mathrm{\ion{H}{i}}}$\,$\in$\,[12.6, 17.0].
We use our own line identification and measurements as a reference in
this section. 

D16 adopts the \ion{H}{i} absorption line parameter from 
a Voigt profile fit at $\log N_{\mathrm{\ion{H}{i}}}$\,$\le$\,14 (no other Lyman
lines can be detected in low-S/N COS spectra) and a curve-of-growth fit at 
$\log N_{\mathrm{\ion{H}{i}}}$\,$\ge$\,14 (high-order Lyman lines
can be detected), respectively. Without including {\it FUSE} spectra, D16
measures \ion{H}{i} line parameters only from a single-line Ly$\alpha$ at $z$\,$<$\,0.1. 
The vast majority ($\sim$86\%) of detected IGM \ion{H}{i} lines at $z$\,$\sim$\,0.15 
have $\log N_{\mathrm{\ion{H}{i}}}$\,$\le$\,14. 
Therefore, the comparison is done for our Ly$\alpha$-only fit and their Ly$\alpha$-only
profile fit and Ly$\alpha$ curve-of-growth 
measurements. 
Both $N$ measurements for a saturated Ly$\alpha$
should be treated as lower limits, although {\tt VPFIT} gives a very reliable
column density for mildly saturated lines.

The two upper panels of Fig.~\ref{fig6} show the comparison of $b$ and $N$ of
136 common \ion{H}{i} components for $\log N_{\mathrm{\ion{H}{i}}}$\,$\in$\,[13, 17]
from the 14 highest-S/N ($S/N$\,$>$\,30) COS AGN
analysed by both studies. Only absorption features to have a similar component
structure, i.e. a single-component or two-component absorption features,
are shown. Among our 173 securely detected \ion{H}{i} at
$\log N_{\ion{H}{i}}$\,$\in$\,[13, 17], 136 components (79\%) have a similar
component structure.
About 5\% (9/173) have unaccounted metal-line blending or are
incorrectly identified as \ion{H}{i} in D16. For example, an absorption at $\sim$1362.4\,\AA\ 
toward PHL\,1811 is identified as \ion{H}{i} at $z$\,$=$\,0.120700 in D16, but we identify it
as \ion{Si}{ii} $\lambda$1260.42 at $z$\,$=$\,0.08093.
The remaining components have a different multi-component structure including
saturated absorption complexes or a different line identification from D16. Since both studies
do not include a continuum fitting error, the errors are comparable and the {\tt VPFIT} errors
are often larger.

While the column density of common lines is mostly in good agreement, their $b$ shows 
a larger difference, especially at larger $b$. Twenty components out of 21 with 
$b$\,$>$\,60\,\kms (12\%, 21/173) have $\log N_{\mathrm{\ion{H}{i}}}$\,$\le$\,14.
At the typical COS S/N in our sample, these broad, weak
lines are highly susceptible to the continuum placement and the line alignment among
individual extractions to coadd, which reflects in the large $b$ errors. The mean 
difference and its standard error ($=$\,$1\sigma/\sqrt{N}$ with $N$ being
the number of common lines) of 173 common \ion{H}{i} lines
is $\Delta b$\,$=$\,$0.6\pm0.4$\,\kms\ 
and $\Delta \log N_{\mathrm{\ion{H}{i}}}$\,$=$\,$0.03\pm0.07$.
The difference in line parameters for the lines not shown due to a different one-to-one
component structure or uncorrected metal blending 
is obviously much larger.

The difference becomes increasingly larger for 88 common weaker lines at 
$\log\!N_{\mathrm{\ion{H}{i}}}$\,$\in$\,[12.6, 13.0] (lower panels). 
As the profile fitting is exclusively based on the absorption profile, the discrepancy
is largely due to the difference in the profile shape of
weak lines in the two studies, likely caused by the different coaddition procedure
and by our improved wavelength re-calibration.
As broader lines are highly sensitive to the local S/N and continuum, only 7\%
(8 out of a total of our 119 secure \ion{H}{i}) have
$b$\,$>$\,60\,\kms. 
About 12\% (14/119) have unaccounted metal contamination 
or are wrongly identified as metals in D16.
The mean difference and its standard error of common lines is
$\Delta b$\,$=$\,$1.2\pm0.6$\,\kms\ 
and $\Delta \log N_{\mathrm{\ion{H}{i}}}$\,$=$\,$0.03\pm0.01$.

The difference is even larger for lower-S/N spectra 
(Fig.~\ref{fig7}), since the coadded profile shape
is more sensitive to the coaddition procedure and 
line alignment. At $\log N_{\mathrm{\ion{H}{i}}}$\,$\in$\,[13.2, 17.0] ([12.8, 13.2]), 
the mean difference and its standard error of 293 (133) common lines noted as filled circles
is $\Delta b$\,$=$\,$0.7\pm0.4$\,\kms\/ ($\Delta b$\,$=$\,$1.2\pm1.0$\,\kms)
and $\Delta \log N_{\mathrm{\ion{H}{i}}}$\,$=$\,$0.01\pm0.01$
($\Delta \log N_{\mathrm{\ion{H}{i}}}$\,$=$\,$0.02\pm0.01$).

\begin{figure}

\vspace{-0.1cm}
\includegraphics[width=8.5cm]{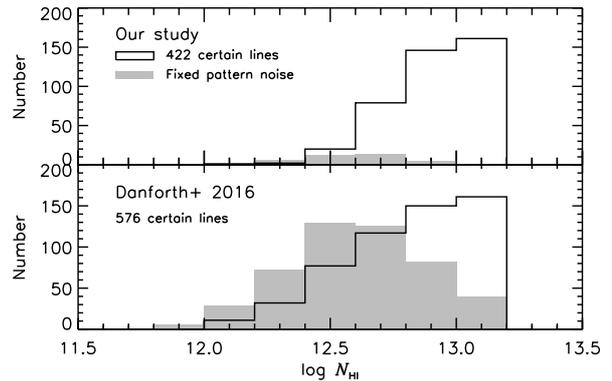}\\
\vspace{-0.4cm}
\caption{Histograms of $N_{\ion{H}{i}}$ of weak \ion{H}{i} from the 44 COS AGN
in common. The number distribution of \ion{H}{i} clearly shows a turnover 
at $\log N_{\mathrm{\ion{H}{i}}}$\,$\le$\,13 due to missed detections by noise.
}
\label{fig8}
\end{figure}

Figure~\ref{fig8} displays the histogram of weak \ion{H}{i} components in both studies.
With wavelength calibration uncertainties at 
5--10\,\kms\ and fixed pattern noise (FPN), the fitted line parameters, in particular $b$,
and identifications of weak lines are not as reliable as for strong lines.
We measured 580 components at $\log N_{\mathrm{\ion{H}{i}}}$\,$\le$\,13.2. 
Certain and uncertain ($\sim$3.5$\sigma$) components
are 73\% (422/580) and 25\% (146/580), respectively. The remaining is FPN features
(gray-shade histogram, Fig.~\ref{fig3}). Real weak absorption
features can be missed easily in noisy spectra and a large fraction of detected weak lines
can be spurious at 
$\log N_{\mathrm{\ion{H}{i}}}$\,$\le$\,12.8. This incompleteness decreases the number
of detected \ion{H}{i} lines toward lower-$N_{\ion{H}{i}}$ end.
We did not attempt to remove any FPN in our coadding
procedure \citep{wakker15}. With a very conservative approach, we flagged weak absorption 
features in coadded spectra as FPN only when we were certain by
examining individual extractions. Not all of flagged fixed pattern noise were fitted.

In the lower panel of Fig.~\ref{fig8}, the distribution of their 576 \ion{H}{i} components
and 468 FPN features from D16 suggests that FPN features become dominant at 
$\log N_{\mathrm{\ion{H}{i}}}$\,$\le$\,12.8. 
D16 strictly measures all the absorption features at $\ge$3$\sigma$, thus
their detection of weak absorption features is likely to be more objective and less biased. 
About 61\% (287/468) of absorption features flagged as FPN in D16 are not measured in our study.
However, their identification of weak lines should be taken with caution. For example, their \ion{H}{i} features
at $\sim$1288\,\AA\ toward HE\,1228+0131 (their Q\,1230+0115) and at $\sim$1292\,\AA\  
toward 1H\,0419$-$577 (their RBS\,542) are likely to be FPN as shown in Fig.~\ref{fig3}.

\section{Transmitted \ion{H}{i} flux statistics}
\label{sect4}

\subsection{The mean flux and the flux PDF}

\label{sect4.1}

The two simplest measurements of the amount of intergalactic \ion{H}{i} 
are the transmitted mean flux  and the transmitted flux probability distribution function (PDF).
Both measurements are motivated by the current picture of the IGM in
which the absorption arises from continuous matter fluctuations instead of
discrete clouds, so are measured from the {\it continuous}
spectrum and often referred to as ``continuous flux statistics".

The mean \ion{H}{i} flux is the average intervening absorption along the sightline and is
proportional to the mean $N_{\mathrm{\ion{H}{i}}}$ through
a combination of the gas density, the number of lines and line widths in redshift space. 
For the highly photoionised IGM, $N_{\mathrm{\ion{H}{i}}}$
is inversely proportional to the UV background intensity. In practice,
the mean flux is used to calibrate simulations to observations
and constrains the combined effect of the
baryon density, the amplitude of the matter
density fluctuation $\sigma_{\mathrm{8}}$, the temperature-density relation and
the UVB \citep{rauch97, kirkman07, becker13, onorbe17}.

The mean \ion{H}{i} flux is
related to
the effective opacity $\tau_{\mathrm{eff}}$,
\begin{equation}
<\!F\!> \,=\, <\!f_{\mathrm{obs}}/f_{\mathrm{cont}}\!> \,= \, <\!e^{-\tau}\!>
\, = e^{-\tau_{\mathrm{eff}}},
\end{equation}

\noindent where $f_{\mathrm{obs}}$ is the observed flux, $f_{\mathrm{cont}}$ is the 
continuum flux, $\tau$ is the optical depth and $\tau_{\mathrm{eff}}$ is the effective
optical depth. The effective optical depth is introduced to account for the fact that when
close to 0 the normalised flux cannot be converted to the correct $\tau$. 
The uncertainty is largely due to the continuum placement and the amount of
unremoved metal lines, but this is not straightforward to determine. 
Based on a visual inspection of each spectrum, we arbitrarlly define the error
as 0.25 times the r.m.s. of the unabsorbed region.

The probability distribution function (PDF or $P(F)$) of the transmitted flux $F$ 
is a higher order
continuous statistic. It is defined as the fraction of pixels having a flux between
$F$ and $F$\,$+$\,$\Delta F$ for a given flux $F$
\citep{jenkins91, rauch97, mcdonald00}. While the mean \ion{H}{i} flux is 
a one-parameter function of $z$, the flux PDF is a two-parameter
function that constrains the amount of
absorptions as a function of $z$ and absorption strength $F$.
Being a higher-order statistic, the PDF is more sensitive
to the profile shape of absorption lines through the density distribution
and thermal state of the IGM than the mean \ion{H}{i} flux \citep{bolton08}.
In practice, the PDF is 
also sensitive to the continuum uncertainties
at $F$\,$\sim$\,1 and to the amount of
unremoved metal lines at $F$\,$\sim$\,0.4 \citep{kim07,  calura12, lee12, rollinde13}.

With a large number of pixels per redshift bin, the conventional
standard deviation significantly
underestimates the actual PDF errors. Therefore, the errors were calculated using a
modified jackknife method as outlined in \citet{lidz06}.
First,
all the individual spectra longer than 50\,\AA\ in each $z$ bin are put together to generate
a single, long spectrum to calculate the averaged PDF, with the
bin size $\Delta F$\,$=$\,0.05 at 0\,$<$\,$F$\,$<$\,1. Pixels with $F$\,$\le$\,0.025 or $F$\,$\ge$\,0.975 
are included in the $F$\,$=$\,0.0 and the $F$\,$=$\,1.0 bins.
Second, this long spectrum was divided into $n_{c}$ chunks with a
length of $\sim$50\,\AA\/. In the $\tilde{z}\!=\!0.08$ bin, $n_{c}$ is 40 from the single long spectrum
composed from 24 individual spectra. If the PDF estimated
at the flux bin $F_i$ is $\widehat{P}(F_i)$ and
the PDF estimated without the $k$-th
chunk at the flux bin $F_i$ is $\widetilde{P}_{k}(F_i)$, then
the variance at a flux bin $F_i$ becomes  

\begin{equation}
\sigma_{i}^2 = \sum_{k=1}^{n_{c}} [\widehat{P}(F_i)-\widetilde{P}_{k}(F_i)]^{2}.
\end{equation}

\noindent This modified jackknife method is not sensitive to the
length of chunks, but the errors become larger when the
number of chunks is too small.

\begin{figure*}

\vspace{-0.1cm}
\includegraphics[width=17.9cm]{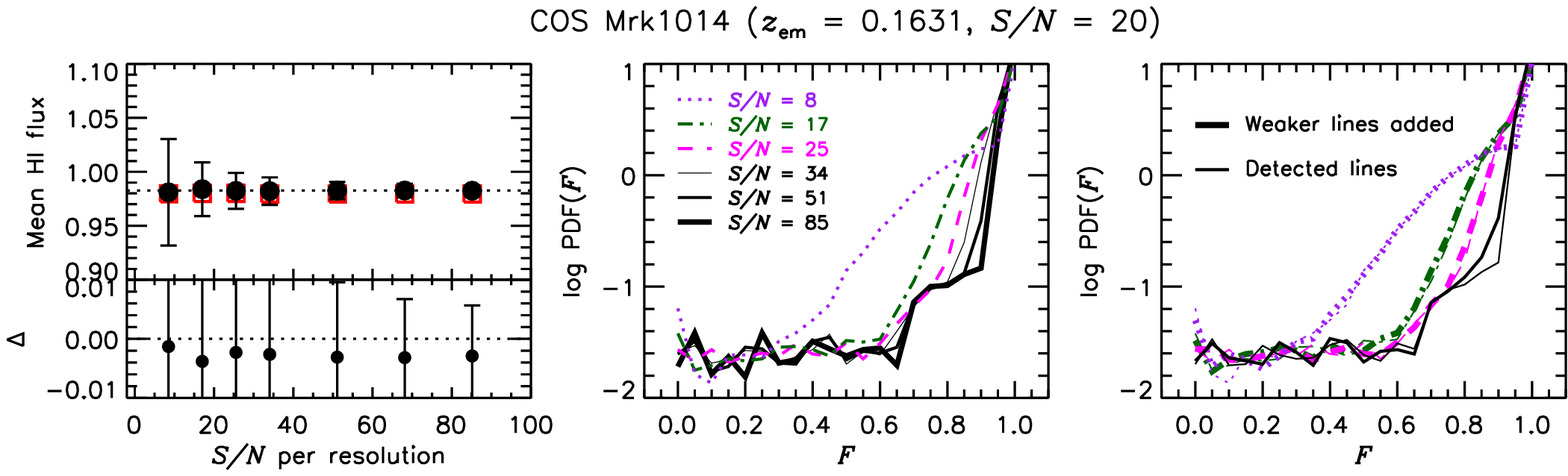}\\
\vspace{-0.5cm}

\caption{The effect of S/N and undetected weak lines on the flux statistics. 
The metal-free spectrum of Mrk\,1014 (COS FUV AGN) was generated from the detected 
lines, assuming the Gaussian LSF of 19\,\kms. We also added 
the several, different artificial Gaussian noise and/or supposedly undetected weak lines
at $\log N_{\mathrm{\ion{H}{i}}}$\,$\in$\,[12.3, 13.0].
{\it Left panel}: The mean \ion{H}{i} flux (filled circles)
does not change with S/N. Undetected weak lines 
due to a low S/N (open red squares without errors for clarity) do not have
any noticeable effect seen from $\Delta\!=\!\bar{F}_{\mathrm{weak\, lines}}\!-\!\bar{F}$
in the lower panel.
{\it Middle panel}: The flux PDF converges at $0.1\!<\!F\!<\!0.7$ if $S/N\!>\!23$.
{\it Right panel}: The PDFs including artificially added
undetected weak lines (thicker curves) are indistinguishable from the PDFs from
detected \ion{H}{i} at $\log N_{\mathrm{\ion{H}{i}}}\!\ge\!13$ (thin curves). The
black curves are for $S/N\!=\!68$, while the colour at other S/N is the same
as used in the middle panel.
}
\label{fig9}
\end{figure*}

\subsection{Data quality on the flux statistics}
\label{sect4.2}

Removing the metal contamination in the  
AGN spectrum is not straightforward, especially when metals can be often blended with
strong \ion{H}{i} complexes over a considerable wavelength range. In addition, 
due to the non-Gaussian LSF of COS and STIS,
the flux statistics directly measured from these spectra cannot be  
compared to the UVES/HIRES spectra (see Section~\ref{sect3.2}).

To avoid these drawbacks, a set of \ion{H}{i}-only spectra was generated
for each AGN. We included the fitted \ion{H}{i} only with $\log N_{\mathrm{\ion{H}{i}}}$\,$<$\,19,
excluding sub-DLAs. 
A Gaussian LSF was assumed to be 19\,\kms\/, 
12\,\kms\/, 10\,\kms\/ and 6.7\,\kms\, for COS FUV, COS NUV, STIS and UVES/HIRES data, 
respectively. Note that the majority of our COS FUV spectra were taken at Lifetime Position 1 
when the approximated Gaussian resolution was $\sim$19\,\kms\/. 
As almost all COS \ion{H}{i} lines are 
resolved, accounting for a degraded resolution by a few km/sec 
at a later Lifetime Position does not make any difference in the generated
spectrum. The wavelength coverage used for
the Ly$\alpha$-only fit is in general larger than for the Lyman series fit, while
both fits reproduce the observed absorption profiles within noise.
Therefore, we used the Ly$\alpha$-only
fit to generate the metal-free spectrum to study the flux statistics. Lowest 
$N_{\mathrm{\ion{H}{i}}}$ included differs for each AGN.
We also added artificial Gaussian noise to each generated spectrum,
using the observed S/N ($S/N$\,$=$\,$1/\sigma$, where 1$\sigma$ is the r.m.s. of 
the unabsorbed region).

The effect of different S/N and undetected weak lines
on the continuous flux statistics is demonstrated in 
Fig.~\ref{fig9}. In the left panel, the filled
circles are $<$$F$$>$ measured from
the generated spectrum of Mrk\,1014 as a function of
artificially added S/N. The mean flux is not sensitive to S/N as expected from 
Gaussian noise being symmetrical at $F$\,$=$\,1, although
the errors (0.25 times the r.m.s. of the unabsorbed region) are larger at lower S/N by definition. 

Mrk\,1014 is one of the 
lowest-S/N COS FUV spectra in this study with a detection limit 
$\log N_{\mathrm{\ion{H}{i}}}$\,$\sim$\,13.0.
However, the highest-S/N COS FUV spectra (3C\,273 and Mrk\,876) show 
\ion{H}{i} at $\log N_{\mathrm{\ion{H}{i}}}$\,$<$\,13.0, indicating that {\it real} weak absorptions
are undetected in low-S/N spectra. We manually add
the expected number of \ion{H}{i} lines at $\log N_{\mathrm{\ion{H}{i}}}$\,$\in$\,[12.3, 13.0] 
by extrapolating from the number of
lines at $N_{\mathrm{\ion{H}{i}}}$\,$>$\,13.0 per $N_{\mathrm{\ion{H}{i}}}$ 
(Section~\ref{sect5.1} for details). 
The red open squares are the mean \ion{H}{i}
flux averaged from 10 generated spectra including artificial weak lines at each S/N. Added weak lines
produce more absorption, but $<$$F$$>$ decreases insignificantly by 
$\sim$0.004, less than 0.5\%. 
The expected decrease becomes even lower for higher-S/N sightlines since they have a 
lower $N_{\mathrm{\ion{H}{i}}}$ detection limit so that the number of added weak lines below
the detection limit down to $\log N_{\mathrm{\ion{H}{i}}}$\,$\sim$\,12.3 is smaller. 
We conclude that undetected weak lines
do not have any meaningful impact on the mean \ion{H}{i} flux.

The S/N has a significant impact on the PDF, as shown in the middle
panel of Fig.~\ref{fig9}. The PDF at
0.1\,$<$\,$F$\,$<$\,0.7 converges if $S/N$\,$>$\,23. 
In the right panel, adding supposedly undetected weak lines 
has a noticeable impact on the PDF only when $S/N$\,$>$\,60 at $F$\,$\ge$\,0.85 since
added weak lines with $\log N_{\mathrm{\ion{H}{i}}}$\,$\le$\,13.0 ($F$\,$\ge$\,0.9) can be 
detected only at high S/N. 
Note that this discrepancy is negligible for COS FUV spectra with observed $S/N$\,$>$\,60, 
since \ion{H}{i} at $\log N_{\mathrm{\ion{H}{i}}}$\,$\in$\,[12.5, 13.0] is detected and included in
the PDF at $F$\,$\ge$\,0.9. 

The PDF at $F$\,$\sim$\,1 is also subject to continuum placement uncertainty,
especially at high redshifts \citep{kim97, calura12, lee12}.
The largest systematic uncertainty comes from the unknown, 
possible overall continuum depression by the Gunn-Peterson effect
\citep{faucher08a}, which is likely to be removed 
during the local continuum fit as we did.
At $z_{\mathrm{em}}$\,$<$\,3.5--3.7, the profile
fit using all the available Lyman lines of the highest-S/N QSO spectra does 
not require a significant Gunn-Peterson depression \citep{calura12}.
Our previous work \citep[their Fig.~2]{kim07} and our experience on high-S/N UVES/HIRES 
QSO spectra suggest that a continuum in general
changes very smoothly over large wavelength ranges. Therefore,
we do not expect our
continuum error is much larger than $\sim$2\% at $z$\,$\sim$\,3 if the S/N is larger than
$\sim$70 per resolution element. Note that 21 out of our 24 UVES/HIRES QSO spectra
have $S/N$\,$\ge$\,70.
Since we apply the same procedure to the continuum placement for our high-$z$
QSO spectra, we assume that a systematic continuum uncertainty is smaller than the statistical
uncertainty at $z$\,$<$\,3.5.

Our approach directly removes the metal contribution from the IGM, 
instead of commonly-used 
masking the metal regions \citep{mcdonald01, kirkman07} or removing statistically using the 
metal contribution above the Ly$\alpha$ emission \citep{faucher08a}. 
At $z$\,$<$\,0.5,
metals are almost fully identified, as line blending is low and the Ly$\alpha$ line is
observed down to $z$\,$=$\,0 so that associated metals are easily identified. 
At $z$\,$>$\,1, most medium-strength/strong metal lines are fully identified, however,
weak narrow lines are not. 
Fortunately, when medium-strength/weak
unidentified metal lines are blended with \ion{H}{i} lines, their contribution 
to the whole blended profile is often negligible. We
empirically conclude that the unremoved metal contamination contributes
$\le$1\% to $<$$F$$>$ at $z$\,$\sim$\,3 and only affect the PDF at $F$\,$\sim$\,1.

The PDF from most COS FUV and STIS spectra (S/N\,$\sim$\,18--40) is sensitive to the
continuum placement at $F$\,$\sim$\,1 and to S/N at $F$\,$>$\,0.7, and the PDF from most 
UVES/HIRES spectra (S/N\,$\ge\!60$) has the largest uncertainty at $F$\,$\sim$\,1 due to 
the continuum error. Out of five 
COS NUV spectra, only one (HE\,1211--1322) has a lower S/N (10--15) than the S/N cut of 18
for COS FUV data. However, its contribution to the total wavelength length at $z$\,$\sim$\,1 is 
only 18\%. 
Therefore, we will consider the PDF only at 0.1\,$<$\,$F$\,$<$\,0.7 at 0\,$<$\,$z$\,$<$\,3.6
in this study.

\subsection{The observed mean \ion{H}{i} flux}

\begin{table}
\caption{Averaged mean \ion{H}{i} flux $<$$F$$>_{\mathrm{ave}}$}
\label{tab4}
{\small{
\begin{tabular}{cccc}
\hline
\noalign{\smallskip}
$\tilde{z}$ & $z$ range & \# of AGN & $<$$F$$>_{\mathrm{ave}}{\bf ^{\mathrm{a}}}$ \\
\noalign{\smallskip}
\hline
\noalign{\smallskip}
 0.08 &  0.00--0.15 &  40 & 0.983$\pm$0.003$\pm$0.006 \\ 
 0.25 &  0.15--0.45 &  24 & 0.978$\pm$0.002$\pm$0.005 \\ 
 0.98 &  0.78--1.29 &    5 & 0.943$\pm$0.006$\pm$0.010 \\
 2.07 &  1.85--2.30 &   17 & 0.872$\pm$0.013$\pm$0.001 \\
 2.54 &  2.30--2.80 &   12 & 0.790$\pm$0.014$\pm$0.001 \\
 2.99 &  2.80--3.20 &    6 &  0.719$\pm$0.017$\pm$0.001 \\
 3.38 &  3.20--3.55 &    2 &  0.642$\pm$0.016$\pm$0.001 \\

\hline
\end{tabular}
}}

\begin{flushleft}
{\small
{\setlength{\itemsep}{1pt}
{Notes -- a:} The first error is the jackknife error of individual $<$$F$$>$ values and the second
error is the standard deviation of their adopted associated error (0.25$\sigma$).
}}
\end{flushleft}

\end{table}

\begin{figure}

\vspace{-0.1cm}
\includegraphics[width=8.2cm]{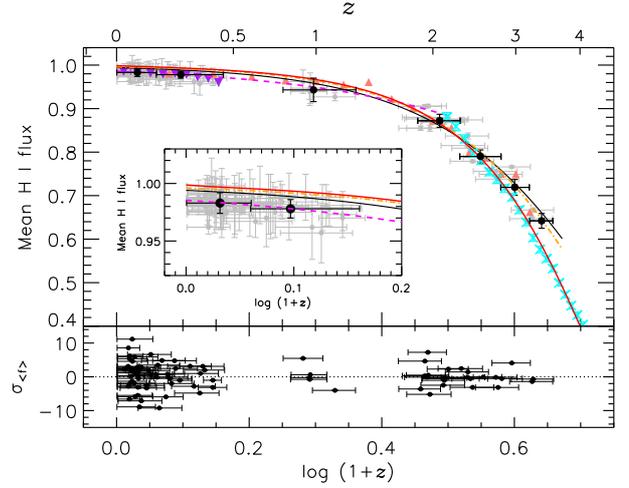}\\
\vspace{-0.6cm}

\caption{{\it Upper panel:} The mean \ion{H}{i} flux as a function of $z$ (upper x-axis)
and $\log (1$\,$+$\,$z)$ (lower x-axis) 
is plotted as gray filled circles for individual sightlines and as
filled circles for the averaged mean flux for each $z$ bin.
The x-axis error is the $z$ range. The y-axis error is the 0.25 r.m.s. of unabsorbed 
regions for individual sightlines and is
the sum of the jackknife error and 
standard deviation of the errors of individual $<$$F$$>$ in each bin for 
the averaged mean flux.
The inset plot shows the sightline variation at low $z$ more
clearly. In both panels, the solid curve 
is a single power-law model for our individual measurements
at 0\,$<$\,$z$\,$<$\,3.6, while
the magenta and orange dashed curves 
are the fit for $z$\,$<$\,1.5 and $z$\,$>$\,1.5, respectively. The red curve is the single 
exponential fit $\tau$\,$=$\,$0.00126 \times
e^{(3.294\times \sqrt{z})}$ suggested by \citet{onorbe17}.
The dark-orange triangles, upside-down purple triangles and cyan crosses are taken from
\citet{kirkman07}, \citet{danforth16} and \citet{becker13}, respectively. 
{\it Lower panel:} Deviation of the individual mean flux from $<$$F$$>_{\mathrm{ave}}$.
}
\label{fig10}
\end{figure}

The upper panel of Fig.~\ref{fig10} plots the mean \ion{H}{i} flux of individual AGN from the 
Ly$\alpha$-only fit as a function of $\log (1$\,$+$\,$z)$ with gray filled circles. The mean flux 
toward each sightline is available as an online table on the MNRAS website (Table~S1). 
The adopted error of 0.25$\sigma$ of unabsorbed regions
does not reflect a {\it true relative} error, but the S/N of each spectrum, and this adopted
error is likely to be over-estimated.
The filled circles are the {\it averaged} mean \ion{H}{i} flux
$<$$F$$>_{\mathrm{ave}}$, listed in Table~\ref{tab4}.  
This is not an arithmetic mean of individual $<$$F$$>$ at each $z$ bin, but 
is estimated from a single long spectrum combined from all the  
generated \ion{H}{i}-only spectra with an appropriate Gaussian noise.
Due to a large number of pixels in each $z$ bin, any standard error estimates
significantly under-estimate a true error. Therefore, 
we used the sum of the two error estimates: the jackknife error of individual 
$<$$F$$>$ values in the $z$ bin and the
standard deviation of the associated error (0.25$\sigma$) of individual $<$$F$$>$ 
to account for a continuum uncertainty.
Our measurement is consistent with the previous observations
within the errors. 

The mean flux from each sightline shows a large scatter (the inset plot). 
This scatter is more clearly seen in the lower panel. The deviation from the
averaged mean flux at each sightline is calculated using the standard error 
($1\sigma_{<F>}$\,$=$\,$1\sigma/\sqrt{N}$ with $N$ being the number of sightlines)
of the {\it arithmetic} mean of all the 
sightlines within a given redshift range $\Delta z$, but excluding the sightline in consideration.
Due to the paucity of data points at higher redshifts, we use
a different $\Delta z$ at different redshifts:
$\Delta z$\,$=$\,0.05 at $z$\,$<$\,0.45,  $\Delta z$\,$=$\,0.51 at $z$\,$\sim$\,1,
$\Delta z$\,$=$\,0.2 at 1.9\,$<$\,$z$\,$<$\,3.0 and $\Delta z$\,$=$\,0.35 at 
3.0\,$<$\,$z$\,$<$\,3.6, respectively. About 71\% of the sightlines have a mean flux at
$\ge$1$\sigma_{<F>}$ and about 55\% have a mean flux at $\ge$2$\sigma_{<F>}$.
This considerable cosmic variance depends largely 
on the occurrence rate of passing through intervening overdense or underdense environments
such as galaxy groups or galaxy voids. Note that the large discrepancy from the Becker
measurement noted as the cyan cross \citep{becker13} is mainly caused by the fact that our sample
does not have enough sightlines at $z$\,$>$\,3, given that the cosmic variance is important.

The overlaid solid black curve is a conventional single power-law fit
to individual measurements at 0\,$<$\,$z$\,$<$\,3.6, 
$\ln <$$F$$>$\,$=$\,$-\tau_{\mathrm{eff}}$\,$=$\,$A_{0} (1+z)^{\alpha}$ with 
$A_{0}$\,$=$\,$-0.0060\pm0.0001$ and $\alpha$\,$=$\,$2.87\pm0.01$. Note that we
used a median 
error $\pm$0.005 of the UVES/HIRES data as the error of both COS/STIS individual $<$$F$$>$
for this fit, since the adopted error of the latter incorrectly gives more weight to
the UVES/HIRES data at $z$\,$>$\,1.5.
This simple single power law over-predicts 
$<$$F$$>$ at $z$\,$<$\,1.5, i.e. less absorption than the observations. 
The suggested single exponential fit
(red curve) by \citet{onorbe17} also overpredicts the observations at $z$\,$<$\,1.5,
more than a simple power law.

In fact, $<$$F$$>$ increases faster (less absorption) from $z$\,$=$\,3.6\,$\rightarrow$\,1.5, 
slows down at $z$\,$\sim$\,1, then becomes almost invariant at $z$\,$<$\,0.5. This
requires a more complicated fitting function. 
If a double power law to individual data points is assumed, 
$A_{0}$\,$=$\,$-0.0145\pm0.0003$ and
$\alpha$\,$=$\,$1.86\pm0.07$ at $z$\,$<$\,1.5 (magenta dashed curve) and
$A_{0}$\,$=$\,$-0.0040\pm0.0001$ and
$\alpha$\,$=$\,$3.18\pm0.02$ at $z$\,$>$\,1.5 (orange dashed curve), respectively.
Note that a single power-law fit at $z$\,$<$\,0.5 is similar to the fit at $z$\,$<$\,1.5:
$A_{0}$\,$=$\,$-0.0142\pm0.0004$ and
$\alpha$\,$=$\,$2.06\pm0.16$ (not shown). This means that the mean flux does not
show any abrupt evolutionary change at $z$\,$<$\,1.5.

\begin{figure*}

\includegraphics[width=17.5cm]{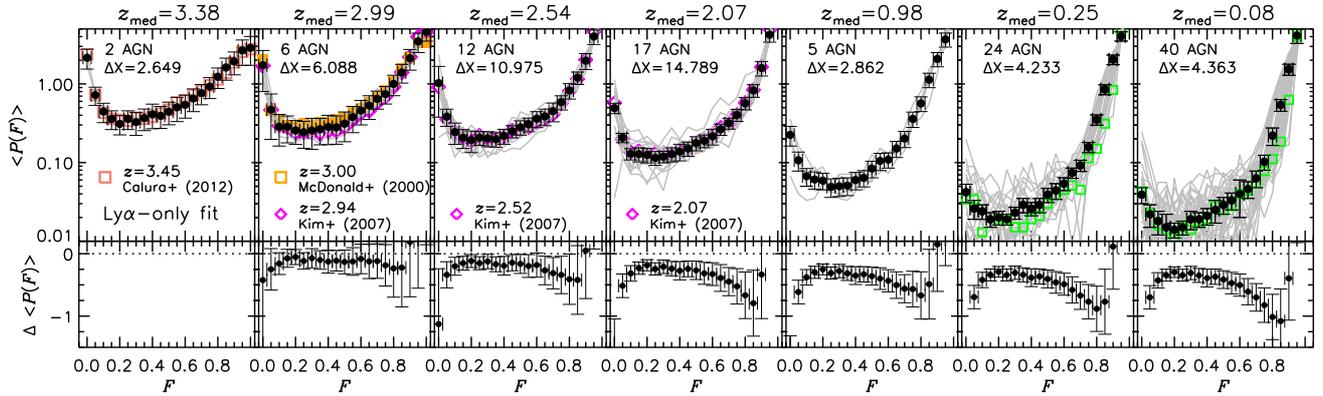}\\
\vspace{-0.3cm}

\caption{{\it Upper panel:}
The observed mean PDF $<$$P(F)$$>$ as a function of $F$ with the filled circles.
The overlaid thin gray curves are the PDF of individual AGN. The x-axis error
shows the $F$ bin size of 0.05, while the y-axis error is from the modified jackknife method.
The green open squares at $z$\,$\sim$\,0.08 and 0.25 are $<$$P(F)$$>$ measured from a 
subset of 14 ($\Delta X$\,$=$\,1.467) and 4
($\Delta X$\,$=$\,0.414) high-S/N COS spectra with $S/N$\,$\ge$\,30. The absorption path length
from the subset is only 30\% and 10\%, which leads to the spiky $<$$P(F)$$>$ as a function of $F$.
In the $\tilde{z}$\,$=$\,2.54 and 2.07 panels, our new measurements
(filled circles) are indistinguishable from our previous ones (open magenta diamonds, \citet{kim07}).
{\it Lower panel:} The difference between the mean
PDF at $\tilde{z}$\,$=$\,3.38 and at the given redshift.
}
\label{fig11}
\end{figure*}

\subsection{The observed flux PDF}  

The upper panel of Fig.~\ref{fig11} shows the mean PDF, $<$$P(F)$$>$, measured from a
single, long spectrum combined from all the \ion{H}{i}-only AGN spectra as filled circles
at each $z$ bin. 
Table~\ref{tab5} lists $<$$P(F)$$>$ and their 
errors estimated from the modified jackknife method.
The absorption path length $\Delta X$ 
noted in each panel provides a relative sample size, 
as the number of included pixels is meaningless due to the
different pixel size for the different data sets. 
The green open squares at $z$\,$\sim$\,0.08 and 0.25 are from a
subset of high-S/N COS spectra. A factor of 10 smaller $\Delta X$ at $z$\,$\sim$\,0.25
causes $<$$P(F)$$>$ from the subset sample to be $\sim$25\% smaller, demonstrating 
importance of a large sample to reduce systematic bias.

At the redshift bin with a large number of AGN,
the individual PDF (thin gray curves) 
varies significantly, 40--50\% at $F$\,$\sim$\,0.5. This sightline 
variance becomes stronger at lower redshifts. This is in part 
caused by the fact that the number of pixels per sightline is on average a 
factor of 18 smaller at $z$\,$<$\,0.5 than at $z$\,$>$\,2.5, i.e. coverage bias, and 
in part by the fact that the forest clustering increases at lower redshifts
\citep{kim97}.

In the $\tilde{z}$\,$=$\,3.38 panel, a noticeable difference exists between
the PDF measured by \citet{calura12} (open dark-orange squares) and our
measurements at $F$\,$>$\,0.5. Although within 1$\sigma$ errors, the amount of 
difference depends on $F$, suggesting that the main cause of the discrepancy
might be the continuum uncertainties at high redshifts \citep{calura12}, in addition to
the small number of sightlines included in both studies and the different redshift
range studied.
In the $\tilde{z}$\,$=$\,2.99 panel, the open orange squares are the PDF at $z$\,$=$\,3.0
measured by \citet{mcdonald00}, about 1.7$\sigma$ larger than our 
present measurements. The discrepancy is in part caused by their imperfect
metal removal as metal contamination increases $<$$P(F)$$>$ especially at 
0.2\,$<$\,$F$\,$<$\,0.6 \citep{kim07}, 
and in part by the sightline variance as their sample size is smaller by a factor of 2.
In the same panel, the open purple triangles are our previous measurement at $z$\,$=$\,2.94
which are $\sim$1.4$\sigma$ smaller \citep{kim07}.
Since we treated the data in a similar manner in both studies,
the discrepancy is likely due 
to the fact that our older sample size is 2 times smaller and the measurement 
was done at a slightly lower $z$.

At each $z$,  the overall shape of $<$$P(F)$$>$ is a convex function with the $z$-independent
minimum at $F$\,$\sim$\,0.2: $<$$P(F)$$>$ rapidly decreases at $F$\,$=$\,0.0\,$\rightarrow$\,0.2, 
then it increases slowly at $F$\,$=$\,0.2\,$\rightarrow$\,0.6 and rapidly at 
$F$\,$=$\,0.6\,$\rightarrow$\,1.0. 
At a given $F$, $<$$P(F)$$>$ decreases rapidly as $z$ decreases 
(the lower panel), consistent with the higher mean flux (lower \ion{H}{i} absorption) 
at lower $z$. If the line width of a typical \ion{H}{i} line is assumed to be $b$\,$\sim$\,25\,\kms,  
$F$\,$=$\,0.3 ($F$\,$=$\,0.7) corresponds to $\log N_{\mathrm{\ion{H}{i}}}$\,$\sim$\,13.7 (13.1). This
approximately translates that only lines with $\log N_{\mathrm{\ion{H}{i}}}$\,$\ge$\,13.7 can
contribute to the PDF at $F$\,$\sim$\,0.3. If we ignore the $b$-dependence on $z$ 
and $N_{\mathrm{\ion{H}{i}}}$,
a factor of 18 lower $<$$P(F$\,$=$\,$0.3)$$>$ at $\tilde{z}$\,$=$\,0.08
than at $z$\,$\sim$\,3.37 indicates that 
the number of \ion{H}{i} absorbers with $\log N_{\mathrm{\ion{H}{i}}}$\,$\ge$\,13.7 is 
a factor of 18 lower at $\tilde{z}$\,$=$\,0.08.


\begin{table*}
\caption{Averaged \ion{H}{i} PDF from the Ly$\alpha$-only fit}
\label{tab5}
{\small{
\begin{tabular}{c c c c c c c c}
\hline
\noalign{\smallskip}
$F$ &  $\tilde{z}$\,$=$\,0.08 & $\tilde{z}$\,$=$\,0.25  &
   $\tilde{z}$\,$=$\,0.98 & $\tilde{z}$\,$=$\,2.07  &
   $\tilde{z}$\,$=$\,2.54 & $\tilde{z}$\,$=$\,2.99 & 
   $\tilde{z}$\,$=$\,3.38 \\
\noalign{\smallskip}

  & $z$\,$=$\,0.00--0.15 &  $z$\,$=$\,0.15--0.45 & $z$\,$=$\,0.78--1.29 &  $z$\,$=$\,1.85--2.30 & $z$\,$=$\,2.30--2.80 
  & $z$\,$=$\,2.80--3.20 &  $z$\,$=$\,3.20--3.55 \\

\hline 
\noalign{\smallskip}

  0.00 &   0.039$\pm$0.009 &   0.042$\pm$0.011 &   0.225$\pm$0.063 &   0.493$\pm$0.074 &   1.024$\pm$0.646 &   1.729$\pm$0.930 &   2.152$\pm$0.609 \\ 
  0.05 &   0.022$\pm$0.007 &   0.026$\pm$0.007 &   0.107$\pm$0.029 &   0.207$\pm$0.028 &   0.383$\pm$0.094 &   0.471$\pm$0.274 &   0.721$\pm$0.187 \\ 
  0.10 &   0.018$\pm$0.006 &   0.024$\pm$0.005 &   0.067$\pm$0.017 &   0.129$\pm$0.021 &   0.244$\pm$0.074 &   0.286$\pm$0.124 &   0.445$\pm$0.123 \\ 
  0.15 &   0.015$\pm$0.007 &   0.019$\pm$0.004 &   0.061$\pm$0.015 &   0.128$\pm$0.026 &   0.207$\pm$0.056 &   0.285$\pm$0.107 &   0.359$\pm$0.096 \\ 
  0.20 &   0.014$\pm$0.003 &   0.020$\pm$0.004 &   0.059$\pm$0.014 &   0.125$\pm$0.026 &   0.193$\pm$0.059 &   0.256$\pm$0.086 &   0.310$\pm$0.085 \\ 
  0.25 &   0.015$\pm$0.003 &   0.019$\pm$0.003 &   0.049$\pm$0.013 &   0.115$\pm$0.023 &   0.207$\pm$0.041 &   0.243$\pm$0.091 &   0.359$\pm$0.096 \\ 
  0.30 &   0.019$\pm$0.005 &   0.023$\pm$0.004 &   0.050$\pm$0.013 &   0.119$\pm$0.020 &   0.203$\pm$0.044 &   0.256$\pm$0.109 &   0.328$\pm$0.107 \\ 
  0.35 &   0.019$\pm$0.004 &   0.029$\pm$0.006 &   0.051$\pm$0.013 &   0.128$\pm$0.026 &   0.198$\pm$0.041 &   0.266$\pm$0.098 &   0.367$\pm$0.112 \\ 
  0.40 &   0.021$\pm$0.005 &   0.026$\pm$0.005 &   0.060$\pm$0.015 &   0.138$\pm$0.033 &   0.218$\pm$0.047 &   0.284$\pm$0.115 &   0.411$\pm$0.142 \\ 
  0.45 &   0.025$\pm$0.008 &   0.029$\pm$0.006 &   0.064$\pm$0.015 &   0.151$\pm$0.031 &   0.250$\pm$0.044 &   0.281$\pm$0.110 &   0.394$\pm$0.122 \\ 
  0.50 &   0.029$\pm$0.007 &   0.039$\pm$0.008 &   0.084$\pm$0.020 &   0.177$\pm$0.036 &   0.281$\pm$0.058 &   0.312$\pm$0.128 &   0.444$\pm$0.142 \\ 
  0.55 &   0.033$\pm$0.008 &   0.044$\pm$0.010 &   0.105$\pm$0.026 &   0.189$\pm$0.035 &   0.307$\pm$0.061 &   0.379$\pm$0.139 &   0.504$\pm$0.172 \\ 
  0.60 &   0.040$\pm$0.020 &   0.054$\pm$0.010 &   0.109$\pm$0.024 &   0.216$\pm$0.049 &   0.361$\pm$0.083 &   0.459$\pm$0.158 &   0.543$\pm$0.176 \\ 
  0.65 &   0.046$\pm$0.011 &   0.074$\pm$0.011 &   0.151$\pm$0.035 &   0.265$\pm$0.043 &   0.386$\pm$0.092 &   0.530$\pm$0.177 &   0.654$\pm$0.210 \\ 
  0.70 &   0.063$\pm$0.015 &   0.092$\pm$0.015 &   0.202$\pm$0.047 &   0.316$\pm$0.050 &   0.451$\pm$0.121 &   0.644$\pm$0.166 &   0.763$\pm$0.220 \\ 
  0.75 &   0.102$\pm$0.022 &   0.157$\pm$0.023 &   0.359$\pm$0.079 &   0.403$\pm$0.070 &   0.586$\pm$0.120 &   0.740$\pm$0.185 &   0.926$\pm$0.275 \\ 
  0.80 &   0.220$\pm$0.055 &   0.351$\pm$0.052 &   0.565$\pm$0.120 &   0.568$\pm$0.101 &   0.825$\pm$0.183 &   0.999$\pm$0.291 &   1.234$\pm$0.359 \\ 
  0.85 &   0.540$\pm$0.092 &   0.851$\pm$0.123 &   1.132$\pm$0.236 &   0.827$\pm$0.151 &   1.199$\pm$0.216 &   1.397$\pm$0.373 &   1.620$\pm$0.506 \\ 
  0.90 &   1.533$\pm$0.246 &   2.043$\pm$0.285 &   2.078$\pm$0.426 &   1.595$\pm$0.335 &   1.977$\pm$0.450 &   2.118$\pm$0.623 &   1.928$\pm$0.622 \\ 
  0.95 &   4.123$\pm$0.852 &   4.049$\pm$0.544 &   3.685$\pm$0.746 &   4.216$\pm$0.819 &   4.007$\pm$0.834 &   3.463$\pm$0.955 &   2.670$\pm$0.948 \\ 
  1.00 &  13.066$\pm$2.807 &  11.988$\pm$1.628 &  10.738$\pm$2.208 &   9.494$\pm$2.014 &   6.494$\pm$1.117 &   4.601$\pm$1.099 &   2.870$\pm$1.145 \\ 

 \noalign{\smallskip}
\hline
\end{tabular}
}}
\end{table*}

\begin{figure}

\includegraphics[width=8.2cm]{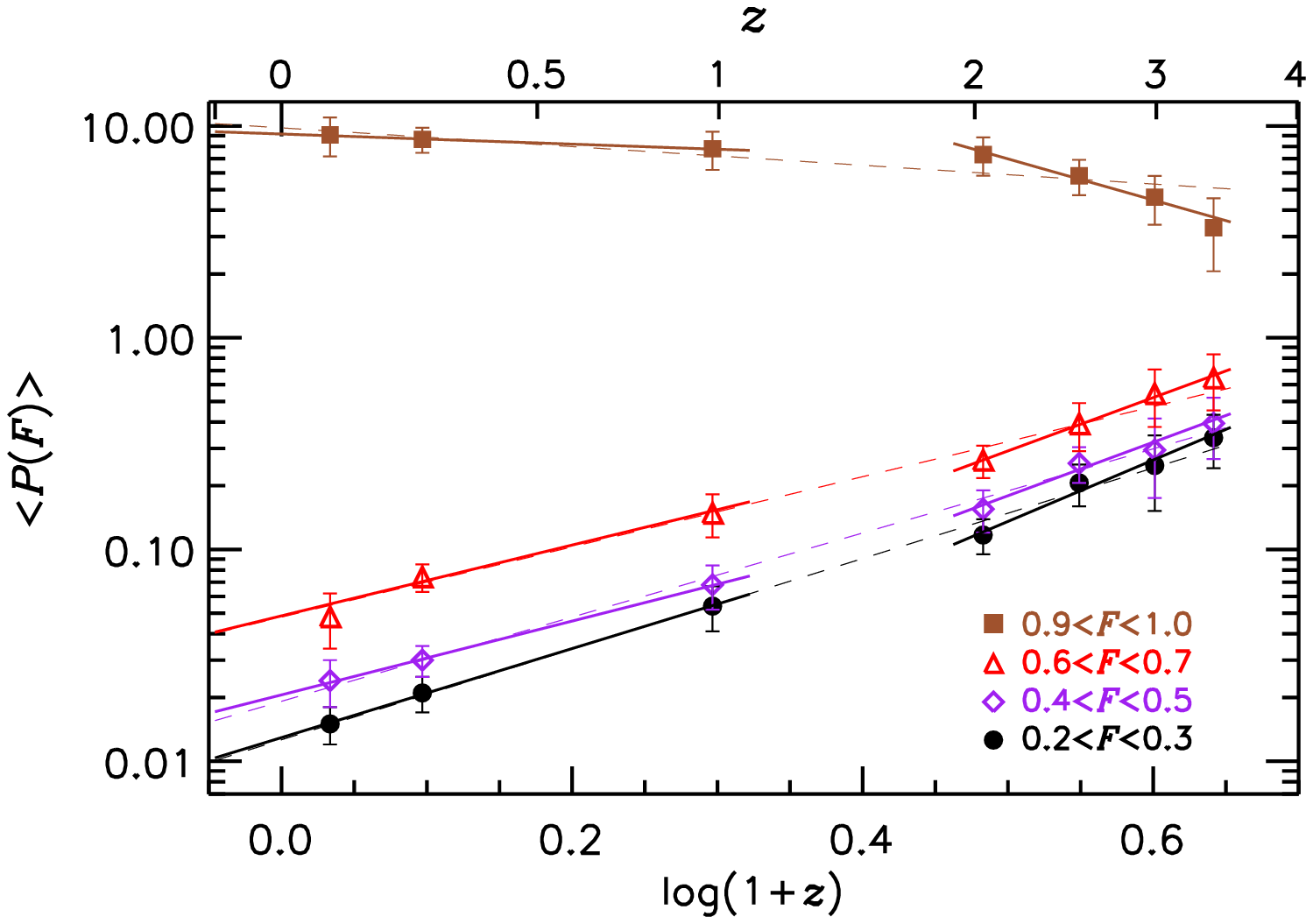}\\
\vspace{-0.5cm}

\caption{The redshift evolution of $<$$P(F)$$>$ at the larger $F$ bin size of 0.1.
The $F$\,$=$\,0.95 bin is included only for a qualitative comparison. 
} 
\label{fig12}
\end{figure}

The $z$-evolution of the PDF is more clearly illustrated in Fig.~\ref{fig12} with a larger $F$ bin 
size $\Delta F$\,$=$\,0.1 to decrease a statistical fluctuation caused by a smaller $F$ range. 
The overlaid dashed line is
a single power-law fit $P(F, z)$\,$=$\,$C_{0}(1$\,$+$\,$z)^{C_{1}}$ at 0\,$<$\,$z$\,$<$\,3.6, 
while the solid line is a double power-law fit at $z$\,$<$\,1.5 and $z$\,$>$\,1.5, respectively,
with the fit parameters listed in online Table~S2 on the MNRAS web site. 

This evolution reflects the fact that Ly$\alpha$ forest absorption typically probes rarer,
higher density gas toward lower redshift due to the evolution of the UVB and the
decrease in the proper density of gas in the IGM \citep{khaire19}. 
Although a different IGM structure corresponds to a different $F$ (or
$N_{\mathrm{\ion{H}{i}}}$) at a different $z$ due to large-scale structure evolution 
\citep{dave99, schaye01, hiss18}, 
the pixels with 0.2\,$<$\,$F$\,$<$\,0.7 and $F$\,$\sim$\,1
can be considered to sample roughly the filaments/sheets and cosmic flux voids
(under-dense regions and regions under enhanced ionisation radiation)
of the low-density IGM structure, respectively.
The $<$$P(F,z)$$>$ measurements shown in Fig.~\ref{fig12} 
qualitatively suggest that the volume fraction of flux voids increases 
rapidly from $z$\,$\sim$\,3.5 down to $z$\,$\sim$\,1.5, reflecting the higher Hubble expansion 
rate and also probably the rapidly increasing number of UV \ion{H}{i} ionising photons 
compared to
lower redshifts \citep{theuns98a, dave99, haardt12}. The volume fraction  
increases slowly at $z$\,$<$\,1.5. In contrast, the volume fraction occupied 
by IGM filaments
and sheets decreases continuously with time, faster at $z$\,$>$\,1.5 and 
slower at $z$\,$<$\,1.5.

\section{Absorption line statistics}
\label{sect5}

\begin{figure*}

\includegraphics[width=18cm]{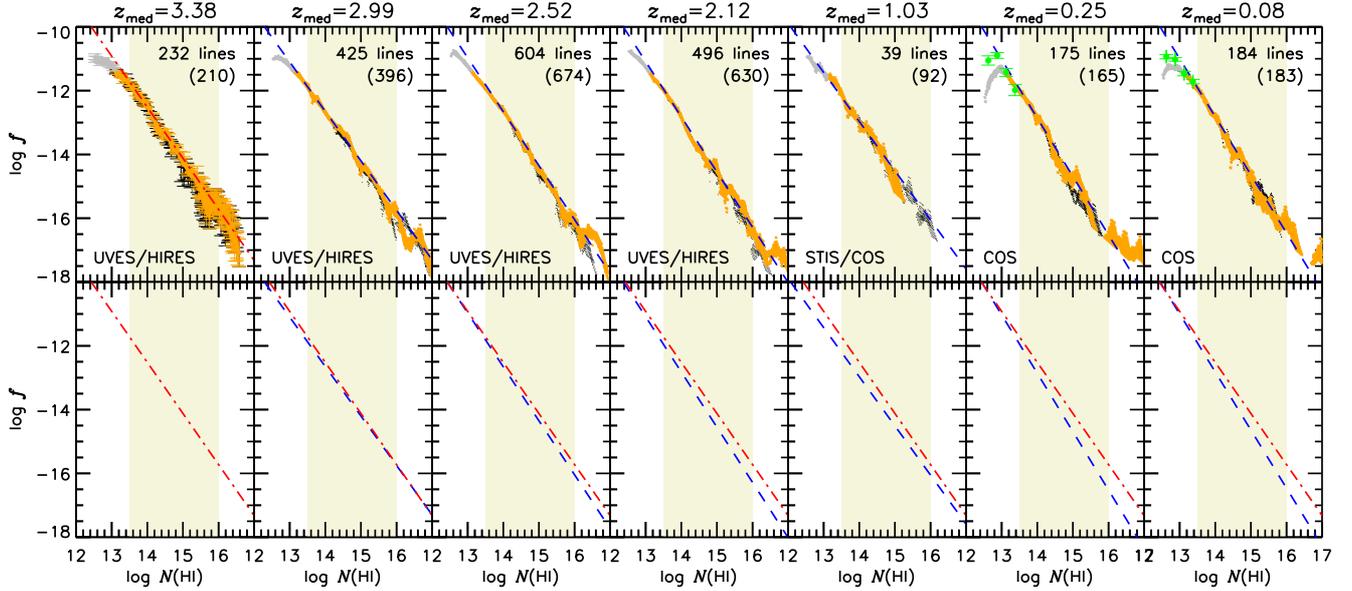}\\
\vspace{-0.25cm}

\caption{{\it Upper panel:} The logarithmic CDDF ($\log f$)
as a function of $\log N_{\mathrm{\ion{H}{i}}}$.
The orange and black dots are the CDDF measured 
from the Lyman series and the Ly$\alpha$-only
fits above the detection limit for each $z$ bin, while
gray dots are measured from the Lyman series fit below the detection limit.
The filled green circles at $\tilde{z}\!=\!0.08$ and $\tilde{z}\!=\!0.25$ plot the CDDF 
at $\log N_{\mathrm{\ion{H}{i}}}$\,$\le$\,13.5
measured from a subset of the high-S/N COS spectra used in Fig.~\ref{fig11}, clearly
demonstrating the impact of incompleteness. 
For clarity, the Poisson errors are shown only in the $\tilde{z}$\,$=$\,3.38 panel 
for selected data points. In the upper right corner the first (second) value is the number of lines at
$\log N_{\mathrm{\ion{H}{i}}}$\,$\in$\,[13.5, 16.0] from the Lyman series (Ly$\alpha$-only) fit.
The red dot-dashed line on the $\tilde{z}$\,$=$\,3.38 panel the is a power-law fit to the lines 
from the Lyman
series fit 
at $\log N_{\mathrm{\ion{H}{i}}}$\,$\in$\,[13.5, 16.0] (beige shaded regions),
while the blue dashed line on every panel is a power-law at the given $z$ for the Lyman-series-fit lines
over the same $N_{\mathrm{\ion{H}{i}}}$ range.
{\it Lower panel:} Comparison between a power-law fit at $\tilde{z}$\,$=$\,3.38 (red dot-dashed line)
and other redshifts (blue dashed lines).
}
\label{fig13}
\end{figure*}

Our three data sets almost fully resolve  the IGM \ion{H}{i} at 
$\log N_{\mathrm{\ion{H}{i}}}\!\le\!17$. Therefore,
the reliability of absorption line statistics
combined from lower-S/N COS/STIS data and higher-S/N UVES/HIRES
data is largely dependent on the chosen \ion{H}{i} column density range
for which each data set provides robust fitted line parameters, i.e.
above the detection limit of $N_{\mathrm{\ion{H}{i}}}$. In order to obtain a
reliable $N_{\mathrm{\ion{H}{i}}}$ of saturated lines, our fiducial line parameter
for absorption line statistics is from the Lyman series fit.

\subsection{The \ion{H}{i} column density distribution function}
\label{sect5.1}

The \ion{H}{i} column density distribution function (CDDF) is 
an analogue of the galaxy luminosity function. It is defined by
the number of absorbers per \ion{H}{i} column density
and per absorption distance path length $dX$ as defined by Eq.~\ref{eq1} 
\citep{rahmati12}:

\begin{equation}
f(N_{\mathrm{\ion{H}{i}}}, dX) \equiv \frac{d^{2} n}{d N_{\mathrm{\ion{H}{i}}} \, d X}
  \equiv  \frac{d^{2} n}{d N_{\mathrm{\ion{H}{i}}}\, d z} \frac{H(z)}{H_{0}}
  \frac{1}{(1+z)^{2}},
\end{equation}

\noindent where $n$ is the number of absorbers in a column density range
$d N_{\mathrm{\ion{H}{i}}}$ centred on $N_{\mathrm{\ion{H}{i}}}$ and in the redshift
range $dz$ centred on $z$. Tables~\ref{tab1}, \ref{tab2} and \ref{tab3} list
$dX$ without excluded regions, i.e. the Galactic ISM-contaminated regions.
Since photons produced by the UVB, stellar and/or AGN feedback affect the
observed $N_{\mathrm{\ion{H}{i}}}$, comparisons between the observed
and simulated CDDFs have been used to probe the importance of these effects
\citep{kollmeier14, shull15, gurvich17, viel17, gaikwad19}. 

At $z$\,$\sim$\,3, the shape of the CDDF at the entire observable range 
$\log N_{\mathrm{\ion{H}{i}}}$\,$\in$\,[12.5, 22.0]
displays various dips and knees due to the 
non-uniform spatial distribution of \ion{H}{i} absorbers, importance of self-shielding,
changes in the UVB and the ionisation state of absorbers and galaxy feedback
\citep{noterdaeme09, dave10, prochaska10, altay11, rahmati12, kim13, omeara13, rudie13}. 
However, it has been customary to fit the CDDF with a power law over a smaller 
$N_{\mathrm{\ion{H}{i}}}$ range, 
$f$\,$=$\,$B N_{\mathrm{\ion{H}{i}}}^{-\beta}$, with $\beta$\,$\sim$\,1.5 at $z$\,$\sim$\,3
for the forest \citep{carswell87, petitjean93, hu95, kim13, rudie13}.

Since the same $N_{\mathrm{\ion{H}{i}}}$ samples a higher overdensity
at lower $z$, i.e. probing different structures at different $z$,
the slope $\beta$ is also expected to change with $z$ due to
structure formation/evolution. Indeed, various simulations have predicted a steepening 
of the CDDF slope from $\sim$1.5 at $z$\,$=$\,2 to $\sim$1.9 at $z$\,$\sim$\,0
\citep{paschos09, dave10, tepper12, nasir17}.
A few existing low-$z$ IGM studies at $z$\,$<$\,2 find indeed a steeper  
$\beta$\,$\sim$\,1.7 at $\log N_{\mathrm{\ion{H}{i}}}$\,$\in$\,[13, 16], without any hint
of dips and knees \citep{lehner07, janknecht06, tilton12, danforth16}. 
\citet{penton04} suggested a deviation from a single power law at $z$\,$\sim$\,0.03. 
However, their \ion{H}{i} column density was converted from 
the equivalent width assuming a fixed $b$ value for all \ion{H}{i} lines, which
may result in an incorrect conclusion.

The upper panel of Fig.~\ref{fig13} shows the logarithmic CDDF, $\log f$,
measured from the Lyman series fit 
(orange dots) and the Ly$\alpha$-only fit (black dots). 
The shown CDDF is measured 
at $\log N_{\mathrm{\ion{H}{i}}}$\,$\in$\,[12.5, 17.0] with a 
$\log N_{\mathrm{\ion{H}{i}}}$ bin size varying randomly between 0.1 and 0.5 to 
capture the various CDDF features in details. 
A total of 53 such measurements were performed with $\sim$500 data
points per unit $\log N_{\mathrm{\ion{H}{i}}}$.
This approach can produce several CDDF measurements at the same 
$N_{\mathrm{\ion{H}{i}}}$, but each CDDF is measured over a different 
$\Delta \log N_{\mathrm{\ion{H}{i}}}$, 
e.g. the number of lines whose $N_{\mathrm{\ion{H}{i}}}$
is in $\log N_{\mathrm{\ion{H}{i}}}$\,$=$\,13.5\,$\pm$\,0.3 vs
$\log N_{\mathrm{\ion{H}{i}}}$\,$=$\,13.5\,$\pm$\,0.5.
A large scatter at a given $N_{\mathrm{\ion{H}{i}}}$
indicates that the lines whose $N_{\mathrm{\ion{H}{i}}}$ is around this 
$N_{\mathrm{\ion{H}{i}}}$ are rare and are not uniformly
distributed in redshift space. At $z$\,$\sim$\,0.25, 
there are no lines with $\log N_{\mathrm{\ion{H}{i}}}$\,$\sim$\,15.7.

When there are no lines at $N_{\mathrm{\ion{H}{i}}}$\,$\pm$\,$\Delta N_{\mathrm{\ion{H}{i}}}$,
the CDDF is not shown. This is more evident at $\log N_{\mathrm{\ion{H}{i}}}$\,$\ge$\,15.5,
as higher-$N_{\mathrm{\ion{H}{i}}}$ lines are rarer, thus requiring more sightlines.
At $z$\,$\sim$\,1, there exist no lines at $\log N_{\mathrm{\ion{H}{i}}}$\,$\in$\,$[15.2, 17.2]$
from the Lyman series fit (only 39 lines at $\log N_{\mathrm{\ion{H}{i}}}$\,$\in$\,$[13.5, 16.0]$
vs 92 lines from the Ly$\alpha$-only fit). Therefore, there is no CDDF measurement at 
$\log N_{\mathrm{\ion{H}{i}}}$\,$\ge$\,15.3 from the Lyman series fit (orange dots).
The difference between the CDDFs measured from the Lyman series and Ly$\alpha$-only fits
is evident at $\log N_{\mathrm{\ion{H}{i}}}$\,$>$\,14.5, where the line parameters of 
saturated lines cannot be reliably measured from Ly$\alpha$ only.

The turnover of the CDDF at $\log N_{\mathrm{\ion{H}{i}}}$\,$\sim$\,12.5--13.0 
is mainly caused by incompleteness as expected from Fig.~\ref{fig8}.
Due to noise including COS fixed pattern noise (FPN), limited S/N and line blending,
all the real absorption lines around the detection limits of $N_{\mathrm{\ion{H}{i}}}$ and $b$
cannot be detected, causing 
a CDDF turnover below the $N_{\mathrm{\ion{H}{i}}}$ detection limit.
In fact, the CDDF measured from a subset of highest-S/N COS data ($S/N$\,$>$\,30) 
at $\tilde{z}$\,$=$\,0.08 shows the higher CDDF at $\log N_{\mathrm{\ion{H}{i}}}$\,$\le$\,13.1 
(green circles). Without a full FPN characterisation, 
the non-Gaussian COS LSF and low-S/N varying along the 
same COS spectrum, we did not attempt to do incompleteness corrections for COS data.
Similarly, without knowing the amount of line blending at higher redshifts in addition
to the continuum uncertainty, we also did not correct incompleteness for STIS/UVES/HIRES data.

In the upper and lower panels, the red dot-dashed line is a power-law fit to the 
Lyman-series-fit \ion{H}{i} lines
at $\log N_{\mathrm{\ion{H}{i}}}$\,$\in$\,[13.5, 16.0] at 
$\tilde{z}$\,$=$\,3.38, while the blue dashed line is a power-law fit at each $z$ (Table~\ref{tab6}).
The fit error is the standard deviation of the 53 sets of the
CDDF measurements shown in Fig.~\ref{fig13}.
The slope $\beta$ of the CDDF is
sensitive to the column density range fitted \citep{kim13}. The fit becomes more
reliable with a larger fitting range because small-scale deviations from the power law
are smoothed out. 
At $\log N_{\mathrm{\ion{H}{i}}}$\,$<$\,14.5, IGM \ion{H}{i} lines are 
more uniformly distributed in the intergalactic space for the CDDF 
to follow a power-law distribution. In contrast, at 
$\log N_{\mathrm{\ion{H}{i}}}$\,$>$\,14.5, the IGM distribution starts to show irregularity.
This is in part due to a stronger clustering of 
higher-$N_{\mathrm{\ion{H}{i}}}$ absorbers \citep[D16]{kim97} and in part due to a lower
number of higher-$N_{\mathrm{\ion{H}{i}}}$ absorbers, i.e. 81 absorbers at 
$\log N_{\mathrm{\ion{H}{i}}}$\,$\in$\,[13.5, 13.8] versus two absorbers at
$\log N_{\mathrm{\ion{H}{i}}}$\,$\in$\,[15.5, 15.8] from the Lyman series fit at $\tilde{z}$\,$=$\,3.38.
Therefore, determining a reliable shape for the CDDF at $N_{\mathrm{\ion{H}{i}}}$\,$\ge$\,14.5
requires a larger total path length to decrease the fluctuations by these effects.
Interestingly, this $N_{\mathrm{\ion{H}{i}}}$ range at $\log N_{\mathrm{\ion{H}{i}}}$\,$\ge$\,14.5 
is also where the intergalactic \ion{H}{i} starts to reside in collapsed regions
and to interact with galaxies through IGM accretion and stellar/AGN feedback. 
The interaction between the IGM and galactic outflows
affects the small-scale distribution of high-$N_{\mathrm{\ion{H}{i}}}$ absorbers
around galaxies, which might result in the stronger clustering and the deviation from a
power-law CDDF. 
In addition, the IGM temperature-density relation starts to break down at 
$\log N_{\mathrm{\ion{H}{i}}}$\,$>$\,14.5 \citep{hui97, theuns98b, dave10, 
peeples10, martizzi19}.

\begin{table}
\caption{The CDDF power-law fit at $\log N_{\mathrm{\ion{H}{i}}}$\,$\in$\,[13.5, 16.0]}
\label{tab6}
{\footnotesize{
\begin{tabular}{p{0.3cm}R{1.5cm}C{1.3cm} p{0.3cm}R{1.5cm}C{1.3cm}}
\hline
\noalign{\smallskip}
\multicolumn{3}{c}{Ly$\alpha$-only fit} & \multicolumn{3}{c}{Lyman series fit}  \\[1pt]
\cmidrule(lr){1-3}  \cmidrule(lr){4-6} \\[-10pt]
$\tilde{z}_{\mathrm{Ly}\alpha}$ & $\log B$ & $\beta$  &  $\tilde{z}_{\mathrm{Ly}\alpha\beta}$  
    & $\log B$ & $\beta$ \\[1pt]
\hline
\noalign{\smallskip}

  0.08 &  12.63$\pm$0.43 &   1.82$\pm$0.03 &   0.08 &  12.66$\pm$0.41 &   1.82$\pm$0.03 \\
  0.25 &  14.37$\pm$0.52 &   1.95$\pm$0.04 &   0.25 &  12.76$\pm$0.28 &   1.83$\pm$0.02 \\
  0.98 &   9.52$\pm$0.65 &   1.61$\pm$0.05 &   1.03 &   8.59$\pm$1.09 &   1.54$\pm$0.08 \\
  2.07 &  12.16$\pm$0.17 &   1.79$\pm$0.01 &   2.12 &  11.40$\pm$0.14 &   1.73$\pm$0.01 \\
  2.54 &  11.78$\pm$0.16 &   1.75$\pm$0.01 &   2.52 &  11.00$\pm$0.13 &   1.69$\pm$0.01 \\
  2.99 &  10.53$\pm$0.16 &   1.66$\pm$0.01 &   2.99 &   9.05$\pm$0.17 &   1.55$\pm$0.01 \\
  3.38 &  10.90$\pm$0.33 &   1.68$\pm$0.02 &   3.38 &   9.88$\pm$0.26 &   1.60$\pm$0.02 \\[1pt]
  
\hline

\end{tabular}
}}
\end{table}

\begin{figure}

\includegraphics[width=8.5cm]{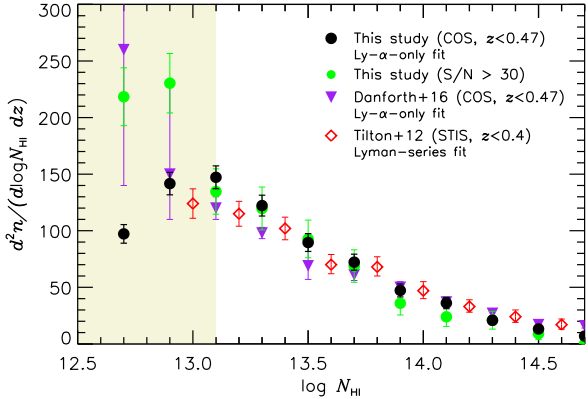}\\
\vspace{-0.6cm}

\caption{Comparisons between the four CDDF measurements 
per unit redshift $dz$ instead of $dX$ at the low-$N_{\mathrm{\ion{H}{i}}}$ end.
Incompleteness causes a turnover in our COS CDDF (filled circles) at 
$\log N_{\mathrm{\ion{H}{i}}}$\,$\sim$\,13.1 (the shaded region), while the 
incompleteness-corrected
D16 COS CDDF continuously increases at $\log N_{\mathrm{\ion{H}{i}}}$\,$\le$\,13.1. 
The CDDF calculated from the high-S/N subsample (green filled circles) used in
Fig.~\ref{fig13} abruptly increases at $\log N_{\mathrm{\ion{H}{i}}}$\,$\sim$\,13.0.
Note that the CDDF plotted depends both on $\log N_{\mathrm{\ion{H}{i}}}$
and $\Delta N_{\mathrm{\ion{H}{i}}}$.}
\label{fig14}
\end{figure}

The impact of incompleteness and the non-uniqueness of fitted line parameters
including spurious lines on the CDDF are more significant at low $N_{\mathrm{\ion{H}{i}}}$ 
as better illustrated in Fig.~\ref{fig14}. Our Ly$\alpha$-fit CDDF (filled circles) shows a turnover at 
$\log N_{\mathrm{\ion{H}{i}}}$\,$\sim$\,13.1.
At $\log N_{\ion{H}{i}}$\,$\le$\,13.5, a typical IGM \ion{H}{i} is too weak 
to produce detectable Ly$\beta$ in COS spectra with $S/N$\,$<$\,25 and there is no significant
difference between the Ly$\alpha$-only and Lyman series fits.
The incompleteness-corrected
D16 COS CDDF (filled purple upside triangles) continuously
increases at $\log N_{\mathrm{\ion{H}{i}}}$\,$\le$\,13.1, while the raw D16 CDDF
is expected to show a similar turnover from Fig.~\ref{fig8}. The STIS CDDF (open red diamonds)
is shown only at $\log N_{\mathrm{\ion{H}{i}}}$\,$\ge$\,13.0 where the impact of incompleteness
becomes negligible \citep{tilton12}. 
Matching the observations and simulations at low-$N_{\mathrm{\ion{H}{i}}}$ end
should be approached with caution.

\begin{figure}

\includegraphics[width=8.5cm]{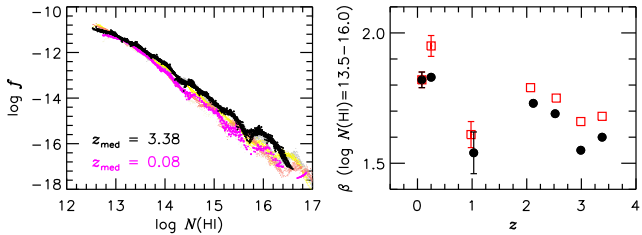}\\
\vspace{-0.8cm}

\caption{{\it Left panel:} The CDDF from the Lyman series fit at the seven redshift bins. The symbol
sizes at $\tilde{z}$\,$=$\,3.38 (highest $z$) and 0.08 (lowest $z$) are a factor of 2 larger to 
contrast the CDDF over the largest $z$ interval as its $z$-evolution is weak. 
The Poisson errors are not shown for clarity.
{\it Right panel:} The CDDF slope $\beta$ measured at 
$\log N_{\mathrm{\ion{H}{i}}}$\,$\in$\,[13.5, 16.0] from the Lyman series fit (filled circles)
and the Ly-$\alpha$-only fit (open red squares). Only errors larger than the symbol size are plotted.
}
\label{fig15}
\end{figure}

The left panel of Fig.~\ref{fig15} displays the redshift evolution of the overall shape
of the CDDF. The CDDF shape at lower redshifts
can be reproduced by a small amount of clockwise rotation of a higher-$z$ CDDF with a 
slightly larger CDDF normalisation $B$. This is caused by a fact that the number of \ion{H}{i} absorbers 
decreases faster at higher $N_{\mathrm{\ion{H}{i}}}$ and at lower $z$ with a self-similar manner
in terms of the evolution of the large-scale structure 
and the degree of the IGM-galaxy interaction as a function of $z$. 

The right panel indicates that the CDDF slope $\beta$ from the Lyman series fit
in general becomes steeper as $z$ decreases, if the $\tilde{z}_{\alpha\beta}$\,$=$\,1.03 CDDF 
is excluded due to the small-number statistics
(Table~\ref{tab6}). At $\tilde{z}_{\alpha\beta}$\,$=$\,1.03, 
a very small $\Delta X$ coverage also decreases a probability of detecting less common
\ion{H}{i} absorbers at $\log N_{\ion{H}{i}}$\,$>$\,15.0. 
These lead to a factor of 3 larger statistical error than at other redshifts.
This trend also holds for $\beta$ estimated from the Ly$\alpha$-only fit.
On the other hand, the lower $\beta$ at $z\!\sim\!3$ compared to at the
adjacent $z$ seems to be real as the number of analysed lines are large enough to obtain
a reliable $\beta$. Due to a lack of data at $z\!>\!3.5$ we cannot discard a possibility
of $\beta$ continuously increasing at $z\!=\!3\!\rightarrow\!4$ with a local minimum at $z\!\sim\!3$, 
which could be 
caused by a change in the IGM $N_{\mathrm{\ion{H}{i}}}$ distribution due to
extra heating and ionisation by \ion{He}{ii} reionisation at $z\!\sim\!3$. 
\citep{reimers97, songaila98, syphers13, worseck16}.

\subsection{The forest gas-phase mass density}

One of the key cosmological parameters constrained by the IGM 
is the gas-phase hydrogen mass density
relative to the critical density of the universe ($\Omega_{\mathrm{H}}$).
This $\Omega_{\mathrm{H}}$ is model-dependent and is  
bound to be revised with an advent of more realistic models and with a better constraint on 
the UVB, the
characteristic size of the IGM geometry and a density profile \citep{schaye01, penton04, tilton12}.
We used a simple method developed by \citet{schaye01} to obtain a qualitative trend over time:

\begin{eqnarray}
\Omega_{\mathrm{H}}  &\!\sim\!& 2.2\!\times\!10^{-9} h^{-1}
   \Gamma_{12}^{1/3}
   \left ({f_{g} \over 0.16}\right )^{1/3}\!T_4^{0.59}  \nonumber \\
   &   &   \times\!\int \!N_{\ion{H}{i}}^{1/3} f(N_{\ion{H}{i}}, dX) \,dN_{\ion{H}{i}},
\label{eq6}
\end{eqnarray} 

\noindent where $f_{g}$ is a mass fraction in gas-phase hydrogen, 
the hydrogen photoionisation rate $\Gamma_{12}\!\equiv\!\Gamma_{\mathrm{\ion{H}{i}}}\!\times\!
10^{-12}$~s$^{-1}$ and the gas temperature $T\! \equiv \!T_{4}\!\times\! 
10^{4}$~K, respectively, for our assumed cosmology $h=0.7$ \citep{schaye01}. 
Strictly speaking, this holds only for overdense regions, i.e. $\log N_{\ion{H}{i}}$\,$\ge$\,13.5
at $z$\,$\sim$\,3 and $\Omega_{\mathrm{H}}$ can be under-estimated by $\sim$20\% at
$\log N_{\ion{H}{i}}$\,$\le$\,13.5 \citep{penton04}.

We directly integrated $f(N_{\mathrm{\ion{H}{i}}}, dX)$ 
as shown in Fig.~\ref{fig13} 
over several different column density ranges with the $\pm$1$\sigma$ Poisson errors.
The model-independent factor 
2.2\,$\times$\,$10^{-9} h^{-1}\!\int\!N_{HI}^{1/3} f(N_{HI}, dX) \,dN_{HI}$ and
the model-dependent $\Omega_{\mathrm{H}}$ are
are tabulated in an online table on the MNRAS website (Table S3).
The model-independent factor is a purely observational quantity and
 will not be likely to be changed significantly within our adopted column density
 range at $\log N_{\ion{H}{i}}$\,$\in$\,[13, 16] in the near future. 
 For the model-dependent $\Omega_{\mathrm{H}}$, 
$\Gamma_{12}$ is interpolated from the HM01 QG UVB at the given $z$, while
$f_{g}$ and $T$ were 
interpolated from the IllustrisTNG simulation
\citep[their Table 1 and Fig. 4, respectively]{martizzi19}. For simplicity, we assume that 
the observed Ly$\alpha$ forest is mostly from the cool diffuse IGM and the halo 
gas in filaments and sheets and that
the minimum temperature of simulated filaments at $\log n_{\mathrm{H}}\!=\!-4$ is
a fair representative of the IGM temperature. An uncertainty of 10\%
in $\Gamma_{12}$, $f_{g}$ and $T_{4}$ changes $\Omega_{\mathrm{H}}$
by $\sim$3\%, $\sim$3\% and $\sim$6\%, respectively, indicating that $T_{4}$
is the most important model-dependent parameter. However, the uncertainties
associated with $T_{4}$ and $\Gamma_{12}$ are likely to be different, especially at
low redshifts. Simulated distributions of \ion{H}{i} line widths which are determined
by gas temperature and non-thermal motion are not in agreement 
with observations by a factor of $\sim$2 at $z$\,$\sim$\,0.1
\citep{viel17}. Several studies also suggest a factor of 2--5 larger $\Gamma_{12}$ than 
the widely-used theoretical prediction by \citet{haardt12} 
at $z$\,$\sim$\,0.2 \citep{kollmeier14, shull15, wakker15, khaire19, faucher20}. 

Being fully consistent with previous studies, 
the low-$N_{\mathrm{\ion{H}{i}}}$
absorbers at $\log N_{\mathrm{\ion{H}{i}}}\!\in\![13, 15]$ contain most baryons at $z$\,$>$\,2.5,
but their contribution decreases down to about 22\% at $z$\,$\sim$\,0 \citep{rauch97, 
shull12b, danforth16}. 
The relative contribution to $\Omega_{\mathrm{b}}$ by absorbers
at $\log N_{\mathrm{\ion{H}{i}}}$\,$\in$\,[13.0, 14.5] and at 
$\log N_{\mathrm{\ion{H}{i}}}$\,$\in$\,[14.5, 16.0] is about 4.5 at $z$\,$\sim$\,0 and 
2 at $z$\,$\sim$\,3.4, which reflects a steeper slope of the CDDF at lower $z$.
Due to incompleteness at $\log N_{\mathrm{\ion{H}{i}}}$\,$\sim$\,13, it is
not currently possible to constrain the contribution to $\Omega_{\mathrm{b}}$ by 
these weaker absorbers.

\begin{figure*}

\hspace{-0.2cm}
\includegraphics[width=18cm]{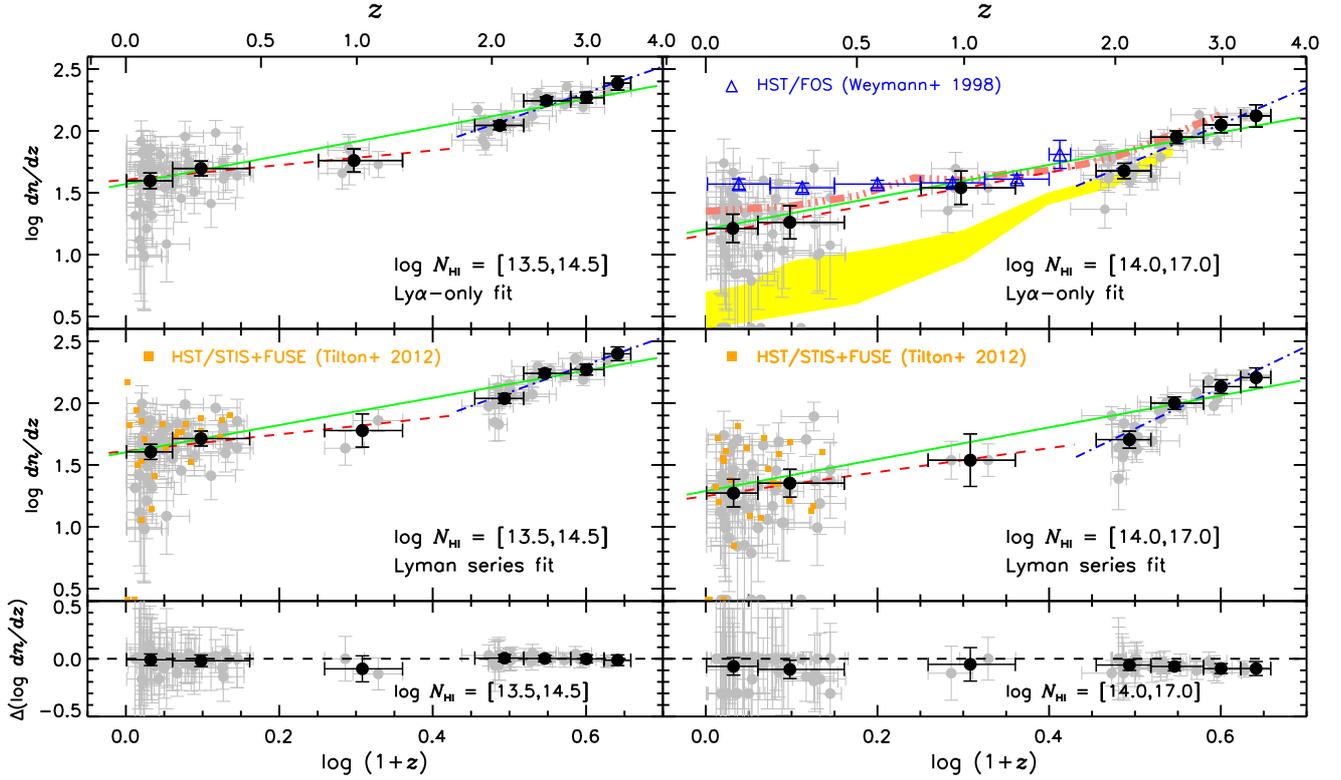}\\
\vspace{-0.75cm}
\caption{{\it Left upper panel:} The redshift evolution of the number of absorbers per unit $z$,
$dn/dz$, from the Ly$\alpha$-only fit at $\log N_{\mathrm{\ion{H}{i}}}$\,$\in$\,[13.5, 14.5].
The individual and averaged $dn/dz$ are shown as filled gray and black circles, respectively. 
The x-axis errors indicate the redshift range, while the y-axis 
errors are the 1$\sigma$ Poisson error accounting for lines with a questionable identification. 
The blue dot-dashed and red dashed lines are a best-fit single power law
to the averaged $dn/dz$ at  $z$\,$>$\,1.5 and $z$\,$<$\,1.5, respectively. The green solid line
is a single power-law fit to the individual $dn/dz$ at 0\,$<$\,$z$\,$<$\,3.6.
{\it Left middle panel:} $dn/dz$ from the Lyman series fit. 
{\it Left lower panel:} The difference in $\log dn/dz$ between the Ly$\alpha$-only and
Lyman series fits. The line number density $dn/dz$ from the Ly$\alpha$-only fit was recalculated 
over the same $z$ range used in the Lyman series fit. 
{\it Right panels:} Same as the left panels except for $\log N_{\mathrm{\ion{H}{i}}}$\,$\in$\,[14, 17].
When there is no line, $\log dn/dz$ is set to be 0.41 without y-axis errors at the bottom of the panel.
The open blue triangles are $dn/dz$
from {\it HST}/FOS spectra \citep{weymann98}, converted from
equivalent width measurements assuming $b$\,$=$\,25\,\kms. 
In the upper panel, the yellow shade outlines the $dn/dz$ range from theoretical predictions by 
\citet{dave10} and \citet{nasir17}, while the pink dot-dot-dot-dashed curve is a prediction
by \citet{dave99}.
}
\label{fig16}
\end{figure*}

\subsection{Absorption line number density $dn/dz$}
\label{sec5.3}

The \ion{H}{i} absorber number density, $dn/dz$, is defined as the number of 
absorbers per unit
redshift. It is proportional to the cross section and comoving
number density of absorbers.
It is usually measured over a specified \ion{H}{i} column density range and its
evolution as a function of $z$ is traditionally described as a single power law, 
$dn/dz$\,$=$\,$n_{0}\!\times\!(1\!+\!z)^{\gamma_{n}}$, where $n_{0}$ is the number density at
$z$\,$=$\,0. 

\begin{table}
\caption{Averaged $dn/dz$}
\label{tab8}
{\scriptsize{
\begin{tabular}{p{0.22cm}p{0.25cm}rp{0.47cm}p{1pt}rp{0.47cm}}
\hline
\noalign{\smallskip}

&   &  \multicolumn{2}{c}{$\log N_{\mathrm{\ion{H}{i}}} \in [13.5, 14.5]$} &
      & \multicolumn{2}{c}{$\log N_{\mathrm{\ion{H}{i}}} \in [14.0, 17.0]$} \\[2pt]
\cline{3-4} \cline{6-7} \\[-3pt]
  \multicolumn{1}{c}{$\log (1$\,$+$\,$\tilde{z})$}    &   \multicolumn{1}{c}{$dz$}     & \# of  &   &  &   \# of  &       \\
&    & lines & $\log dn/dz$  &  & lines & $\log dn/dz$  \\[0.05cm]

\hline \\[-0.2cm]
\multicolumn{7}{c}{Ly$\alpha$-only fit} \\[0.05cm]
\hline \\[-0.15cm] 

 0.032$_{-0.031}^{+0.029}$ & 3.926 & 155 & 1.60$\pm$0.03$\pm$0.03 &  &  64 & 1.21$\pm$0.05$\pm$0.06 \\
 0.098$_{-0.037}^{+0.063}$ & 3.018 & 150 & 1.70$\pm$0.04$\pm$0.02  &  & 55 & 1.26$\pm$0.06$\pm$0.08 \\
 0.297$_{-0.046}^{+0.063}$ & 1.267 &   73 & 1.76$\pm$0.05$\pm$0.04  &  & 44 & 1.54$\pm$0.07$\pm$0.07  \\
 0.487$_{-0.032}^{+0.031}$ & 4.804 & 532 & 2.04$\pm$0.02$\pm$0.02  &  & 228 & 1.68$\pm$0.03$\pm$0.03 \\
 0.548$_{-0.030}^{+0.031}$ & 3.281 & 575 & 2.24$\pm$0.02$\pm$0.01  &  & 292 & 1.95$\pm$0.03$\pm$0.03 \\
 0.600$_{-0.021}^{+0.023}$ & 1.700 & 316 & 2.27$\pm$0.02$\pm$0.02   & & 190 & 2.05$\pm$0.03$\pm$0.03 \\
 0.641$_{-0.018}^{+0.017}$ & 0.703 & 171 & 2.39$\pm$0.03$\pm$0.02   &  & 93 & 2.12$\pm$0.05$\pm$0.05 \\[2pt]
 
\hline \\[-0.2cm]
\multicolumn{7}{c}{Lyman series fit} \\[0.05cm]
\hline \\[-0.15cm]

0.032$_{-0.032}^{+0.028}$ & 3.792 & 153 & 1.61$\pm$0.04$\pm$0.03 &  & 71 & 1.27$\pm$0.05$\pm$0.06 \\
0.098$_{-0.037}^{+0.063}$ & 3.018 & 156 & 1.71$\pm$0.04$\pm$0.03 &  & 70 & 1.37$\pm$0.05$\pm$0.06 \\
0.308$_{-0.049}^{+0.052}$ & 0.550 &   33 & 1.78$\pm$0.08$\pm$0.06 &  & 19 & 1.54$\pm$0.10$\pm$0.11 \\ 
0.494$_{-0.039}^{+0.025}$ & 3.709 & 404 & 2.04$\pm$0.02$\pm$0.02 & & 188 & 1.70$\pm$0.03$\pm$0.04 \\
0.546$_{-0.028}^{+0.034}$ & 2.893 & 503 & 2.24$\pm$0.02$\pm$0.01 & & 289 & 2.00$\pm$0.03$\pm$0.03 \\
0.600$_{-0.021}^{+0.023}$ & 1.700 & 317 & 2.27$\pm$0.02$\pm$0.02 & & 231 & 2.13$\pm$0.03$\pm$0.03 \\
0.641$_{-0.018}^{+0.017}$ & 0.703 & 176 & 2.40$\pm$0.03$\pm$0.02 & & 113 & 2.21$\pm$0.04$\pm$0.04 \\[2pt]
\hline
 
\end{tabular}
}}
\end{table}

Due to the growth of structure, the same \ion{H}{i} column density 
corresponds to a higher overdensity at lower $z$ (overdensity $\delta$\,$=$\,$\rho/\rho_{\mathrm{o}}$,
where $\rho_{\mathrm{o}}$ is the cosmic mean matter density):
$\log N_{\mathrm{\ion{H}{i}}}$\,$=$\,15 corresponds to $\delta$\,$\sim$\,100 
(inside halos) at $z$\,$=$\,0 and $\delta$\,$\sim$\,6 (the diffuse IGM) at $z$\,$=$\,3 \citep{dave99}. 
In addition, star formation and feedback is predicted to affect the \ion{H}{i} absorbers close to
galaxies \citep{dave10, nasir17}.
Therefore, $dn/dz$ is expected to change with $N_{\mathrm{\ion{H}{i}}}$ and $z$,
constraining structure evolution \citep{theuns98a, schaye01, dave10, williger10, kim13}.

Figure~\ref{fig16} displays the $dn/dz$ evolution from the Ly$\alpha$-only (upper panels)
and Lyman series (middle panels) fits at two different
$N_{\mathrm{\ion{H}{i}}}$ ranges, at $\log N_{\mathrm{\ion{H}{i}}}$\,$\in$\,[13.5, 14.5] (left panels)
and at $\log N_{\mathrm{\ion{H}{i}}}$\,$\in$\,[14, 17] (right panels), respectively.
The criterion of $S/N\!>\!18$ for COS/STIS spectra enables 
detection of \ion{H}{i} at $\log N_{\mathrm{\ion{H}{i}}}$\,$\le$\,13. However, since the 
$N_{\mathrm{\ion{H}{i}}}$ detection limit varies with $b$ and the two sightlines at
$z$\,$\sim$\,1 have $S/N$\,$\sim$\,10--18, to be conservative 
we use a lower $N_{\mathrm{\ion{H}{i}}}$ limit of
$\log N_{\mathrm{\ion{H}{i}}}$\,$=$\,13.5.

\begin{figure}

\hspace{-0.2cm}
\includegraphics[width=9cm]{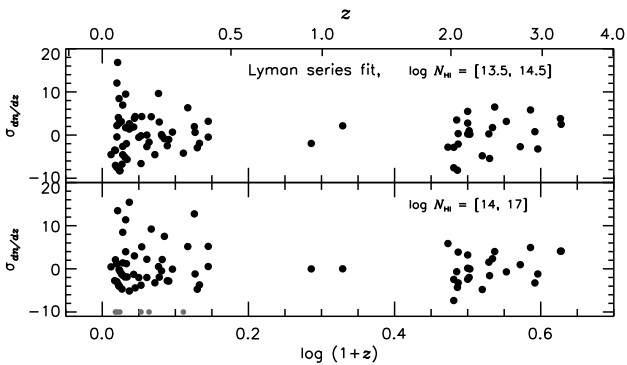}\\
\vspace{-0.65cm}
\caption{Deviation of the individual $dn/dz$ from the averaged $dn/dz$ for the Lyman
series fit. The deviation
is calculated using the Poisson error of the averaged $dn/dz$ within a given redshift range 
$\Delta z$ excluding the sightline in consideration:
$\Delta z$\,$=$\,0.05 at $z$\,$<$\,0.45,
$\Delta z$\,$=$\,0.48 at $z$\,$\sim$\,1, $\Delta z$\,$=$\,0.2 at 1.9\,$<$\,$z$\,$<$\,3.0 
and  $\Delta z$\,$=$\,0.35 at 3.0\,$<$\,$z$\,$<$\,3.6, respectively. For a sightline 
without \ion{H}{i} absorbers in a given column density range, $\sigma_{dn/dz}$ is
assigned to be $-10$ with gray circles. 
The positive deviation indicates that the sightline contains more \ion{H}{i}
absorbers than the averaged $dn/dz$. 
}
\label{fig17}
\end{figure}

The most striking feature of $dn/dz$ in the upper and middle panels of
Fig.~\ref{fig16} is a large scatter in
individual $dn/dz$ (gray filled circles, tabulated in the supplementary online
Tables S4 and S5 on the MNRAS webpage)
at any given redshift for both $N_{\mathrm{\ion{H}{i}}}$
ranges. The scatter becomes larger 
at lower redshifts, spanning about an order of magnitude at $z$\,$\sim$\,0 (Fig.~\ref{fig17}).
About half the COS AGN sightlines at $z$\,$<$\,0.5 do not contain an absorber at
$\log N_{\mathrm{\ion{H}{i}}}$\,$\ge$\,14.5. At the same time, Fig.~\ref{fig17} indicates that
9\% (5/55) of sightlines at $z$\,$<$\,0.5 contain more absorbers with $\ge$8$\sigma$ 
at $\log N_{\mathrm{\ion{H}{i}}}$\,$\in$\,[13.5, 14.5] than the averaged $dn/dz$,
compared to none at $z$\,$>$\,1.5. The contrast between extremely high and low $dn/dz$ 
becomes more prominent at a higher column density range and at low redshifts.
A $dn/dz$ study based on 27 STIS/{\it FUSE} spectra
at $z$\,$>$\,0.02 \citep[filled orange squares]{tilton12} is consistent with our individual COS
$dn/dz$: although the STIS resolution is 3 times higher, its S/N is much lower and a different
method was used for estimating $N_{\mathrm{\ion{H}{i}}}$. Even though not shown, the D16 individual
$dn/dz$ also shows a large scatter.

This large scatter is in part intrinsic caused by
a stronger clustering of stronger absorbers (a large positive $\sigma_{dn/dz}$ combined with
a sightline without strong absorbers)
toward lower $z$ as a result of structure
evolution, cooled-down galactic outflows near star-forming galaxies and enhanced \ion{H}{i} ionizing
photons \citep{dobrzycki02, dallaglio08, dave10, nasir17}. 
The scatter is also in part caused by the different $z$ coverage for each sightline. This {\it redshift 
coverage bias}
is especially significant at $z$\,$\sim$\,0 and $z$\,$\sim$\,2 
as the wavelength coverage 
is smaller due to the rest-frame Ly$\alpha$ and atmospheric cutoffs, respectively.
The large scatter due to
both cosmic variance and redshift coverage bias implies that the $dn/dz$ study
requires many sightlines, especially at lower $z$. A small sample size is the primary 
reason of the earlier discrepancy between the FOS and STIS $dn/dz$ studies 
as the STIS $dn/dz$ was measured using only a few sightlines
\citep{lehner07, williger10}.

While the parameter space occupied by individual $dn/dz$ measurements on the $z$--$dn/dz$
plane is important for constraining the inhomogeneity of \ion{H}{i} distribution, the
{\it averaged} $dn/dz$ (filled circles) is a better quantity to directly compare to simulations which 
usually average thousands of sightlines. To reduce redshift coverage bias, the {\it averaged} 
$dn/dz$ is measured from the combined line lists of all the AGN per 
$N_{\mathrm{\ion{H}{i}}}$ and per $z$ instead of an arithmetic mean. 
Considering the large scatter in the individual $dn/dz$,
the commonly-used Poisson errors seem to underestimate the real errors. 
Therefore, we include the bootstrap error measured from the combined lines for each $z$ bin  
in addition to the Poisson errors.
The averaged $dn/dz$ is tabulated in Table~\ref{tab8} with the first $dndz$ error being the 
Poisson error and the second error being 0.5 times the standard deviation. 

In Fig.~\ref{fig16}, the green solid line is a single power-law fit to the {\it individual} $dn/dz$ at 
$0$\,$<$\,$z$\,$<$\,3.6. Due to the large scatter at a given $z$, this single power-law fit 
roughly describes the overall individual $dn/dz$ evolution, although a underprediction
of $dn/dz$ is suggested at $z$\,$>$\,3.6. 
The blue dot-dashed line shows a best-fit single power law to the averaged $dn/dz$ at 
$z$\,$>$\,1.5. At both $N_{\mathrm{\ion{H}{i}}}$ ranges for 
the Ly$\alpha$-only and Lyman series fits, 
this fit underpredicts the $dn/dz$ at $z$\,$<$\,0.5 \citep{weymann98, kim13}. 
The discrepancy is larger at the higher-$N_{\mathrm{\ion{H}{i}}}$ range, since stronger
absorbers are expected to disappear more rapidly at lower $z$ when extrapolated from high $z$. 
The red dashed lines represents 
a best-fit single power law to the {\it averaged} $dn/dz$ at $z$\,$<$\,1.5.
This fit underpredicts the observed
$dn/dz$ at $z$\,$>$\,1.5. The fit parameters are listed in Table~\ref{tab9}. 
Note that there is no significant difference between the
single power-law fits to the averaged $dn/dz$ (not shown) and 
the individual $dn/dz$ (green solid line).

For both $N_{\mathrm{\ion{H}{i}}}$ ranges, the inadequacy of a single power-law fit is
consistent with the evolution of $<$$F$$>$ and the PDF -- there exists an
IGM evolutionary break at 
$z$\,$\sim$\,1.5--1.7 and the stronger absorbers evolve more strongly, i.e. a larger $\gamma_{n}$
\citep{theuns98a, scott00a, kim13}. 
\citet{ribaudo11} find that the {\it averaged} $dn/dz$ 
of Lyman limit systems at 
$\log N_{\mathrm{\ion{H}{i}}}$\,$\ge$\,17.5 at $0.0$\,$<$\,$z$\,$<$\,2.6 is well
described with $\gamma_{n}$\,$=$\,1.33$\pm$0.61. Although 
the errors are large
for both studies and their $dn/dz$ does not show any evolutionary break at $z$\,$\sim$\,1.5, 
their $\gamma_{n}$ combined with ours at 0\,$<$\,$z$\,$<$\,3.6  
at $\log N_{\mathrm{\ion{H}{i}}}$\,$\in$\,[13.5, 14.0] ($\gamma_{n}$\,$\sim$\,1.10) 
and [14.0, 17.0] ($\gamma_{n}$\,$\sim$\,1.27)
suggests that $\gamma_{n}$ increases with $N_{\mathrm{\ion{H}{i}}}$.

At $\log N_{\mathrm{\ion{H}{i}}}$\,$\in$\,[14, 17] (right upper panel), 
the yellow shade represents the predicted $dn/dz$ evolution in terms of $N_{\mathrm{\ion{H}{i}}}$
instead of the equivalent width, compiled from various simulations
from outflow models to no-wind models under a quasars+galaxies UVB \citep{dave10, nasir17}. 
As outflows eject the processed gas into halos, 
which subsequently cools down and produces strong absorbers, outflow models tend to 
predict higher $dn/dz$. However, it is clear that these models significantly underpredict
the observed $dn/dz$ by a factor of $\sim$3--5, suggesting that saturated Ly$\alpha$ absorbers
at low redshift are not yet correctly simulated. 

In the same panel, the pink dot-dot-dot-dashed
line is a predicted $dn/dz$ by \citet{dave99} under the quasars-only UVB.
Their prediction is in a
strikingly good agreement with our measurement. This can indicate that the quasars-only
UVB is more preferable than the quasars+galaxies UVB under which latest simulations
including more recent models by \citet{dave10} do not reproduce the observations.
However, this comparison is potentially
complicated by the fact that their $\Lambda$-CDM model is based
on outdated cosmological parameters such as $\Omega_{\Lambda}$\,$=$\,0.6 without incorporating
any extra heating source such as \ion{He}{ii} photo-heating at $z$\,$>$\,2
\citep{syphers13, worseck16, nasir17} nor any stellar/AGN feedback. 
The simulation includes only $64^{3}$ particles in a small box of side length 11$h^{-1}$ 
comoving Mpc so that
it does not resolve the IGM gas particles as well as some of current 
IGM simulations \citep{dave10, nasir17, martizzi19}.

The lower left panel displays the difference in $dn/dz$ between the 
Ly$\alpha$-only and Lyman series fits. At $\log N_{\mathrm{\ion{H}{i}}}$\,$\in$\,[13.5, 14.5],
a few individual sightlines display a difference up to $\sim$20\%. However,
there is no noticeable difference in the average except at $z$\,$\sim$\,1 where the averaged $dn/dz$
suffers from small number statistics. 
At $\log N_{\mathrm{\ion{H}{i}}}$\,$\ge$\,14.5, Ly$\alpha$ lines 
start to saturate in the UVES/HIRES spectra. Since some saturated
lines can be resolved into several weaker components in higher order lines
and since the Ly$\alpha$-only fit in general gives a lower $N_{\mathrm{\ion{H}{i}}}$ limit for saturated
lines, the difference between the two sets becomes more noticeable at higher $N_{\mathrm{\ion{H}{i}}}$
(lower right panel). At $\log N_{\mathrm{\ion{H}{i}}}$\,$\in$\,[14, 17], the $dn/dz$ from the
Lyman series fit is about a factor of 1.2 larger than from the Ly$\alpha$-only fit.

\begin{table}
\caption{Power-law fit parameters to the number density $dn/dz$}
\label{tab9}
{\small{
\begin{tabular}{ccccc}

\hline\\[-0.3cm]
   & \multicolumn{2}{c}{Ly$\alpha$-only fit} & \multicolumn{2}{c}{Lyman series fit} \\
\cmidrule(lr){2-3}  \cmidrule(lr){4-5}  \\[-10pt]
$\log N_{\mathrm{\ion{H}{i}}}$ &   $\log n_{\mathrm{0}}$   &  $\gamma_{n}$ & $\log n_{\mathrm{0}}$   &  $\gamma_{n}$ \\

\hline\\[-0.3cm]
& \multicolumn{4}{c}{0\,$<$\,$z$\,$<$\,3.6 for the individual $dn/dz$} \\
\hline\\[-0.3cm]
13.5--14.5 &  1.57$\pm$0.03 &  1.15$\pm$0.05 &             1.60$\pm$0.03 &  1.10$\pm$0.05 \\
14.0--17.0 &  1.20$\pm$0.04 &  1.30$\pm$0.08 &             1.29$\pm$0.04 &  1.28$\pm$0.07 \\

\hline\\[-0.3cm]

& \multicolumn{4}{c}{$z$\,$>$\,1.5 for the averaged $dn/dz$} \\
\hline\\[-0.3cm]
 13.5--14.5  &  1.04$\pm$0.20 &  2.12$\pm$0.37 &        0.97$\pm$0.23 &  2.23$\pm$0.41 \\
 14.0--17.0  &  0.27$\pm$0.34 &  2.96$\pm$0.61 &        0.13$\pm$0.36 &  3.33$\pm$0.64 \\

\hline\\[-0.3cm]

& \multicolumn{4}{c}{$z$\,$<$\,1.5 for the averaged $dn/dz$} \\
\hline\\[-0.3cm]
 13.5--14.5  &  1.61$\pm$0.06 &  0.59$\pm$0.42 &       1.61$\pm$0.06 & 0.67$\pm$0.53 \\
 14.0--17.0  &  1.16$\pm$0.11 &  1.26$\pm$0.65 &       1.25$\pm$0.11 & 0.97$\pm$0.87 \\

\hline
\end{tabular}
}}
\end{table}

\section{Conclusions}
\label{sect6}

We performed a new uniform, consistent Voigt profile fitting analysis on the
84 high-quality AGN spectra from the {\it HST}/COS, {\it HST}/STIS,
VLT/UVES and Keck/HIRES archives in order to characterise the redshift 
evolution of the transmitted flux $F$ and column
density of neutral hydrogen \ion{H}{i} of the low-density IGM 
at 0\,$<$\,$z$\,$<$\,3.6. Although this data set does not sample the IGM
continuously in redshift space, the selected redshift ranges are the best compromise 
within the capabilities of currently available ground-based and space-based 
spectrographs: 

\begin{itemize}

\item {\bf VLT-UVES/Keck I-HIRES set} consists of 24 QSO spectra with
a resolution of $\sim$6.7\,\kms\/ and a S/N ratio per resolution element
of 40--250, sampling the IGM 
at 1.7\,$<$\,$z$\,$<$\,3.6 with the total $z$ coverage $\Delta z$\,$=$\,11.59
for the Ly$\alpha$-only fit range.
The typical $N_{\mathrm{\ion{H}{i}}}$
detection limit is $\log N_{\mathrm{\ion{H}{i}}}$\,$\sim$\,12.5.

\item {\bf {\it HST}/STIS+COS NUV set} covers the IGM 
at $z$\,$\sim$\,1 with $\Delta z$\,$=$\,1.27 (the Ly$\alpha$-only fit). The set includes 
two QSO spectra from the {\it HST}/STIS archive supplemented with our new
observations of three QSO spectra taken with the {\it HST}/COS NUV G225M grating.
The approximated Gaussian 
resolution of STIS E230M and COS NUV spectra is $\sim$10\,\kms\/ 
and $\sim$12\,\kms\/, respectively, with a non-Gaussian wing.
The $S/N$ range is $\sim$13--46. The $\log N_{\mathrm{\ion{H}{i}}}$
detection limit is $\sim$13.

\item {\bf {\it HST}/COS FUV set} has 55 AGN spectra with $S/N$\,$\sim$\,18--85
covering the IGM at 0\,$<$\,$z$\,$<$\,0.5 with $\Delta z$\,$=$\,7.20
(the Ly$\alpha$-only fit). The resolution can be approximated to
$\sim$19\,\kms\/ with a non-Gaussian wing. 
The $\log N_{\mathrm{\ion{H}{i}}}$ detection limit is $\sim$13.

\end{itemize}

For the continuous flux statistics, we used artificial spectra generated from the Ly$\alpha$-only fit 
since it can use a larger wavelength range than the Lyman series fit for which the 
useful wavelength is sometimes shortened due to the need for coverage of high-order Lyman lines.
The generated spectra also enable to 
combine the COS/STIS spectra 
having a non-Gaussian line spread function with the UVES/HIRES data having a Gaussian 
one and to remove the metal contamination. Our consistent analysis based on the best data currently
available confirms previous findings qualitatively \citep{weymann98, penton04, lehner07, tilton12, danforth16} 
and provides more robust quantitative results.
We have found:
 
\begin{enumerate}

\item The mean transmitted \ion{H}{i} flux is not sensitive to S/N, nor supposedly
undetected weak lines due to noise. While the flux PDF (probability distribution function, 
the fraction of pixels having a given normalized flux $F$) is not sensitive to undetected
weak lines, the flux PDF is directly comparable
among different S/N data only at 0.1\,$<$\,$F$\,$<$\,0.7.

\item The mean \ion{H}{i} flux 
increases fast at $z$\,$=$\,3.6\,$\rightarrow$\,1.5, slows down at $z$\,$\sim$\,1, then does not show 
any significant change at $z$\,$=$\,0.5\,$\rightarrow$\,0.0. 
A best-fit double power-law to the individual $<$$F$$>$ measurements is
$\ln <$$F$$>$\,$=$\,$(-0.0145\pm0.0003) \times (1+z)^{1.86\pm0.07}$ at $z$\,$<$\,1.5
and $\ln <$$F$$>$\,$=$\,$(-0.0040\pm0.0001) \times (1+z)^{3.18\pm0.02}$ at $z$\,$>$\,1.5,
respectively.

\item The mean PDF as a function of $F$ and $z$, $<$$P(F,z)$$>$,
qualitatively suggests that the volume fraction occupied by flux voids ($F$\,$\sim$\,1) increases 
rapidly at $z$\,$=$\,3.6\,$\rightarrow$\,1.5, then increases slowly at $z$\,$<$\,1.5.
With no absorption defined as $F$\,$=$\,1, this evolution reflects the thinning of the forest toward lower 
redshift, due to the evolution in the gas proper density and the intensity of the UV background.

\end{enumerate}

For the $N_{\mathrm{\ion{H}{i}}}$ distribution, we
used the Lyman series fit for more reliable determination of $N_{\mathrm{\ion{H}{i}}}$ for
saturated \ion{H}{i} Ly$\alpha$. For the UV spectra taken with the {\it HST}, a corresponding 
non-Gaussian LSF provided for each instrument setting and observation date is used.
At $\log N_{\mathrm{\ion{H}{i}}}$\,$\in$\,[13.5, 16.0] where incompleteness
is negligible, 24 UVES/HIRES spectra at 1.9\,$<$\,$z$\,$<$\,3.6, two STIS+three COS NUV spectra
at $z$\,$\sim$\,1
and 55 COS FUV spectra at 0\,$<$\,$z$\,$<$\,0.45 provide 
1798 (2043), 39 (93) and 371 (360) \ion{H}{i} lines, respectively,
for the Lyman series (Ly$\alpha$-only) fit.
We have found:

\begin{enumerate}

\item The redshift evolution of the column density distribution function (CDDF), 
albeit weak over a small $z$ range, is 
such that the overall shape of the CDDF at lower redshifts can be reproduced by 
a small amount of clockwise rotation of a higher-$z$ CDDF with a slightly larger normalisation
(bottom panels of Fig.~\ref{fig13} and left panel of Fig.~\ref{fig15}). 

\item For a conventional fit to the CDDF, $f$\,$\propto$\,$N_{\mathrm{\ion{H}{i}}}^{-\beta}$,
the slope $\beta$ at $\log N_{\mathrm{\ion{H}{i}}}$\,$\in$\,[13.5, 16.0] 
in general becomes steeper at lower $z$: 
$\beta$\,$=$\,1.60$\pm$0.02 at $z$\,$\sim$\,3.4 and $\beta$\,$=$\,1.82$\pm$0.03
at $z$\,$\sim$\,0.1. This reflects that higher-$N_{\mathrm{\ion{H}{i}}}$ 
absorbers disappear more rapidly and decrease in number or cross-section over time.

\item The slope $\beta$ is lower than the overall
trend at $z$\,$\sim$\,1 where
an evolutionary break in the flux statistics is seen and at $z$\,$\sim$\,3. 
The deviation at $z$\,$\sim$\,1 could be spurious due to
the small sample size, while the deviation at 
$z$\,$\sim$\,3 could be caused by a change in the $N_{\mathrm{\ion{H}{i}}}$ distribution 
due to extra heating and ionisation by the hypothetical \ion{He}{ii} reionisation at $z$\,$\sim$\,3. 
A further study with more data at $z$\,$\sim$\,1 and at $z$\,$>$\,3.6 is required to 
confirm the $\beta$ deviation. 

\item The {\it individual} $dn/dz$ (the number of absorbers per unit $z$)
shows a large scatter at a given $z$. The scatter 
increases toward lower $z$ and spans about an order of magnitude at $z$\,$\sim$\,0,
possibly caused by a combination of
a stronger clustering at lower $z$, outflows near star-forming galaxies, 
locally enhanced \ion{H}{i} ionization rates and a shorter redshift coverage of some sightlines. 

\item The {\it averaged} $dn/dz$
($dn/dz$\,$\propto$\,$(1\!+\!z)^{\gamma_{n}}$) is described better with 
a double power-law fit with an evolutionary break at $z$\,$\sim$\,1.5,
consistent with the evolution of transmitted flux. 
For the more reliable Lyman series fit, at $N_{\mathrm{\ion{H}{i}}}$\,$\in$\,[13.5, 14.5], 
$\gamma_{n}$\,$=$\,2.23$\pm$0.41
at $z$\,$>$\,1.5 and $\gamma_{n}$\,$=$\,0.67$\pm$0.53 at $z$\,$<$\,1.5, while
at $N_{\mathrm{\ion{H}{i}}}$\,$\in$\,[14, 17], $\gamma_{n}$\,$=$\,3.33$\pm$0.64
at $z$\,$>$\,1.5
and $\gamma_{n}$\,$=$\,0.97$\pm$0.87 at $z$\,$<$\,1.5, consistent with
the rapid disappearance of higher-$N_{\mathrm{\ion{H}{i}}}$ absorbers
with time.

\end{enumerate}

\section{Data availability}

The data underlying this article are available in the article and in its online supplementary
material. When the fitted line parameters will be completely analysed in our future papers, 
the entire fitted line list will be available online.

\section*{Acknowledgments}

We are grateful to everyone in the COS, STIS, UVES and HIRES instrument teams 
for building such superb spectrographs, which makes possible this study. 
This research has made use of the services of the ESO Science Archive Facility.
This also has made use of the Keck Observatory Archive (KOA), which is operated by 
the W. M. Keck Observatory and the NASA Exoplanet Science Institute (NExScI), under contract 
with the National Aeronautics and Space Administration. The authors wish to recognize and 
acknowledge the very significant cultural role and reverence that the summit of Mauna Kea 
has always had within the indigenous Hawaiian community.  We are most fortunate to have 
the opportunity to conduct observations from this mountain.
BPW acknowledges funding support provided by 
NASA through grant number 
HST-AR-12842.001-A from the Space Telescope Science Institute, which is operated
by AURA, Inc., under NASA contract NAS 5-26555.
TSK acknowledges funding support by
{\it HST} GO grant HST-GO-14265.001-A from STScI
and the European Research Council Starting Grant ``Cosmology with the IGM"
through grant GA-257670. 
MV is supported by INFN-PD51 grant INDARK and  
from the agreement ASI-INAF n.2017-14-H.0. JCC was supported by 
HST-GO-14265.004-A from STScI. 

\bibliography{literature}

\end{document}